%% file: 00_main.tex
\documentclass[acmtog,screen]{acmart}

\newif\ifmarked

\author{He Chen}
\orcid{0000-0002-5819-3453}
\email{ankachan92@gmail.com}
\affiliation{%
  \institution{University of Utah}
  \city{Salt Lake City}
  \state{UT}
  \country{USA}
}

\author{Elie Diaz}
\orcid{0009-0002-9493-1684}
\email{elie.diaz@utah.edu}
\affiliation{%
  \institution{University of Utah}
  \city{Salt Lake City}
  \state{UT}
  \country{USA}
}

\author{Cem Yuksel}
\orcid{0000-0002-0122-4159}
\email{cem@cemyuksel.com}
\affiliation{%
  \institution{University of Utah \& Roblox}
  \city{Salt Lake City}
  \state{UT}
  \country{USA}
}

\usepackage{booktabs} 
\newtheorem{theorem}{Theorem}
\newtheorem{definition}{Definition}
\usepackage{tabularx}
\usepackage{booktabs}
\usepackage[labelfont=bf,format=plain,singlelinecheck=false]{caption}
\usepackage{float}
\usepackage[ruled,linesnumbered]{algorithm2e}

\usepackage{amsmath}
\setcopyright{acmlicensed}
\acmJournal{TOG}
\acmYear{2023} \acmVolume{42} \acmNumber{4} \acmArticle{1} \acmMonth{8} \acmPrice{15.00}\acmDOI{10.1145/3592136}

\DeclareGraphicsExtensions{.pdf,.jpg,.png}  

\usepackage{subcaption}
\newcommand{\commentText}[1]{#1}
\newcommand{\todo}[1]{\commentText{{\color{red}[\textbf{\textsc{TODO}}: \textit{#1}]}}}

\newcommand{\cem}[1]{\commentText{{\color{orange}[\textbf{\textsc{Cem}}: \textit{#1}]}}}

 \newcommand{\add}[1]{#1}
 \newcommand{\del}[1]{\ignorespaces}

\begin{document}
\title{Shortest Path to Boundary for Self-Intersecting Meshes}



\begin{abstract}
We introduce a method for efficiently computing the exact shortest path to the boundary of a mesh from a given internal point in the presence of self-intersections. We provide a formal definition of shortest boundary paths \del{in the context of}\add{for} self-intersecting objects and present a robust algorithm for computing the actual shortest boundary path. The resulting method offers an effective solution for collision and self-collision handling \del{for}\add{while} simulating deformable volumetric objects, using fast simulation techniques that provide no guarantees on collision resolution. Our evaluation includes complex self-collision scenarios with a large number of active contacts, showing that our method can successfully handle them by introducing a relatively minor computational overhead.
\end{abstract}

%
%


%
%


\newcommand{\TODO}[1]{\textcolor{red}{ToDo:#1}}
\renewcommand{\vec}[1]{\mathbf{#1}}
\newcommand{\rest}[1]{\overline{#1}}
\newcommand{\bound}[1]{\partial#1}
\newcommand{\inter}[1]{#1^\circ}

\newcommand{\restbf}[1]{\overline{\mathbf{#1}}}

\setlength{\fboxsep}{0pt}
\setlength{\fboxrule}{0.2pt}

\begin{teaserfigure}
\centering
\begin{minipage}{0.6\linewidth}\centering
\includegraphics[width=\linewidth,trim=100 330 180 420,clip]{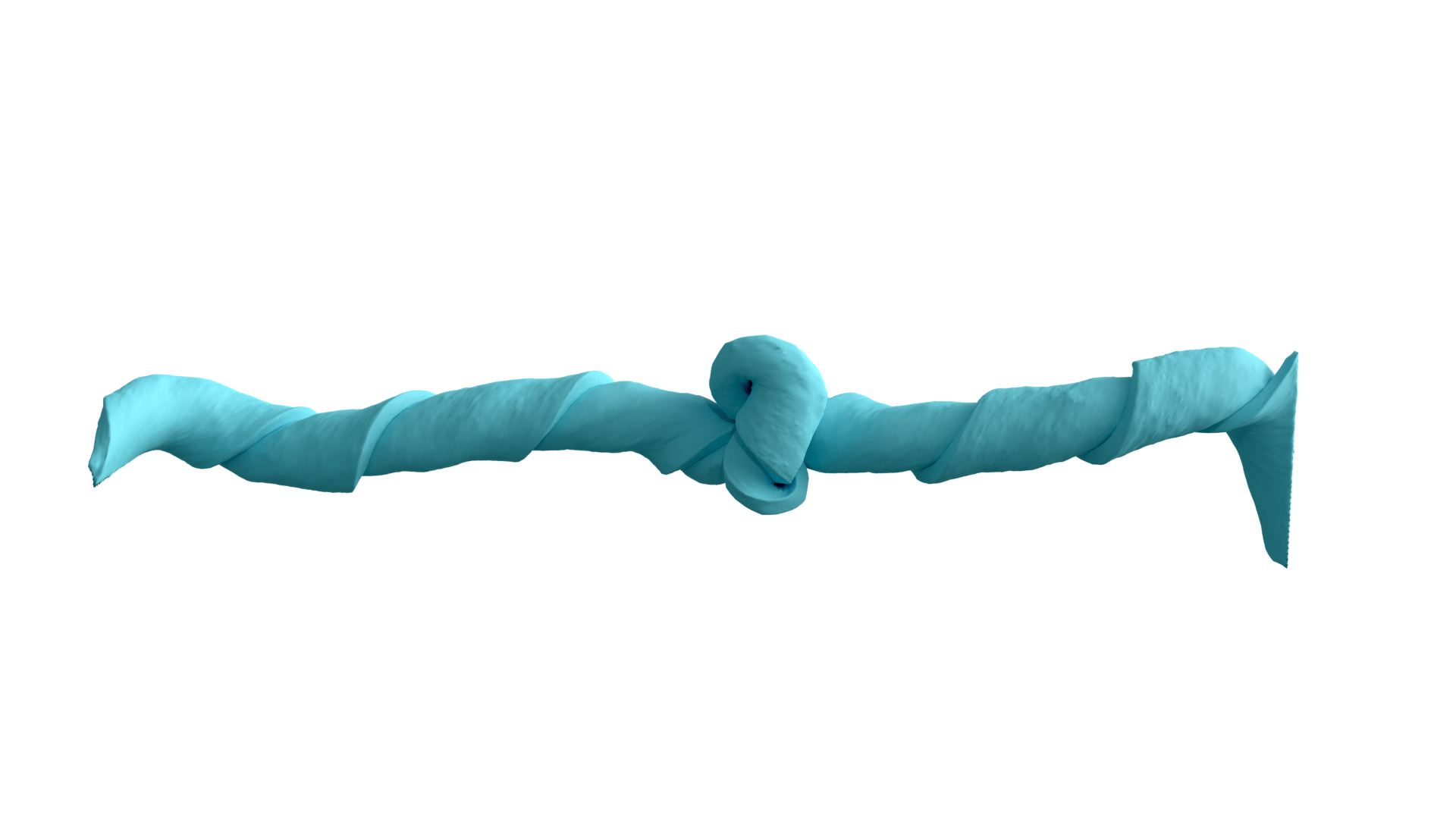}
\includegraphics[height=0.3\linewidth,trim=340 200 300 200,clip]{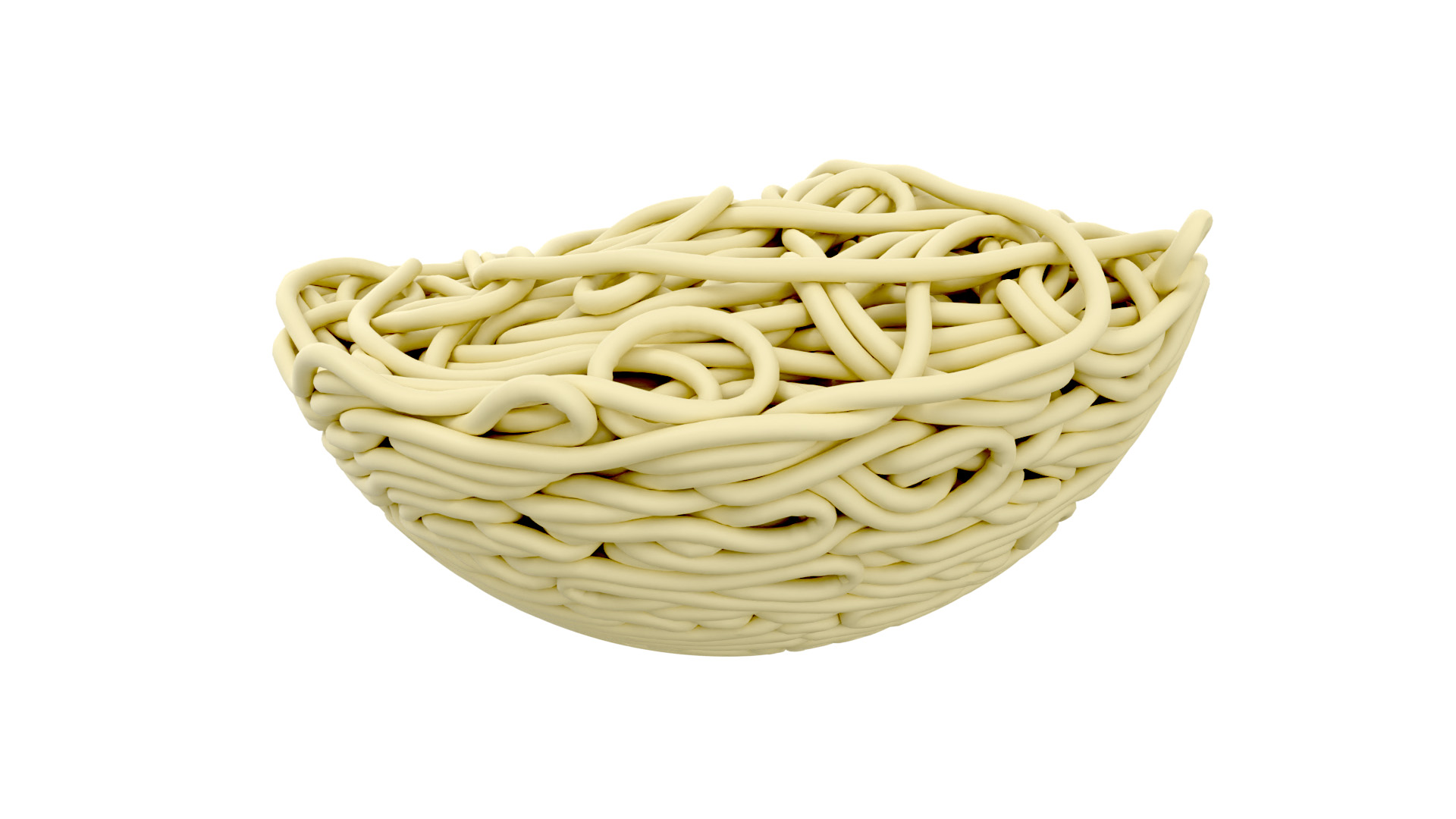}%
\includegraphics[height=0.3\linewidth,trim=340 200 500 200,clip]{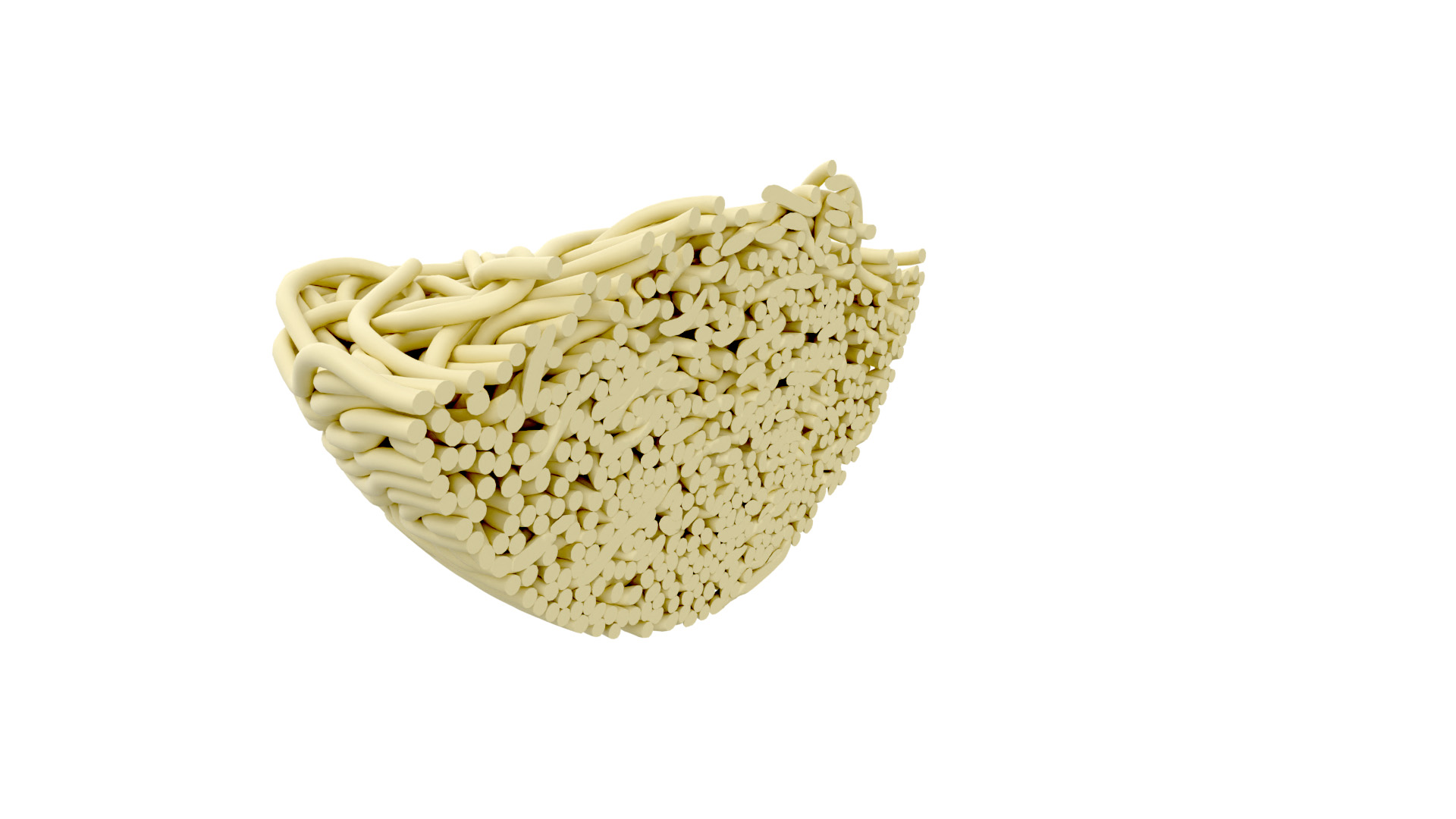}
\end{minipage}%
\begin{minipage}{0.4\linewidth}\centering
\includegraphics[width=\linewidth,trim=550 150 450 250,clip]{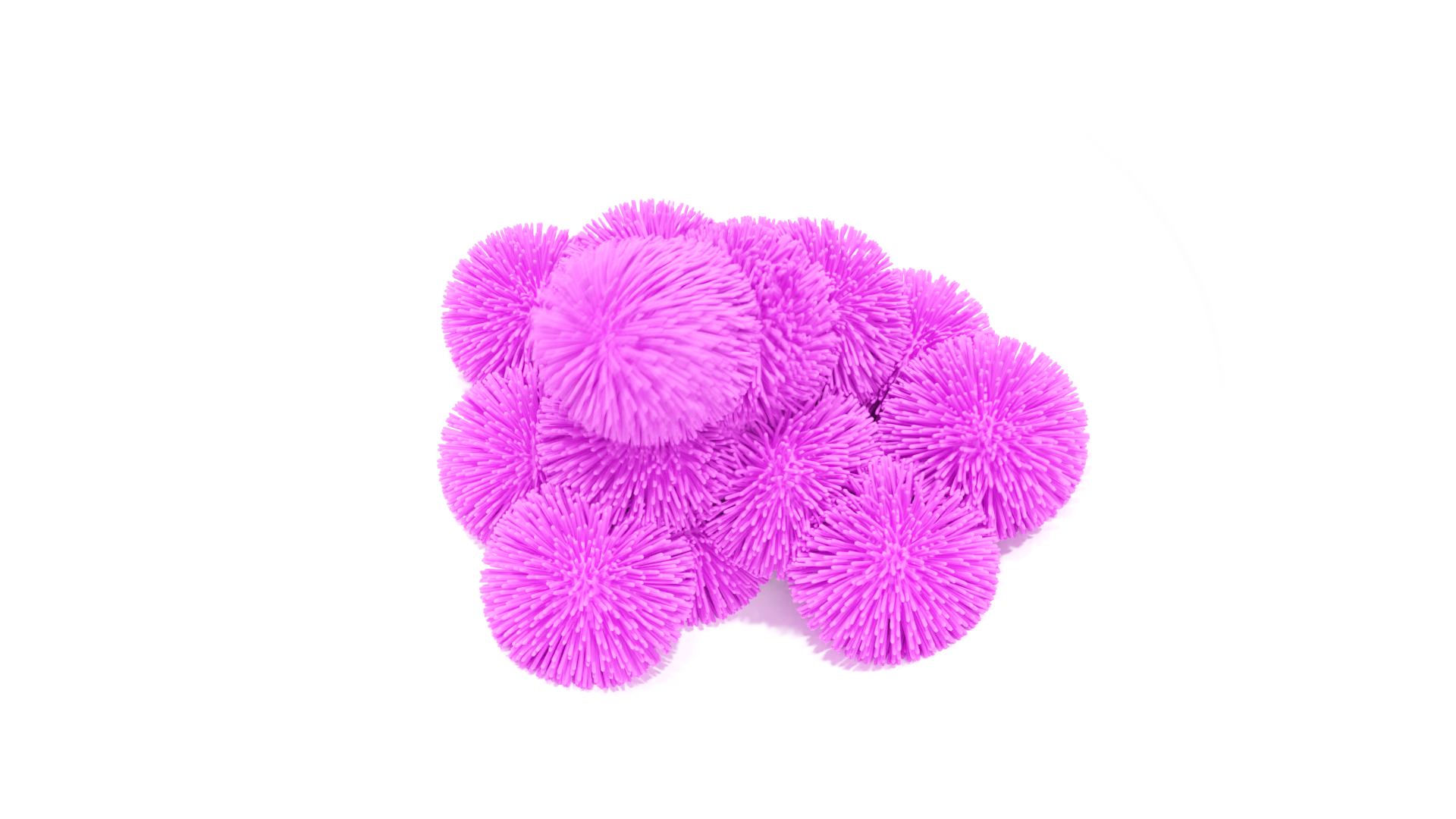}
\end{minipage}
\captionof{figure}{Example simulation results involving complex self-collision scenarios, generated using our method with XPBD \cite{macklin2016xpbd}.}
\label{fig:teaser}
\end{teaserfigure}
\maketitle
\input{01_Intoduction}
\input{02_RelatedWork}
\input{03_ShortestPathToSurface}
\input{04_FeasibleRegionAceleration.tex}

\input{05_CollisionHandling}

\input{06_Results}
\input{07_Limitation.tex}

\section{Conclusion}

We have presented a formal definition of the shortest path to boundary in the context of self-intersections and introduced an efficient and robust algorithm for finding the exact shortest boundary paths for meshes. We have shown that this approach provides an effective solution for handling both self-collisions and inter-object collisions using DCD in combination with CCD, using a simulation system that does not provide any guarantees about resolving the collision constraints. Our results show highly complex simulation scenarios involving collisions and rest-in-contact conditions that are properly handled with our method with a relatively small computational overhead.

\begin{acks}
We thank Alper Sahistan, Yin Yang, and Kui Wu for their helpful comments and suggestions. 
We also thank Alec Jacobson and Tiantian Liu for providing online volumetric mesh datasets.
This project was supported in part by NSF grant \#1764071.
\end{acks}

\bibliographystyle{ACM-Reference-Format}
\bibliography{references}
\input{Appendix}

\end{document}

%% file: 01_Intoduction.tex
\section{Introduction}

Self-intersecting meshes, though they are often highly undesirable, are commonplace in computer graphics. They can appear due to the limitations of modeling techniques, animation methods, or manual editing operations. Even physics-based simulations with self-collision handling are not immune to self-intersections, as most of them cannot guarantee an intersection-free state.

Notwithstanding the amount of work on self-intersection handling within physics-based simulations, it still remains a challenge in most cases. \emph{Continuous collision detection} techniques \cite{li2020incremental} require starting with and maintaining an intersection-free state; therefore, they must be used with computationally-expensive methods that can always resolve all self-intersections and they fail when combined with cheaper techniques that are unable to do so.
Methods that split an object into pieces \cite{macklin2020local} turn the self-intersection problem into intersections of these separate pieces, entirely avoiding the self-intersection problem, and they fail to resolve self-intersections within a piece.
Methods that solve self-intersections using an intersection-free pose \cite{mcadams2011efficient} not only require such a pose, but also become inaccurate as the objects deform and fail with sufficiently large deformations and deep penetrations.
Therefore, none of these methods provides a robust and general solution for self-intersections.


In this paper, we present a method that 
\del{can be used for solving arbitrary self-intersections after they appear, which works with tetrahedral meshes in 3D (with boundaries forming triangular meshes) and triangular meshes in 2D (with polyline boundaries). Our method}
robustly and efficiently finds the \emph{exact shortest internal path} of a point inside a mesh to its boundary, even in the presence of self-intersections and some inverted elements.
We achieve this by introducing a precise definition of the shortest path to the mesh boundary, including points that are both on the boundary and inside the mesh at the same time, an unavoidable condition with self-intersections.
\add{Our approach works with tetrahedral meshes in 3D (with boundaries forming triangular meshes) and triangular meshes in 2D (with polyline boundaries). We demonstrate that one important application of our method is solving arbitrary self-intersections after they appear in deformable simulations, allowing the use of cheaper integration techniques that do not guarantee complete collision resolution.}

Our method is based on the realizations that (1)~the shortest path 
must be fully contained within the geodesic embedding of the mesh and (2)~it must be a line segment under Euclidean metrics. Based on these, given a candidate boundary point, our method quickly checks if the line segment to this point is contained within the mesh. Combined with a spatial acceleration structure, we can efficiently find and test the candidate closest boundary points until the shortest path is determined. We also describe a fast and robust tetrahedral traversal algorithm that avoids infinite loops, needed for checking if a path is within the mesh.
Furthermore, we propose an additional acceleration that can quickly eliminate candidate boundary points based on local geometry without the need for checking their paths.

\del{Our solution can be used as the sole method for}\add{One application of our method is} resolving intersections between separate objects and self-intersections alike within a fast physics-based simulation system that cannot guarantee intersection-free states. \del{Alternatively, it}\add{It} can be used \add{alone or} as a backup for continuous collision detection to handle cases when the simulation system fails to resolve a previously-detected collision. In either case, we achieve a robust collision handling method that can solve extremely challenging cases, involving numerous deep self-intersections, using a fast simulation system that does not provide any guarantees about collision resolution. As a result, we can simulate highly complex scenarios with a large number of self-collisions and rest-in-contact conditions, as shown in \autoref{fig:teaser}.

%% file: 02_RelatedWork.tex
\section{Related Work}


\del{An immediate}\add{One important} application of our method is collision handling (\autoref{sec:prior:collision}), though we actually introduce a method for certain types of geodesic distances and paths (\autoref{sec:prior:geodesic}). A core part of our method is tetrahedral ray traversal (\autoref{sec:prior:tettraverse}). In this section, we overview the prior in these areas and briefly present how our approach compares to them.

\subsection{Collision Handling}
\label{sec:prior:collision}

Collision handling is directly related to how they are detected, which can be done using either
\emph{continuous collision detection} (CCD) or \emph{discrete collision detection} (DCD).

Starting with an intersection-free state, CCD can detect the first time of contact between elements \cite{canny1986collision}, but requires maintaining an intersection-free state.
Through the use of a strong 
barrier function, \emph{incremental potential contact} (IPC) \cite{li2020incremental} provides guaranteed 
collision resolution 
combined with a CCD-aware line search. 
This idea was later extended to rigid
\cite{ferguson2021intersection} and 
almost rigid
bodies \cite{lan2022affine}.
Incorporating projective dynamics into IPC offers 
performance improvement \cite{lan2022penetration}, but resolving all collisions still remains expensive.
Even when the simulation system is able to resolve all collisions, 
CCD itself can fail due to numerical issues,
in which case, it can no longer help with resolving the collision, resulting in objects linking together \cite{wang2021large}.

In contrast, DCD allows the simulation framework to start and recover from a state with existing intersections. 
DCD detects collisions at a single point in time, after they happen.
That is why, extra computation is needed to determine how to resolve the collisions.

Collisions can be resolved by 
minimizing the penetration volume \cite{allard2010volume, wang2012adaptive}
or by applying
constraints \cite{bouaziz2014projective, muller2007position, macklin2016xpbd, verschoor2019efficient}, penalty forces \cite{ding2019penalty, hunvek1993penalty, belytschko1991contact, drumwright2007fast}, or impulses \cite{kavan2003rigid, o1999real, mirtich1995impulse}
that involve computing the \emph{penetration depth},
the minimum translational distance to resolve the penetration \cite{terzopoulos1987elastically, platt1988constraints, hirota2000simulation}. 
The exact penetration depth can be computed
using analytical methods based on geometric information of polygonal meshes \cite{moore1988collision, hahn1988realistic, baraff1994fast, cameron1997enhancing},
or it can be approximated using a volumetric mesh \cite{fisher2001deformed}, \add{mesh partitioning \cite{Redon2006AFM}}, tracing rays \cite{hermann2008ray}, or solving an optimization problem \cite{je2012polydepth}. \citet{heidelberger2004consistent} proposed a consistent penetration depth by propagating penetration depth through the volumetric mesh. 
These methods, however, struggle with handling self-intersections.
\add{Starting with a self-intersecting shape, \citet{li2018immersion} proposed a method to separate the overlapping parts and create a bounding case mesh that represents the underlying geometry to allow "un-glued" simulation.}

Using a \emph{signed distance fields} (SDF) is a more popular alternative for recent methods. They 
can be defined either on a volumetric mesh \cite{fisher2001deformed} or a regular grid \cite{macklin2020local, koschier2017hp, gascuel1993implicit}. 
Once built, both the penetration depth and the shortest path to the surface can be directly queried from the volumetric data structure.
This provides an efficient solution at run time as long as the SDF does not need updating, though the returned penetration depth and shortest path are approximations (formed by interpolating pre-computed values). Also, the SDF is not well defined when there are self-intersections, as they cannot represent immersion, so it must be built using an intersection-free pose.


For handling self-intersections, SDFs of an intersection-free pose can be used \cite{mcadams2011efficient}. This can provide sufficient accuracy for handling minor deformations, but quickly becomes inaccurate with large deformations and deep penetrations. Using a deformable embedding helps \cite{macklin2020local}, but requires splitting the object into pieces \cite{fisher2001deformed, fisher2001fast, mcadams2011efficient, macklin2020local, teng2014simulating}. An alternative approach is bifurcating the SDF nodes during construction when a volumetric overlap, which can be formed by self-intersection, is detected \cite{mitchell2015non}. These solutions entirely circumvent the self-intersection problem by only considering intersections of separate pieces and self-intersections within a piece are ignored. Such approaches are particularly problematic with complex models and in cases when determining where to split is unclear ahead of time, 
since the splitting or bifurcation is usually pre-computed and expensive to update at run time. 
Also, the closest boundary point found within a piece is not necessarily the actual one for the entire mesh, as it might be contained in a separate piece.
Even for cases they can handle with sufficient accuracy, SDFs have a significant pre-computation and storage cost. 

In comparison, our solution can find the \emph{exact} penetration depth for models with arbitrary complexity and the accurate shortest path to the boundary regardless of the type or severity of self-intersections. In addition, we do not require costly pre-computations or volumetric storage.


\subsection{Geodesic Path and Distances}
\label{sec:prior:geodesic}

Following the categorization of \citet{crane2020survey}, our method falls into the category of \emph{multiple source geodesic distance/shortest path} (MSGD/MSSP) problems. Actually, the problem we solve is a special case of MSSP, where the set of sources is the collection of all the boundary points of the mesh. Also, 
ours is an exact polygonal method that can compute global geodesic paths. MMP algorithm \cite{mitchell1987discrete} is the first practical algorithm that can compute geodesic path between any two points on a polygonal surface. Succeeding methods \cite{chen1990shortest, surazhsky2005fast, liu2013exact, xin2009improving} focus on optimizing its computation time and memory requirements. Yet, all of these method only aim at solving the \emph{single source geodesic distance/shortest path} (SSGD/SSSP) problems. 
For solving the \emph{all-pairs geodesic distances/shortest paths} (APGD/APSP) problem, a vertex graph that encodes the minimal geodesic distances between all pairs of vertices on the mesh can be built \cite{balasubramanian2008exact}. 
These methods are general enough for handling 2D manifolds in 3D, but they do not offer an efficient solution for our MSSP problem. 
Our solution for MSSP, however, is limited to planar (2D, triangular) or volumetric (3D, tetrahedral) meshes, where we can rely on Euclidean metrics. 

\subsection{Tetrahedral Ray Traversal}
\label{sec:prior:tettraverse}

For handling tetrahedral meshes in 3D, our method uses a topological ray traversal. Tetrahedral ray traversal
has been used in volumetric rendering 
\cite{marmitt2006fast, parker2005interactive, csahistan2021ray}. 
Methods that improve their computational cost include using scalar triple products \cite{lagae2008accelerating} and Plucker coordinates \cite{maria2017efficient}. More recently, \citet{aman2022compact} introduced a highly-efficient dimension reduction approach.

A common problem with tetrahedral ray traversal is that numerical inaccuracies can lead to infinite loops when a ray passes near an edge or vertex. Many rendering problems can safely terminate when an infinite loop is detected. In our case, however, we must detect and resolve such cases, because failing to do so would result in returning an incorrect shortest path, which can have catastrophic effects in simulation. Therefore, we introduce a robust variant of tetrahedral ray traversal.

%% file: 03_ShortestPathToSurface.tex
\section{Shortest Path to Boundary}

A typical solution for resolving intersections (detected via DCD) is finding the closest boundary point for each intersecting point and then applying corresponding forces/constraints along the line segment toward this point, i.e. the shortest path to boundary. The length of this path is the penetration depth.

When two separate objects intersect, finding the closest boundary point is a trivial problem: it is the closest boundary point on the other object. In the case of self-intersections, however, even the definition of the shortest path to boundary is somewhat ambiguous.

Consider a point on the boundary and also inside the object due to self-intersections. Since this point is already on the boundary, its Euclidean closest boundary point would be itself. Yet, this information is not helpful for resolving the self-intersection.

In this section, we provide a formal definition of the shortest path to boundary based on the geodesic path of the object in the presence of self-intersections (\autoref{sec:derivation}). Then, we present an efficient algorithm to compute it for triangular/tetrahedral meshes in 2D/3D, respectively, (\autoref{sec:ShortestPathToBoundaryForMeshes}). We also describe how to handle meshes that contain some inverted elements, (\autoref{sec:inversion}). The resulting method provides a robust solution for handling self-collisions that can be used with various simulation methods and collision resolution techniques (using forces or constraints).

\begin{figure}
    \centering
    \includegraphics[width=\columnwidth]{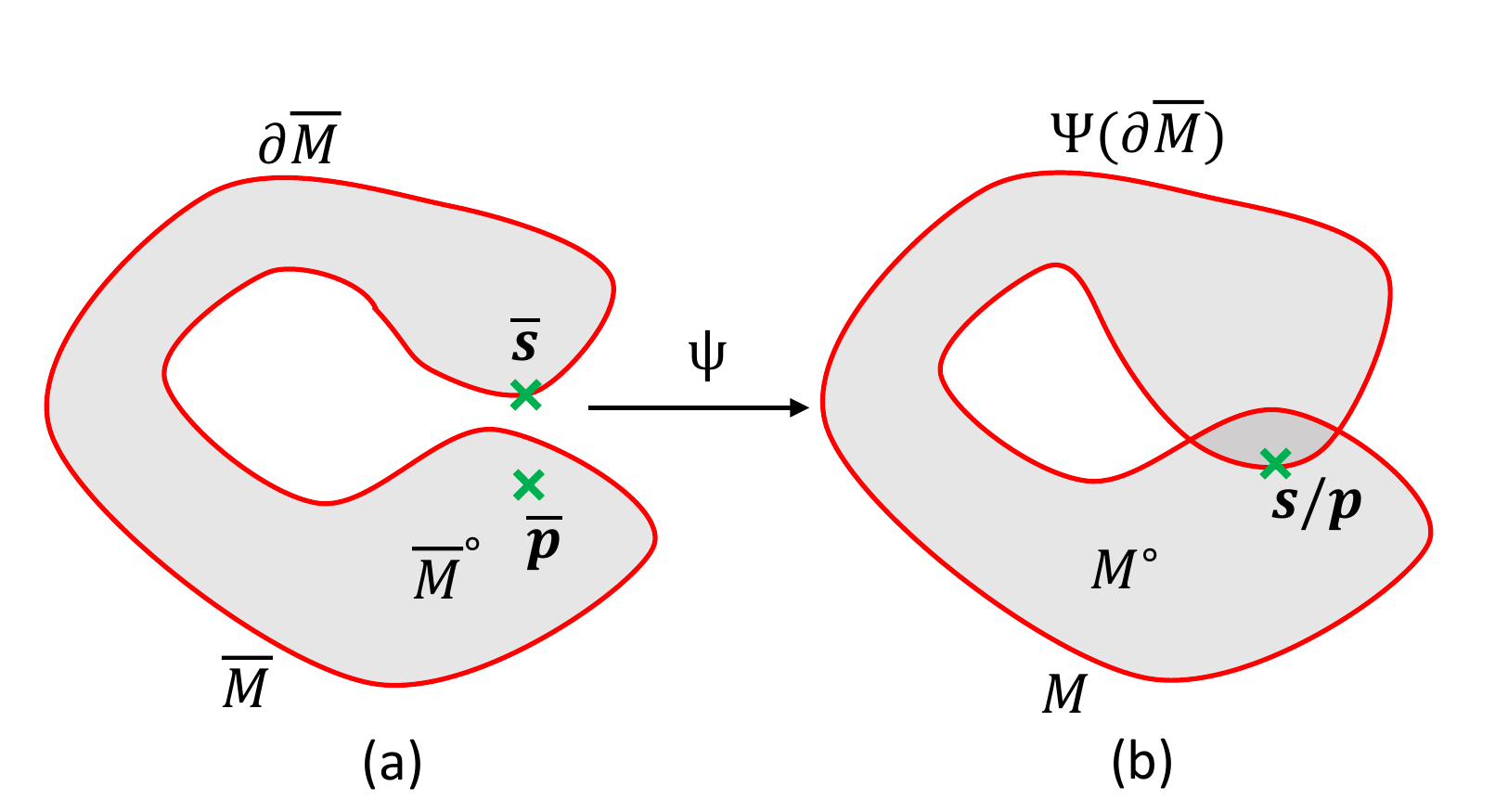}
    \caption{Illustrations of the notations. (a)~Notations on the undeformed pose. (b)~Notations on the deformed model. The image of the undeformed pose boundary $\Psi(\partial\rest{M})$ is marked as the \del{blue}\add{red} curve\del{, and the boundary of the deformed model $\partial lM$ is marked as the dotted blue line}.}
    \label{fig:NotationIllustration}
\end{figure}

\begin{figure*}[ht!]
    \centering
    \includegraphics[width=\textwidth]{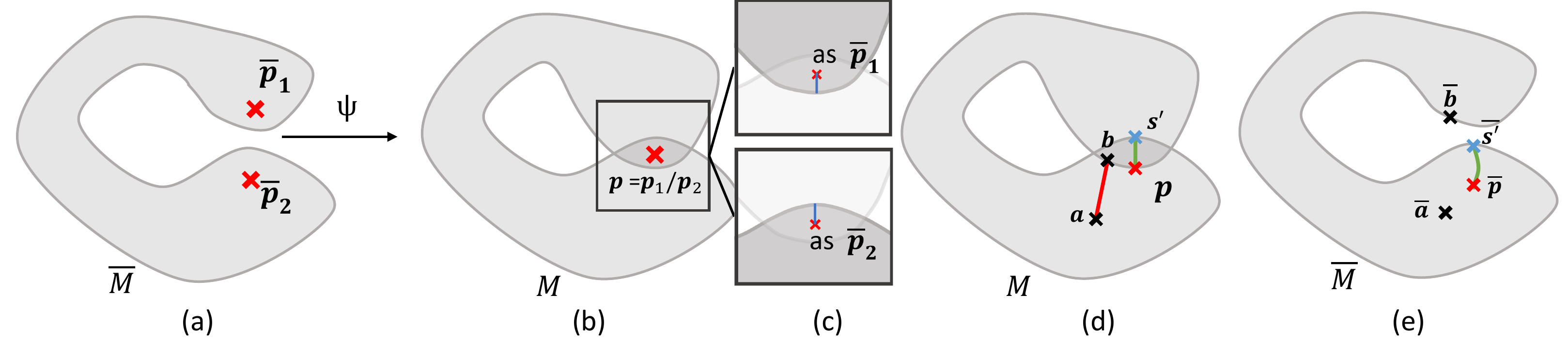}
    \caption{(a) An intersection-free pose of the deformable model $\rest{M}$. $\rest{\vec{p}}_1, \rest{\vec{p}}_2\in \rest{M}^\circ$. (b) $\rest{M}$'s image under $\Psi$, where $\rest{\vec{p}}_1, \rest{\vec{p}}_2$ are mapped to the same point $\vec{p}$. (c) Treated as different pre-image, $\vec{p}$ has different shortest paths to the boundary (blue line). (d) Two paths are contained by $M$. (e) Only $\vec{ps'}$ is a valid path.}
    \label{fig:GlobalGeoDesicPath}
\end{figure*}

\subsection{Shortest Path to Boundary}
\label{sec:derivation}



Consider a self-intersecting model $M$, such that a boundary point $\vec{s}$ coincides with an internal point $\vec{p}$. \autoref{fig:NotationIllustration}b shows a 2D illustration, though the concepts we describe here apply to 3D (and higher dimensions) as well. In this case, $\vec{s}$ and $\vec{p}$ have the same geometric positions, 
but topologically they are different points. In fact, to fix the self-intersection, we need to apply a force/constraint that would move $\vec{s}$ along $\vec{p}$'s geodesic shortest path to boundary.

To provide a formal definition of this geodesic shortest path, we consider a self-intersection-free form of this model as $\rest{M}$ which we call undeformed pose, and a deformation $\Psi$ that maps all points in $\rest{M}$ to its current shape $M$, such that $M=\Psi(\rest{M})$.
Note that our algorithm (explained in \autoref{sec:ShortestPathToBoundaryForMeshes}) does not actually need computing $\rest{M}$ or $\Psi$.
For any point $\rest{\vec{p}}$  in $\rest{M}$, we represent its image under $\Psi$ as $\vec{p}\in M$, such that ${\vec{p}=\Psi(\rest{\vec{p}})}$.
In the following, we assume that $\rest{M}$ is a path-connected (i.e. a single piece) manifold, though the concepts below can be trivially extended to models with multiple separate pieces. 

%
To cause self-intersection, $\Psi$ should not be injective. In this case, $\Psi$ is an immersion of $\rest{M}$ but not embedding, meaning multiple points from $\rest{M}$ are mapped to the same position $\vec{p}$ inside $M$.
To differentiate such points that coincide in $M$, we label them using their unique positions in $\rest{M}$. 
For simplicity, we say $\vec{p}$ \emph{as} $\rest{\vec{p}}$, when we are referring $\Vec{p}$ as the image of $\rest{\vec{p}}$.

For simplicity, let us consider non-degenerate $\Psi$ that forms no inversion, i.e. $\det(\nabla\Psi)>0$. We discuss inversions later in \autoref{sec:inversion}.
Note that under this $\Psi$, the boundary of the undeformed model $\bound{\rest{M}}$ does not completely overlap with the boundary of the deformed model $\bound{M}$, i.e. ${\Psi(\bound{\rest{M}})\neq\bound{M}}$, see \autoref{fig:NotationIllustration}b. 
We use $\rest{M}^\circ$ to denote the set of interior points of $\rest{M}$, such that ${\rest{M}=\bound{\rest{M}}\cup\inter{\rest{M}}}$.



Let $\vec{s}$ as  be a point on the boundary, i.e. $\vec{s}\in\Psi(\bound{\rest{M}})$ and we refer to it as an undeformed pose boundary point $\rest{\vec{s}}$. For a given point $\vec{p}$ (as $\rest{\vec{p}}$), we can construct a path $\vec{c}(t): [0,1] \mapsto M$  as a continuous curve that connects $\vec{p}=\vec{c}(0)$ to $\vec{s}=\vec{c}(1)$.


\begin{definition}[Valid path]
The path $\vec{c}(t)$ from  $\vec{p}$ (as $\rest{\vec{p}}$) to $\vec{s}$ (as $\rest{\vec{s}}$) is a \emph{valid path} if there exists a continuous curve $\rest{\vec{c}}(t): [0,1] \mapsto \rest{M}$ such that $\vec{c}(t)=\Psi(\rest{\vec{c}}(t))$, $\rest{\vec{c}}(0)=\rest{\vec{p}}$, $\rest{\vec{c}}(1)=\rest{\vec{s}}$.
\end{definition}

Based on this definition, a \emph{valid path} must be the image of a path that is fully contained within $\rest{M}$, which connects the two points on the undeformed pose we are referring to. Any path that moves outside of $\rest{M}$ is considered an \emph{invalid path}, see \autoref{fig:GlobalGeoDesicPath}de. Our goal is to find the shortest valid path from a given point $\vec{p}$ (as $\rest{\vec{p}}$) to the boundary.

\begin{definition}[Shortest path to boundary]
For an interior point $\vec{p}$ (as $\rest{\vec{p}}$), the \emph{shortest path to boundary} is the shortest curve $\vec{c}(t)$ in $M$ that connects $\vec{p}$ to a boundary point $\vec{s}$ (as $\rest{\vec{s}}$) that is a valid path between $\vec{p}$ 
and $\vec{s}$.
\end{definition}

\begin{definition}[Closest boundary point]
For an interior point $\vec{p}$ (as $\rest{\vec{p}}$), the \emph{closest boundary point} is the boundary point $\vec{s}$ (as $\rest{\vec{s}}$) 
at the other end of $\vec{p}$'s shortest path to boundary $\vec{c}(t)=\Psi(\rest{\vec{c}}(t))$,
such that $\vec{s}=\vec{c}(1)$ and $\rest{\vec{s}}=\rest{\vec{c}}(1)$.
\end{definition}

Here we must emphasize that the definition of the shortest path is dependent on the pre-image point we are referring to. For a point located at the overlapping part of $M$, referring to it as a different point on the undeformed pose may lead to a different shortest path to the boundary (see \autoref{fig:GlobalGeoDesicPath}c).
Also, this definition is equivalent to the image of $\rest{\vec{p}}$'s global geodesic path to boundary in $\rest{M}$ evaluated under the metrics pulled back by $\Psi$. Thus the shortest path we defined is a special class of geodesics.

To construct an efficient algorithm for finding the shortest path, we rely on two properties:
\begin{itemize}
    \item First, by definition, the shortest path must be a continuous curve that is fully contained inside undeformed model $\rest{M}$.
    \item Second, the shortest path (under the Euclidean distance metrics) that connects two points in the deformed model $M$ must be a line segment.
\end{itemize}

Based on these properties, we can construct and prove the fundamental theorem of our algorithm:
\begin{theorem}
\label{thm:shortestPath}
For any point $\vec{p}\in M$ (as $(\rest{\vec{p}}$), its shortest path to the boundary is the shortest line segment from $\vec{p}$ \del{and}\add{to} a boundary point $\vec{s} \in  \Psi(\partial{\rest{M}})$ (as $\rest{\vec{s}}$), that is a valid path.
\end{theorem}
Here we verbally prove the theorem, we also provide a formal proof in the appendix. 
If the shortest path is not a line segment, we can continuously deform it into a line segment, while keeping the end points fixed. This procedure can induce a deformation on the undeformed pose, which continuously deforms the pre-image of that curve to the pre-image of the line segment, while keeping the end points fixed. This is always achievable because the curve cannot touch the boundary of the undeformed pose during the deformation, otherwise, we will form an even shorter path to the boundary. Thus the line segment is also a valid path.

Based on these properties, our algorithm investigates a set of candidate boundary points $\vec{s}$ and checks if the line segment from the interior point $\vec{p}$ to $\vec{s}$ is a valid path. This is accomplished without having to construct $\rest{M}$ or determine the deformation $\Psi$ by relying on the topological connections of the given discretized model.

\subsection{Shortest Path to Boundary for Meshes}
\label{sec:ShortestPathToBoundaryForMeshes}
In practice, models we are interested in are discretized in a piecewise linear form. These are triangular meshes in 2D and tetrahedral meshes in 3D. We refer to each piecewise linear component as an \emph{element} (i.e. a triangle in 2D and a tetrahedron in 3D) and the one-dimension-lower-simplex shared by two topologically-connected elements as a \emph{face} (i.e. an edge between two triangles in 2D and a triangular face between two tetrahedra in 3D).
This discretization makes it easy to test the validity of a given path, without constructing a self-intersection-free $\rest{M}$ or the related deformation $\Psi$.

We propose the concept of \textit{element traversal} for meshes, as a sequence of topologically connected elements:
\begin{definition}[Element traversal]
For a mesh $M$, and two-point $\vec{a}\in e_{\vec{a}}, \mathbf{b}\in e_{\vec{b}}$, we define a element traversal from $\mathbf{a}$ to $\mathbf{b}$ as a list of elements $\mathcal{T}(\mathbf{a}, \mathbf{b}) = (e_0, e_1, e_2, \dots, e_k)$, where $e_i$ is a element of $M$, $e_0= e_{\vec{a}}$, $e_{k}=e_{\vec{b}}$, and $e_i \cap e_{i+1}$ must be a  face.
\end{definition}
Specifically, we call it \textit{tetrahedral traversal} for 3D meshes, and \textit{triangular traversal} for 2D meshes. 


Let $\vec{c}(t)$ be a line segment from a point $\vec{p}$ inside an element $e_\vec{p}$ to a boundary point $\vec{s}$ of a boundary element $e_\vec{s}$ (with a boundary face that contains $\vec{s}$). If $\vec{c}(t)$ is a valid path, there must be a corresponding piecewise linear path $\rest{\vec{c}}(t)$ in $\rest{M}$ from $\rest{\vec{p}}$ to $\rest{\vec{s}}$ that passes through an element traversal of $\rest{M}$. Actually,  an element traversal containing $\vec{c}(t)$ is the sufficient and necessary condition for  $\vec{c}(t)$ being a valid path. Please see the appendix for a rigorous proof.

Thus, evaluating whether $\vec{c}(t)$ is a valid path, is equivalent to searching for an element traversal from $\restbf{s}$ to $\restbf{p}$, and a piece-wise linear curve $\rest{\vec{c}}(t): I \mapsto \rest{M}$ defined on it, such that $\vec{c}(t)=\Psi(\rest{\vec{c}}(t))$.
Such an element traversal and piece-wise linear curve can be efficiently constructed in $M$. 

\begin{figure}
    \centering
    \includegraphics[width=\columnwidth]{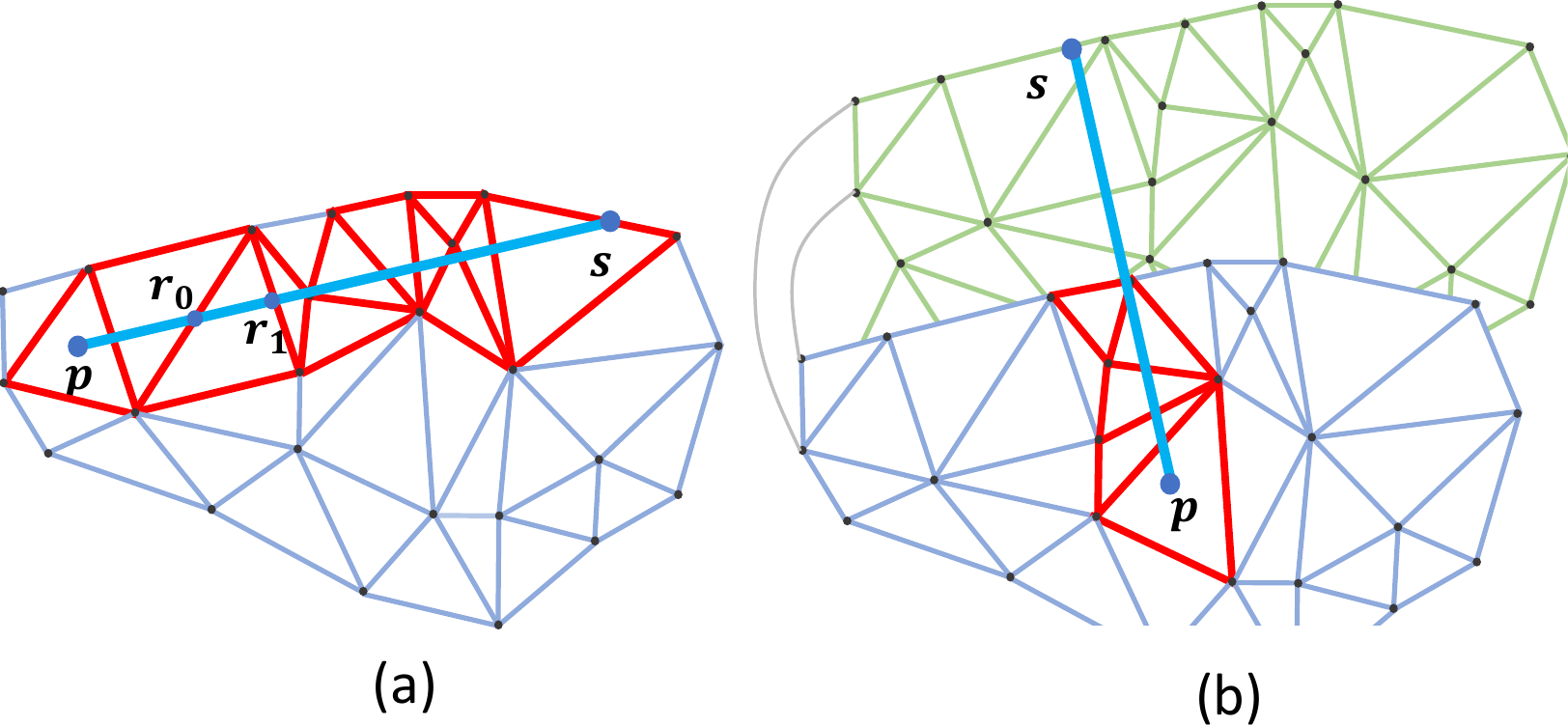}
    \caption{(a) An example of a triangular traversal, marked by red triangles. A line segment connecting $\mathbf{p}$ and $\mathbf{s}$ is included in this triangular traversal.  (b) An example of a line segment being an invalid path when there are self-intersections, the triangular traversal (marked by the red triangles) stops at the boundary of the mesh but the line segment penetrates the boundary and continues going. }
    \label{fig:02_TetrahedralTraverse}
\end{figure}

Going through the element traversal, $\rest{\vec{c}}(t)$ must pass through faces shared by neighoring elements at points $\rest{\vec{r}}_i \in e_i \cap e_{i+1}$, where $i=0,1,2,\dots,k-1$. When $\Psi$ forms no inversion, corresponding face points $\vec{r}_i$ must be along the line segment $\vec{c}(t)$, i.e. $\vec{r}_i=\vec{c}(t_i)$ for some $t_i\in[0,1]$, see \autoref{fig:02_TetrahedralTraverse}a.
If we can form such an element traversal using the topological connections of the model, we can safely conclude that the path is valid.

This gives us an efficient mechanism for testing the validity of the shortest path from $\vec{p}$ to $\vec{s}$. Starting from $e_\vec{p}$, we trace a ray from $\vec{p}$ towards $\vec{s}$ and find the first face point $\vec{r}_0$. If $\vec{r}_0$ is not on the boundary, this face must connect $e_\vec{p}$ to a neighboring element $e_1$. Then, we enter $e_1$ from $\vec{r}_0$ and trace the same ray to find the exit point $\vec{r}_1$ on another face. We continue traversing until we reach $e_\vec{s}$, in which case we can conclude that this is a valid path, see \autoref{fig:02_TetrahedralTraverse}a. This also includes the case $e_\vec{p}=e_\vec{s}$. If we reach a face point $\vec{r}_i$ that is on the boundary (see \autoref{fig:02_TetrahedralTraverse}b) or we pass-through $\vec{s}$ without entering $e_\vec{s}$, $\vec{s}$ cannot be the closest boundary point to $\vec{p}$. 

This process allows us to efficiently test the validity of a path to a given boundary point, but we have infinitely many points on the boundary to test. Fortunately, we are only interested in the shortest path and we can use the theorem below to test only a single point per boundary face.

\begin{theorem}\label{thm:single_point}
For each interior point $\vec{p}$ (as $\rest{\vec{p}}$), if its closest boundary point $\vec{s}$ (as $\rest{\vec{s}}$) is on the boundary face $f$, $\vec{s}$ must also be the Euclidean closest point to $\vec{p}$ on $f$.
\end{theorem}
The proof is similar to \autoref{thm:shortestPath}, \del{please see the supplementary material}\add{which is included in the appendix}.
%
Based on \autoref{thm:single_point}, we only need to check a single point (the Euclidean closest point) on each boundary face to find the closest boundary point. If we test these boundary points in the order of increasing distance from the interior point $\vec{p}$, as soon as we find a valid path to one of them, we can terminate the search by returning it as the closest boundary point. In practice, we use a BVH (bounding volume hierarchy) to test these points, which allows testing them approximately (though not strictly) in the order of increasing distance and, once a valid path is found, quickly skipping the further away bounding boxes.

\subsection{Robust Topological Ray Traversal}

The process we describe above for testing the validity of the linear path to a candidate boundary point involves traversing a ray through the mesh. This ray traversal is significantly simpler than typical ray traversal algorithms used for rendering with ray tracing. This is because it directly follows the topological connections of the mesh.


At each step, the ray enters an element through one of its faces and must exit from one of its other faces. Therefore, we do not need to rely on an acceleration structure to quickly determine which faces to test ray intersections, as they are directly known from the mesh topology.
In fact, 
we do not need to check each one of the other faces individually, since the ray exits from exactly one of them. Therefore, we can quickly test all possible exit faces together.

For example, \citet{aman2022compact} present such a tetrahedral traversal algorithm in 3D. Yet, due to limited numerical precision, this algorithm is prone to forming infinite loops. Such infinite loops are easy to detect and terminate (e.g. using a maximum iteration count), but such premature terminations are entirely unacceptable in our case. This is because incorrectly deciding on the validity of a path would force our algorithm to pick an incorrect shortest path to boundary, which can be arbitrarily far from the correct one. Therefore, the simulation system that relies on this shortest path to boundary can place strong and arbitrarily incorrect forces/constrains in an attempt to resolve the self-intersection. 

Our solution for properly resolving such cases that arise from limited numerical precision is three fold:
\begin{enumerate}
    \item We allow ray intersections with more than one face by effectively extending the faces using a small tolerance parameter $\epsilon_i$ in the intersection test. This forms branching paths when a ray passes between multiple faces and, therefore, intersects (within $\epsilon_i$) with more than one of them.
    \item We keep a list of traversed elements and terminate a branch when the ray enters an element that was previously entered.
    \item We keep a stack containing all the candidate intersecting faces from the intersection test. After a loop is detected, we pick the latest element from it and continue the process.
    
\end{enumerate}
Please see our appendix for the pseudo-code and more detailed explanations of our algorithm.

In practice such branching happens rarely, but solution ensures that we never incorrectly terminate the ray traversal. Note that $\epsilon_i$ is a conservative parameter for extending the ray traversal through branching to prevent problems of numerical accuracy issues. It does not introduce any error to the final shortest paths we find. Using an unnecessarily large $\epsilon_i$ would only have negative, though mostly imperceptible, performance consequences. We verified this by making the $\epsilon_i$ ten times larger, which did not result in a measurable performance difference.

One corner case is when the internal point $\vec{p}$ (as $\rest{\vec{p}}$) and the boundary point $\vec{s}$ (as $\rest{\vec{s}}$) coincide, such that $\vec{p}=\vec{s}$ (within numerical precision). This forms a line segment with zero length and, therefore, does not provide a direction for us to traversal. This happens when testing self-intersections of boundary points, which pick themselves as their first candidate for the closest boundary point. This zero-length line segment cannot be a valid path.
Fortunately, since we know we are testing self-intersection for $\vec{s}$, when the BVH query returns the boundary face includes $\vec{s}$, we can directly reject it.


\subsection{Intersections of Different Objects}
\label{sec:inter-object}

\begin{figure}
    \centering
    \includegraphics[width=0.9\columnwidth]{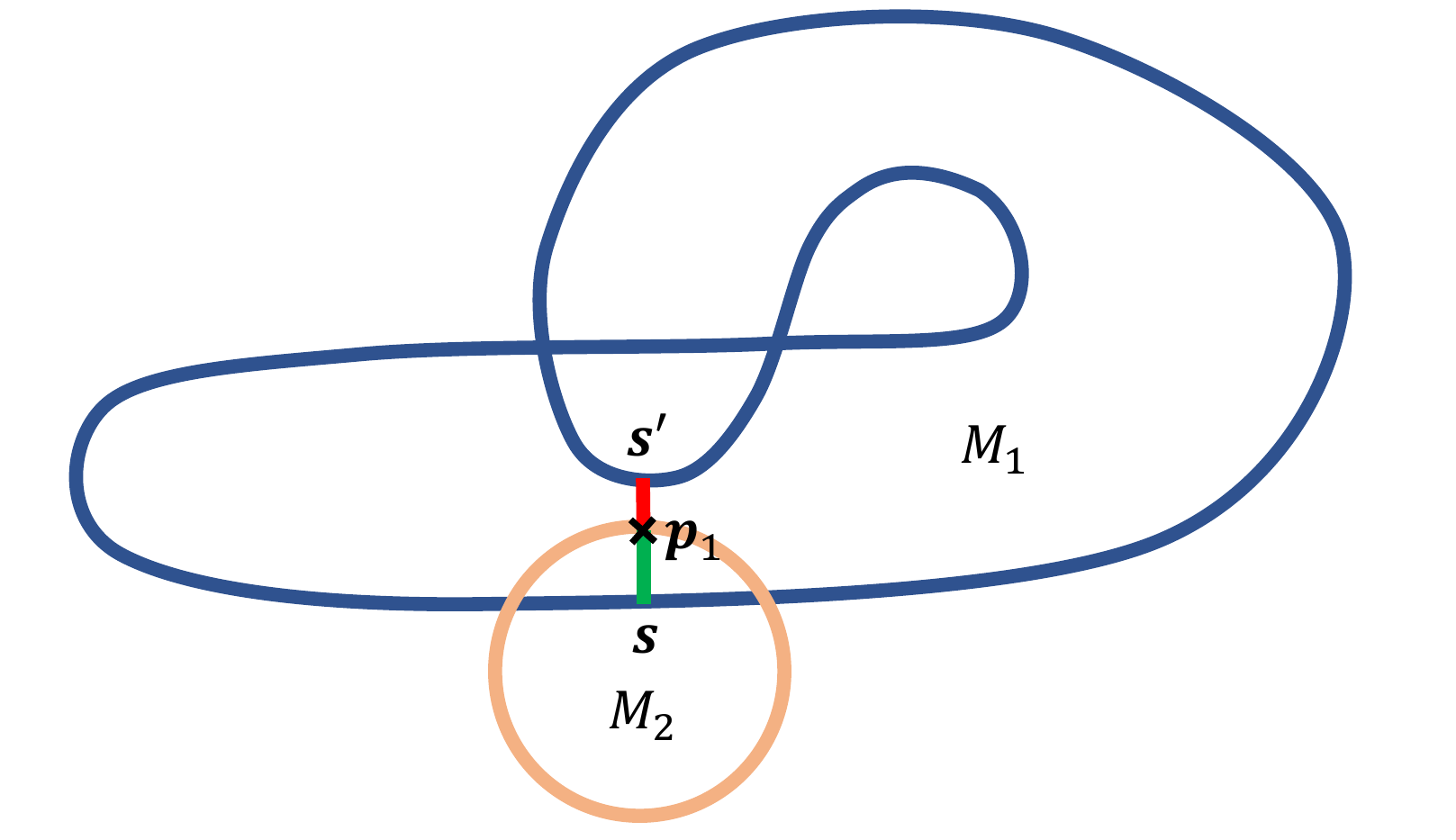}
    \caption{An object $M_2$ intersects with a self-intersecting object $M_1$. A surface point of $M_2$ is overlapping with an interior point $\vec{p}_1 \in M_1$.  $\vec{s}$ and $\vec{s}^\prime$ are $\vec{p}_1$'s closest boundary point by our definition and Euclidean closest boundary point, respectively.}
    \label{fig:InterObjIntersection}
\end{figure}
Although our method is mainly designed for solving self-intersections, it is still needed for handling intersections of different objects when they may have self-intersections as well. 
As shown in \autoref{fig:InterObjIntersection}, an object $M_2$ intersects with a self-intersecting object $M_1$, where a surface point of $M_2$ is overlapping with an interior point $\vec{p}_1 \in M_1$.
Simply querying for $\vec{p}_1$'s Euclidean closest boundary point in $M_1$ will give us $\vec{s}^\prime$, which does not help resolve the penetration. 
This is because $\vec{p}_1 \vec{s}^\prime$ is not a valid path between $\vec{p}_1$ (as $\rest{\vec{p}}_1 \in \rest{M}^\circ_{\add{1}}$) and $\vec{s}_1^\prime$ as ($\rest{\vec{s}}^\prime \in \partial \rest{M}_1$ ). 
What is actually needed is $\vec{p}_1$'s shortest path to  boundary as $\rest{\vec{p}}_1 $, which is the same problem as the self-intersection case, a surface point of $M_1$ is overlapping with an interior point $\vec{p}_1 \in M_1$.



\begin{figure*}[ht!]
    \centering
    \includegraphics[width=\textwidth]{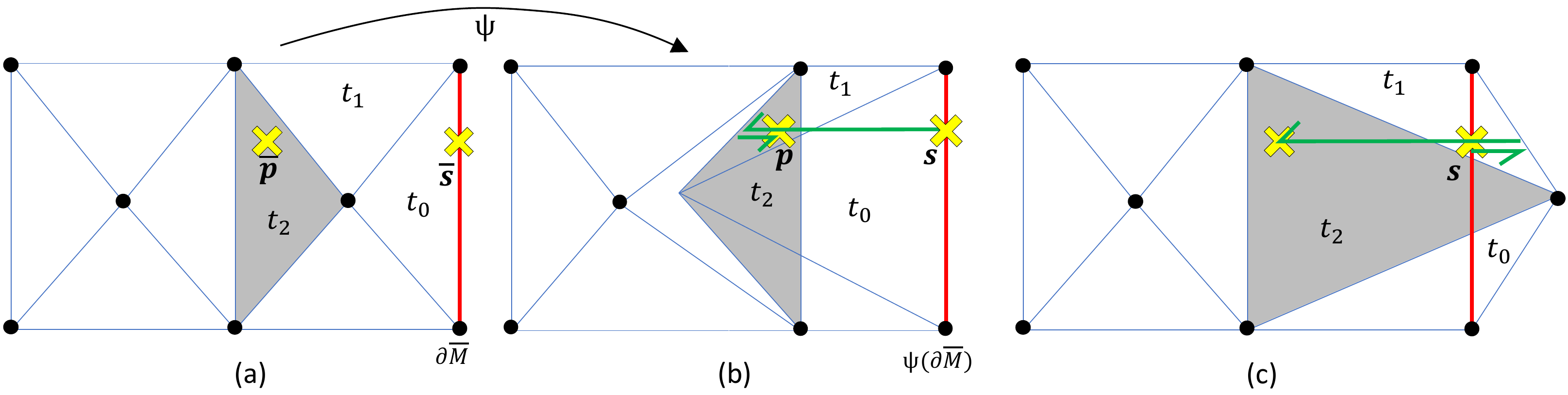}
    \caption{(a) A part of the undeformed pose of a triangular mesh $\rest{M}$, which is inversion free. $\restbf{p}\in \rest{M}$, $\restbf{s}\in \partial\rest{M}$. A surface edge is marked with red color. (b) The image of $\rest{M}$ under $\Psi$, the tetrahedron $t_2$ (colored with gray), is inverted by $\Psi$. The green line illustrates $\mathbf{p}$'s global geodesics to the surface, it has a self-overlapping part, which is marked by the two-sided arrow. (c) An interior tetrahedron is inverted and got out of the surface. In this case, the global geodesics to the surface path can go backward.  }
    \label{fig:InteriorInvertedTetrahedron}
\end{figure*}

\begin{figure}
    \centering
    \includegraphics[width=\columnwidth]{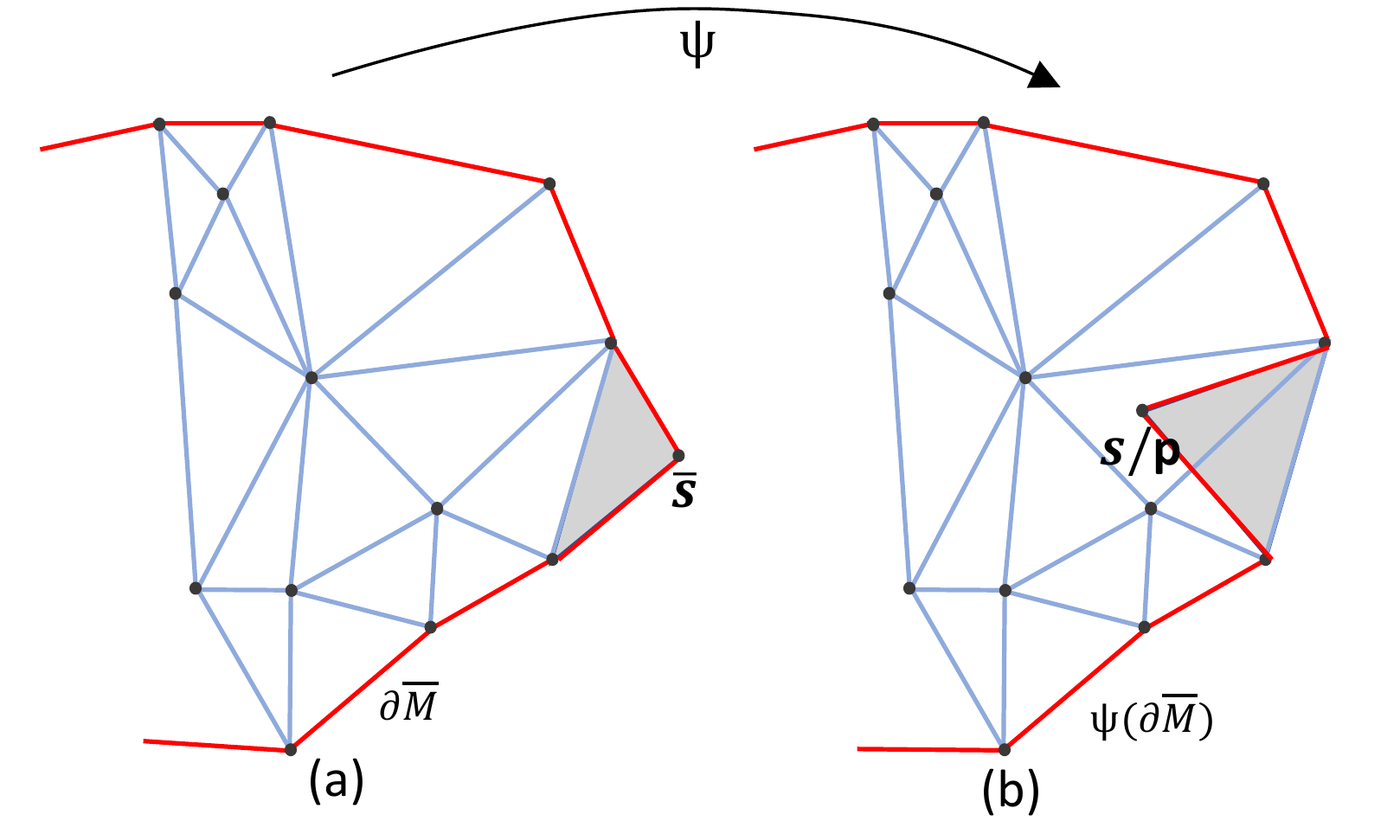}
    \caption{(a) A part of the undeformed pose of a triangular mesh $\rest{M}$, which is inversion free. The surface edges are marked with red color. (b) After deformation, a triangle (marked by gray color), is inverted and folded into the interior of the mesh. A deformed surfaces point $\mathbf{s}$ overlaps with the interior point $\mathbf{p}$. }
    \label{fig:InvertedSurfaceTetrahedron}
\end{figure}

\subsection{Inverted Elements}
\label{sec:inversion}

Our derivations in \autoref{sec:derivation} assume that $det(\nabla \Psi)>0$ everywhere. For a discrete mesh, this would mean no inverted or degenerate elements. Unfortunately, though inverted elements are often highly undesirable, they are not always unavoidable. Fortunately, the algorithm we describe above can be slightly modified to work in the presence of certain types of inverted elements.

If the inverted elements are not a part of the mesh boundary, we can still test the validity of paths by allowing the ray traversal to go backward along the ray.   This is because the ray would need to traverse backward within inverted elements. In addition, we cannot simply terminate the traversal once the ray passes through the target point, because an inverted element further down the path may cause backward traversal to reach (or pass through) the target point, see \autoref{fig:InteriorInvertedTetrahedron}b. Therefore, ray traversal must continue until a boundary point is reached. We also need to allow the ray to go behind the starting point, see \autoref{fig:InteriorInvertedTetrahedron}c.

A consequence of this simple modification to our algorithm is that, when we begin from an internal point $\vec{p}$ toward a boundary point $\vec{s}$, it is unclear if we would reach $\vec{s}$ by beginning the traversal toward $\vec{s}$ or in the opposite direction. While one may be more likely, both are theoretically possible.

To avoid this decision, in our implementation we start the traversal from the target boundary point $\vec{s}$. In this case, there is no ambiguity, since there is only one direction we can traverse along the ray. This also allows using the same traversal routine for the first element and the other elements along the path by always entering an element from a face. Therefore, it is advisable even in the absence of inverted elements.


Nonetheless, our algorithm is not able to handle all possible inverted elements.
For example, if the inverted element is on the boundary, as shown in \autoref{fig:InvertedSurfaceTetrahedron}, the inversion itself can cause self-intersection. In such a case, a surface point $\vec{s}$ is overlapping with an interior point $\vec{p}$ (as $\rest{\vec{p}}$). Our algorithm will not be able to try to construct a tetrahedral traversal between those two points because we cannot determine a ray direction for a zero-length line segment.  Actually, in this case, the very definition of the closest boundary point can be ambiguous.

Our solution is to skip the self-intersection detection of inverted boundary elements.
As a result, the only way for us to solve such self-intersections caused by inverted boundary elements is to resolve the inversion itself.
Fortunately, inverted elements are undesirable for most simulation scenarios, and they are often easier to fix for boundary elements.
Unfortunately, if the inverted boundary elements have global self-intersections with other parts of the mesh, our solution ignores them. Though this does not form a complete solution, because the inverted boundary elements are rare, the other boundary elements surrounding the inverted elements are often enough to solve the global self-intersection.

%% file: 04_FeasibleRegionAceleration.tex
\subsection{Infeasible Region Culling}

In a lot of cases, it is possible to determine that a given candidate boundary point $\vec{s}$ cannot be the closest boundary point to an interior point $\vec{p}$, purely based on the local information about the mesh around $\vec{s}$, without performing any ray traversal. 
%
%
For this test we construct a particular region of space, i.e. the \emph{feasible region}, 
around $\mathbf{s}$.
When $\vec{p}$ is outside of this region of $\vec{s}$, thus in its \emph{infeasible region}, we can safely conclude that $\vec{s}$ is not the closest boundary point.

The formulation of the feasible region can be viewed as a discrete application of the well-known Hilbert projection theorem. The construction of this feasible region depends on whether $\vec{s}$ is on a vertex, edge, or face.

\begin{figure}
    \centering
    \includegraphics[width=\linewidth]{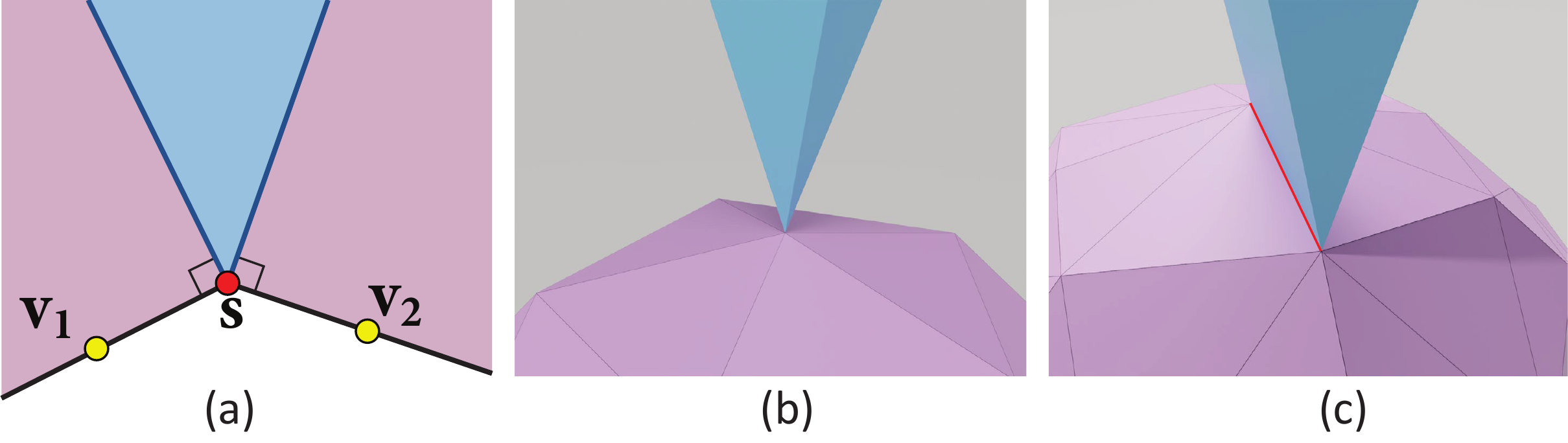}
    \caption{The feasible region, shaded in blue, for (a)~a boundary vertex in 2D, (b)~a boundary vertex in 3D, and (c)~a boundary edge in 3D. Note that the 3D meshes in (b) and (c) are observed from the inside.}
    \label{fig:FeasibleRegionnIllustration}
\end{figure}

\paragraph{Vertex Feasible Region} In 2D, when $\vec{s}$ is on a vertex, the feasible region is bounded by the two lines passing through the vertex and perpendicular to its two boundary edges, as shown in \autoref{fig:FeasibleRegionnIllustration}a. For a neighboring boundary edge of $\vec{s}$ and its perpendicular line that passes through $\vec{s}$, if $\vec{p}$ is on the same side of the line as the edge, based on \autoref{thm:single_point}, there must be a closer boundary point on the face. More specifically, for any neighboring boundary vertex $\vec{v}_i$ connected to $\vec{s}$ by a edge, if the following inequality is true, $\vec{p}$ is in the infeasible region:
\begin{align}
    (\vec{p}-\vec{s})\cdot(\vec{s}-\vec{v}_i) < 0
\end{align}
The same inequality holds in 3D for all neighboring boundary vertices $\vec{v}_i$ connected to $\vec{s}$ by an edge (\autoref{fig:FeasibleRegionnIllustration}b).
The 3D version of the vertex feasible region is actually the space bounded by a group of planes perpendicular to its neighboring edges.

\paragraph{Edge Feasible Region} In 3D, when $\mathbf{s}$ is on the edge of a triangle, its feasible region is the intersection of 4 half-spaces defined by four planes: two planes that contain the edge and perpendicular to its two adjacent faces, and two others that are perpendicular to the edge and pass through its two vertices, as shown in \autoref{fig:FeasibleRegionnIllustration}c. Let $\vec{v}_0$ and $\vec{v}_1$ be the two vertices of the edge and $\vec{n}_0$ and $\vec{n}_1$ be the two neighboring face normals (pointing to the interior of the mesh). $\vec{p}$ is in the infeasible region if any of the following is true:
\begin{align}
    (\vec{p}-\vec{v}_0)\cdot(\vec{v}_1-\vec{v}_0) &< 0 \\
    (\vec{p}-\vec{v}_1)\cdot(\vec{v}_0-\vec{v}_1) &< 0 \\
    (\vec{p}-\vec{s})\cdot(\vec{n}_0\times(\vec{v}_1-\vec{v}_0)) &< 0 \\
    (\vec{p}-\vec{s})\cdot(\vec{n}_1\times(\vec{v}_0-\vec{v}_1)) &< 0 
\end{align}
note that $\vec{n}_0$ is from the face whose orientation accords to $\vec{v}_0 \rightarrow \vec{v}_1$.

\paragraph{Face Feasible Region} We can similarly construct the feasible region when $\vec{s}$ in on the interior of a face as well. Nonetheless, this particular feasible region test is unnecessary, because when $\vec{s}$ is the closest point on the face to $\vec{p}$, which is how we pick our candidate boundary points (based on \autoref{thm:single_point}), $\vec{p}$ is guaranteed to be in the feasible region.



Our \emph{infeasible region culling} technique performs the tests above and skips the ray traversal if $\vec{p}$ is determined to be in the infeasible region, quickly determining that $\vec{s}$ cannot be the closest boundary point. 
Due to numerical precision, the feasible region check can return false results when $\vec{p}$ is close to the boundary of the feasible region. There are two types of errors: false positives and false negatives. A false positive is not a big problem: it will only result in an extra traversal. But if a false negative happens, there is a risk of discarding the actual closest surface point. 
In practice, however, we replace the zeros on the right-hand-sides of the inequalities above with a small negative number $\epsilon_r$ to avoid false-positives due to numerical precision limits. In our tests, we have observed that infeasible region culling can provide more than an order of magnitude faster shortest path query.

%% file: 05_CollisionHandling.tex
\begin{figure}
\centering
 \newcommand{\fig}[2]{%
 \begin{subfigure}{.33\linewidth}
 \includegraphics[width=\linewidth,trim=600 0 700 0,clip]{Figures/19_CCD_DCD_Comparison/#11}\vspace{-0.5em}\\
 \includegraphics[width=\linewidth,trim=600 30 700 400,clip]{Figures/19_CCD_DCD_Comparison/#12}\vspace{-0.5em}
 \includegraphics[width=\linewidth,trim=450 50 450 0,clip]{Figures/19_CCD_DCD_Comparison/#13}\vspace{-0.5em}
 \caption{#2}\end{subfigure}}
\fig{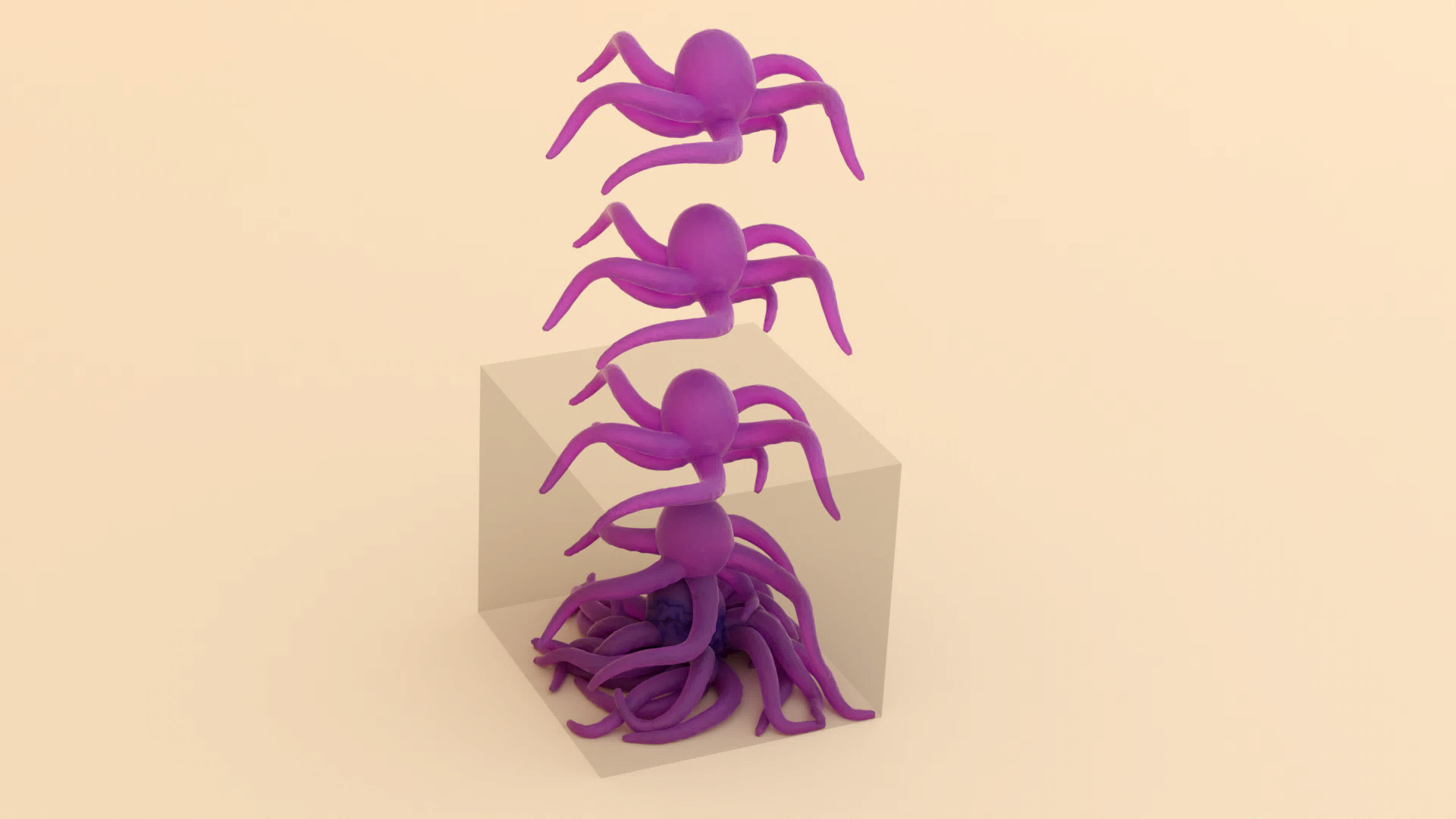}{CCD only}\hfill%
\fig{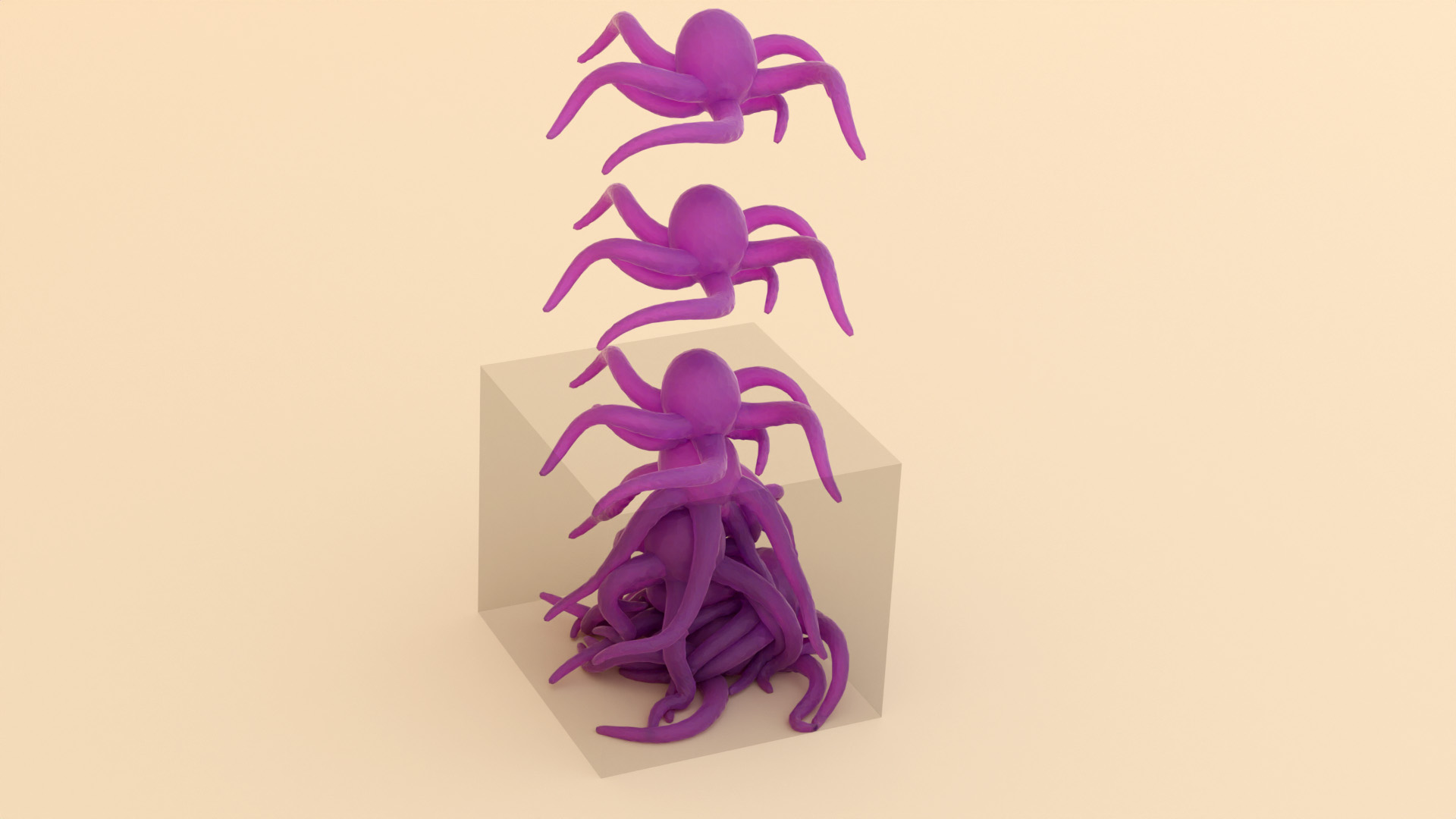}{DCD only}\hfill%
\fig{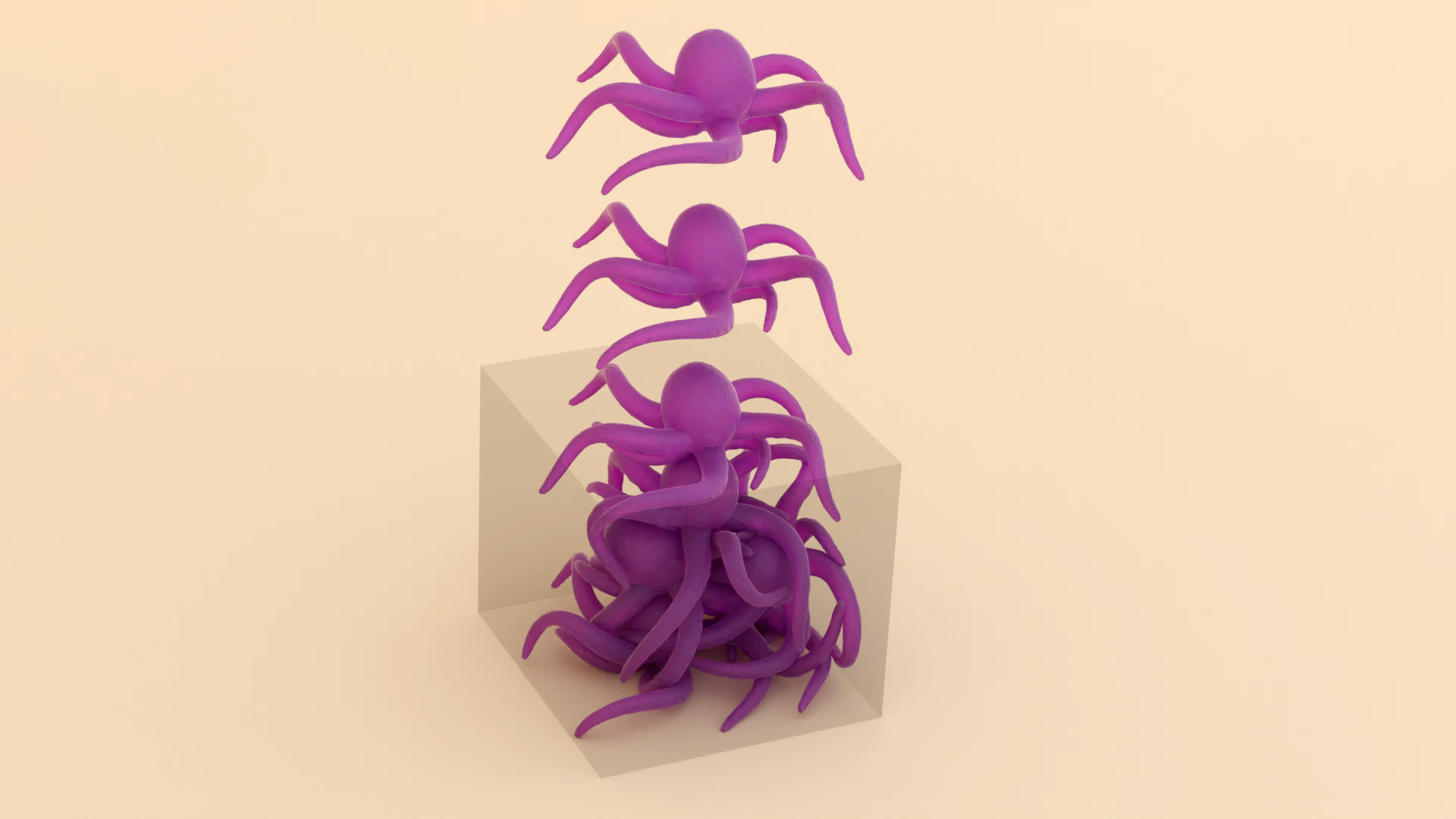}{CCD and DCD}
\caption{Dropping 8 octopi to a box simulated with (a)~CCD only, (b)\add{~DCD with our shortest path query only, and (c)}~CCD and DCD with our shortest path query. \add{The bottom row shows the bottom view of the final state. The blue tint highlights the intersecting geometry. The octopus model is from \citet{Thingi10K}}.
}
\label{fig:19_CCD_DCD_Comparison}
\end{figure}

\add{
\section{Collision Handling Application}

As mentioned above, an important application of our method is collision handling with DCD. When DCD finds a penetration, we can use our method to find the closest point on the boundary and apply forces or constraints that would move the penetrating point towards this boundary point.

In our tests with tetrahedral meshes, we use two types of DCD: vertex-tetrahedron and edge-tetrahedron collisions. For vertex-tetrahedron collisions, we find the closest surface point for the colliding vertex. For edge-tetrahedron collisions, we find the center of the part of the edge that intersects with the tetrahedron and then use our method to find the closest surface point to that center point.
If an edge intersects with multiple tetrahedra, we choose the intersection center that is closest to the center of the edge. The idea is by keep pushing the center of the edge-tetrahedron intersection towards the surface, which eventually resolves the intersection.


This provides an effective collision handling method with XPBD \cite{macklin2016xpbd}.
Once we find the penetrating point $\vec{x}$ we use the standard PBD collision constraint \cite{muller2007position}
\begin{equation}
    c(\Vec{x},\vec{s}) = (\Vec{x} - \Vec{s}) \cdot \Vec{n}
    \label{eq:CollisionConstraint}
\end{equation}
where $\Vec{s}$ is the closest surface point computed by our method when this collision is from DCD, or the colliding point when it is from CCD,  and  $\vec{n}$ is the surface normal at $\vec{s}$. If $\Vec{s}$ is on a surface edge or vertex, we use the area-weighted average of its neighboring face normals. The XPBD integrator applies projections on  each collision constraint $c$ to satisfy $c(\Vec{x}, \Vec{s})\geq0$. We also apply friction, following \citet{bender2015position}. Please see the supplementary material for the pseudocode of our XPBD framework.
}


\begin{figure}
\centering
\newcommand{\fig}[1]{\includegraphics[width=0.33\columnwidth,trim=600 0 700 300,clip]{Figures/23_ImplicitEuler/#1}}
\fig{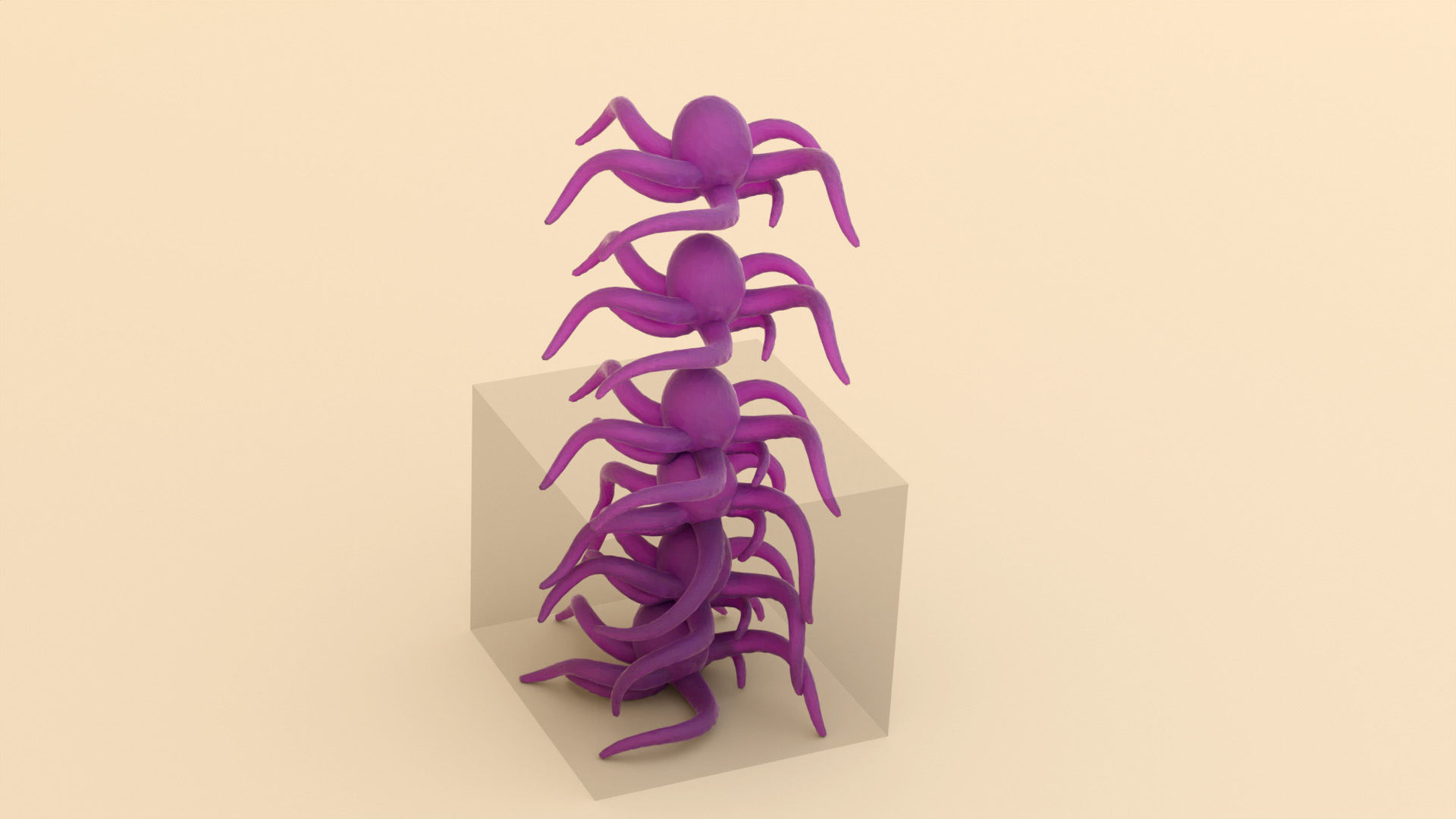}\hfill%
\fig{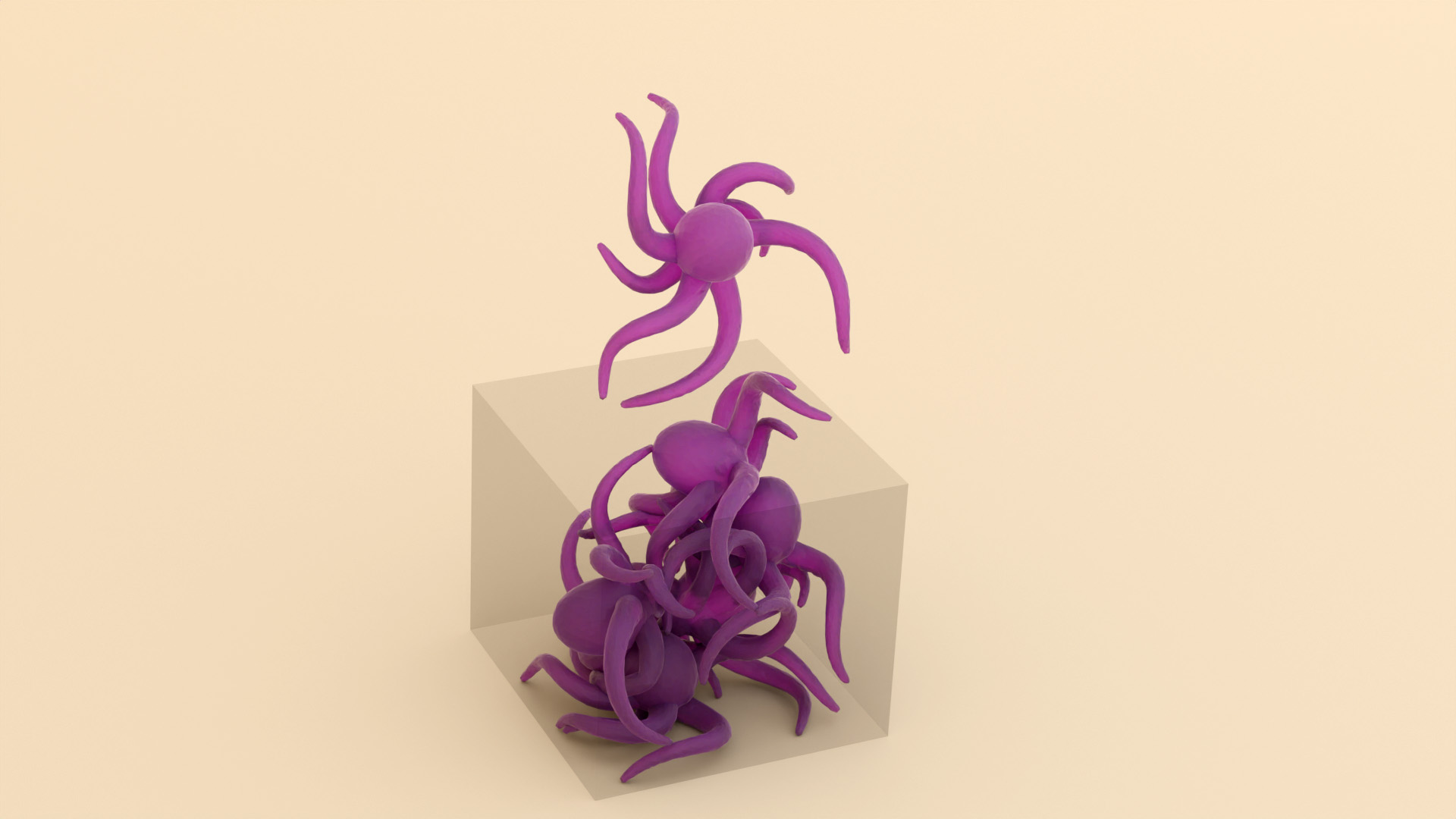}\hfill%
\fig{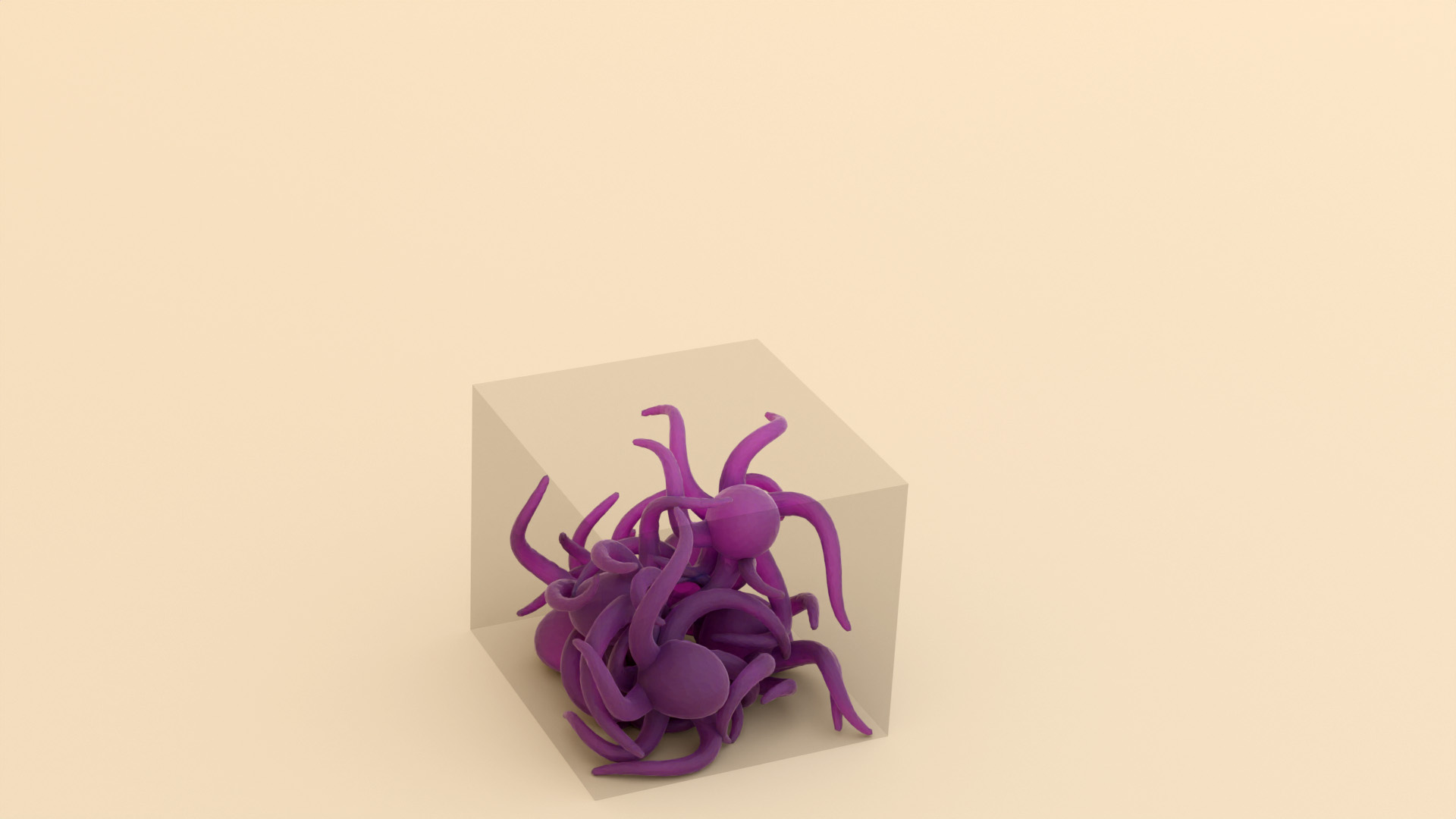}
\caption{\add{Dropping 6 octopi into a box simulated using implicit Euler. This simulation contains 30K vertices and 88K tetrahedra and it takes an average of 15s to simulate each frame. Please see the supplementary material for the details of our implicit Euler framework.}} 
\label{fig:ImplicitEuler}
\end{figure}

\begin{figure*}
\centering
\newcommand{\scap}[2]{\begin{minipage}{#2\linewidth}~\small\textbf{(#1)}\end{minipage}}
\fbox{\includegraphics[height=0.21\linewidth,trim=200 0 400 0,clip]{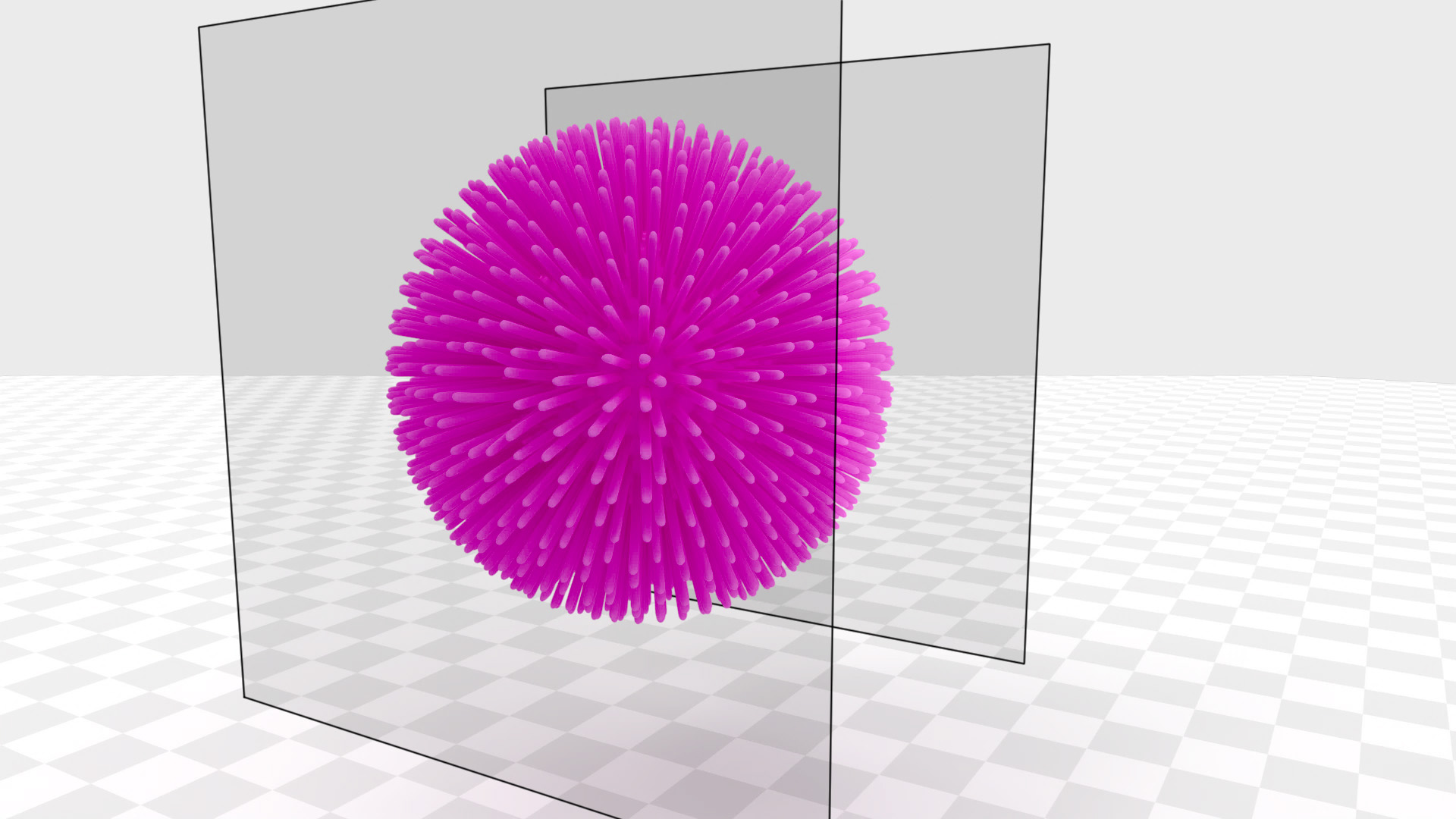}}\hfill%
\fbox{\includegraphics[height=0.21\linewidth,trim=300 0 450 0,clip]{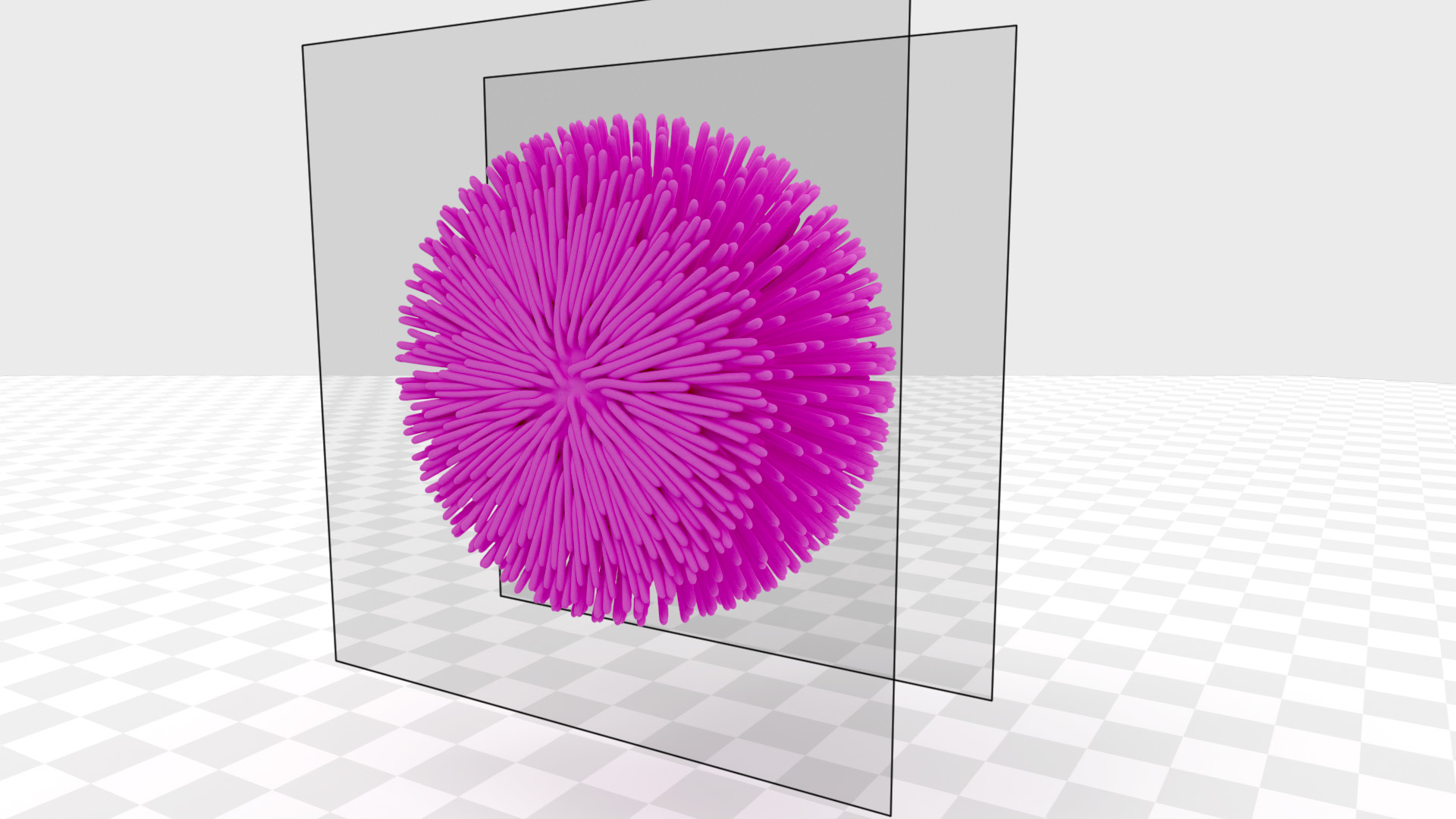}}\hfill%
\fbox{\includegraphics[height=0.21\linewidth,trim=400 0 500 0,clip]{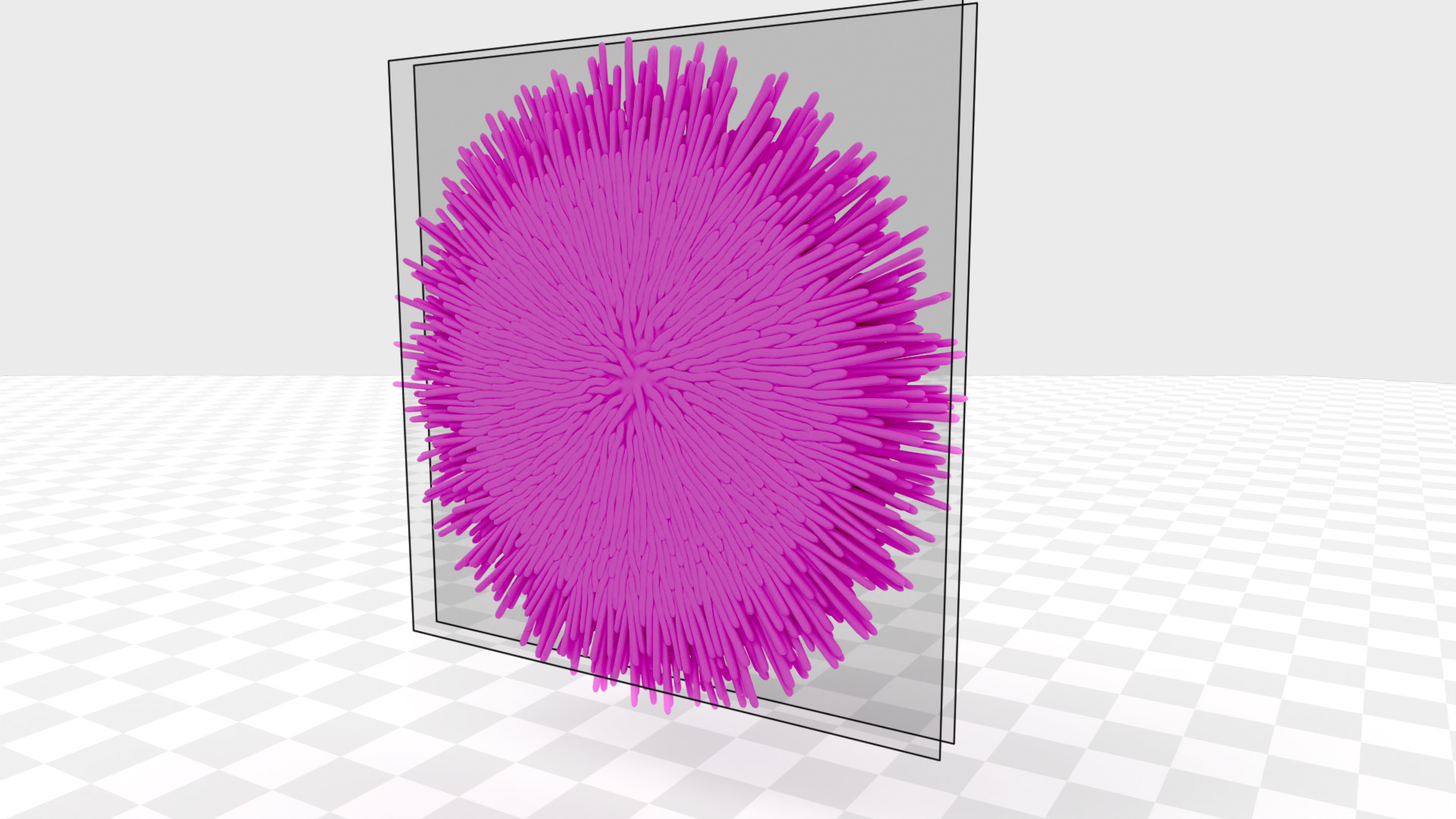}}\hfill%
\fbox{\includegraphics[height=0.21\linewidth,trim=700 0 700 0,clip]{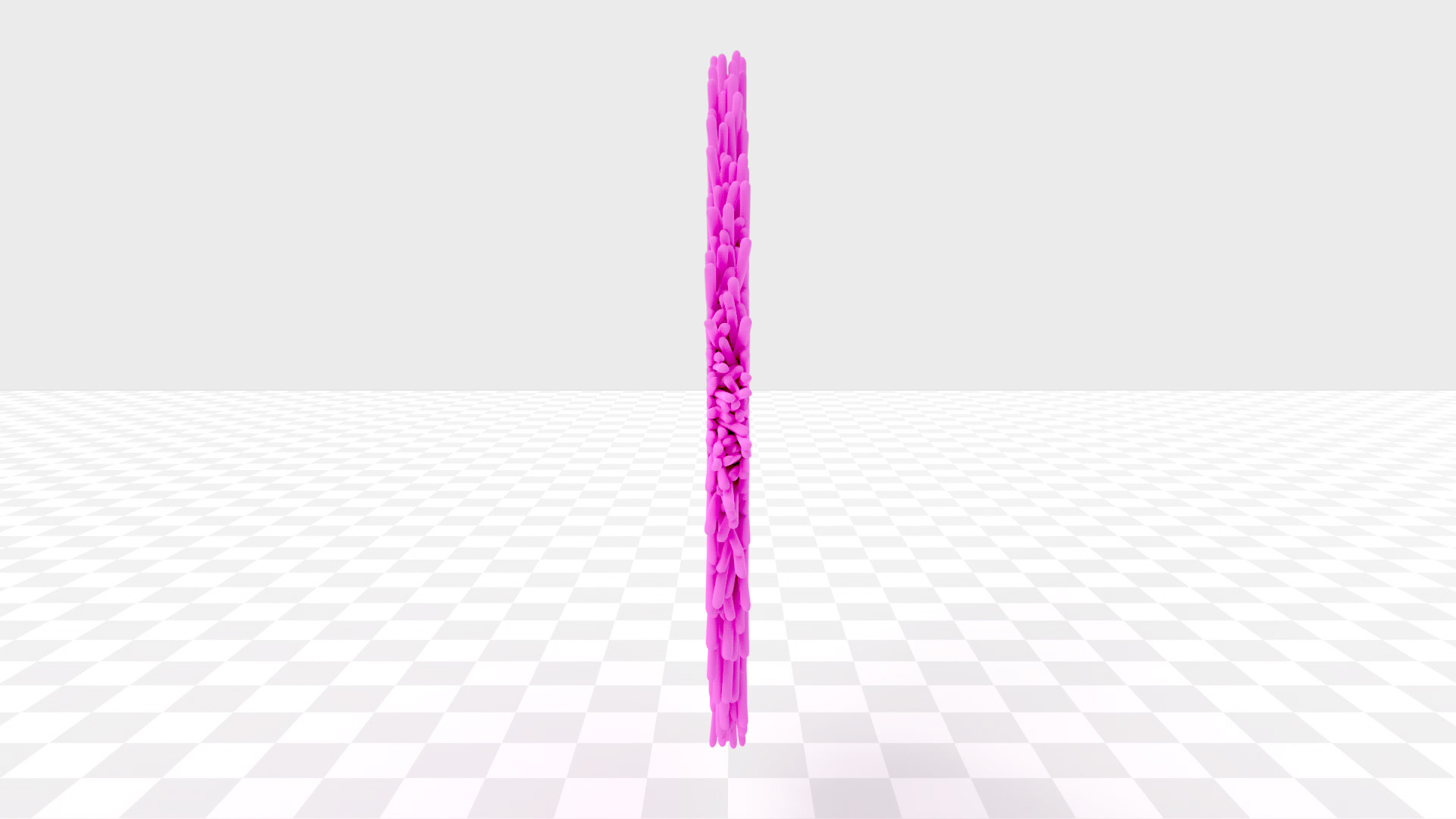}}\hfill%
\fbox{\includegraphics[height=0.21\linewidth,trim=450 0 550 0,clip]{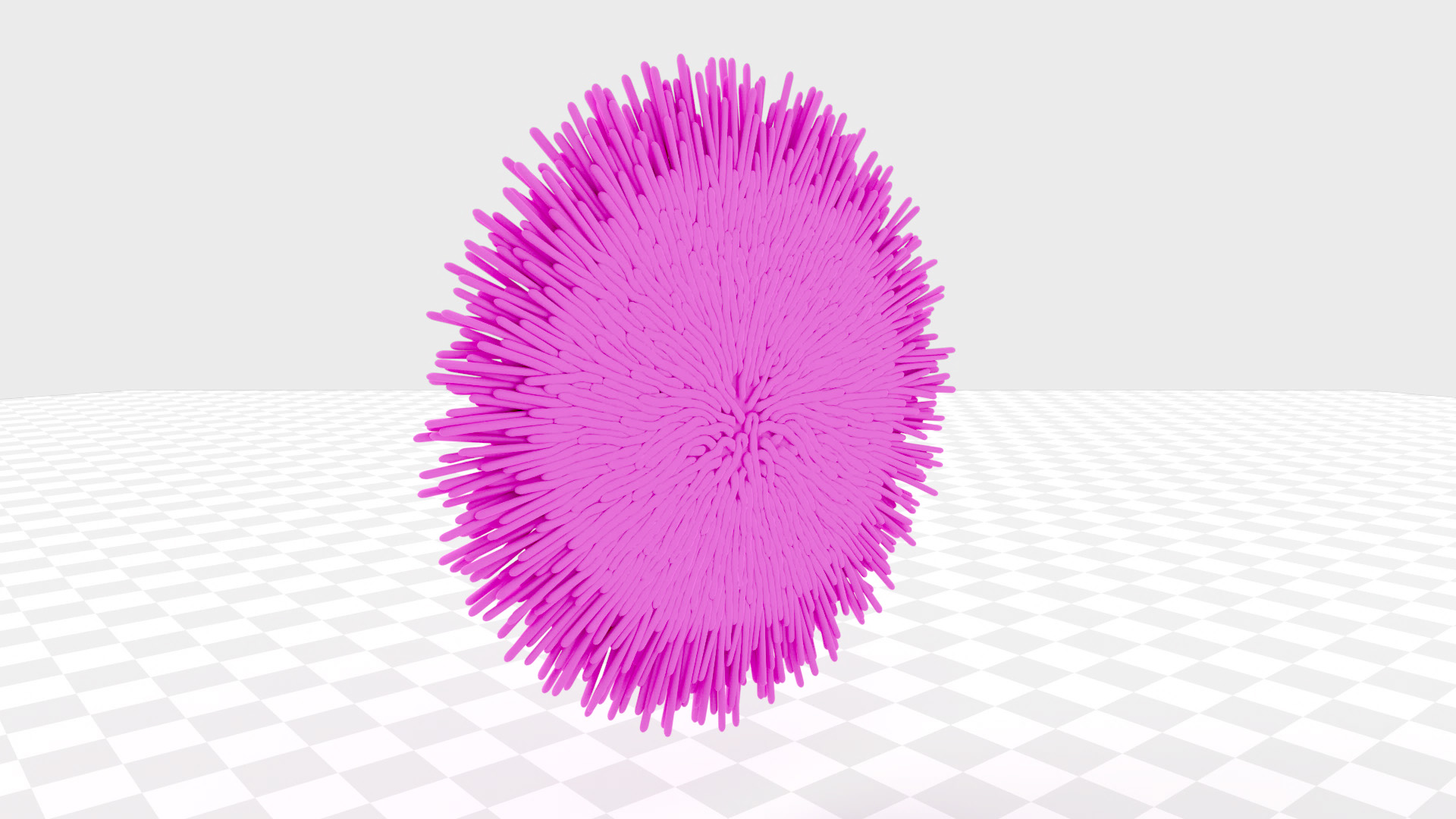}}\vspace{-1.6em}
\scap{a}{0.265}%
\scap{b}{0.24}%
\scap{c}{0.205}%
\scap{d}{0.115}%
\scap{e}{0.175}\vspace{0.2em}
\caption{Flattening a squishy ball (774K vertices, 2.81M tetrahedra) using two planes. (a-c)~the flattening process, (d)~Side view of the flattened ball to to $1/20$ of its radius, and (e)~the other side of the flattened squishy ball.}
\label{fig:SqueezingBall}
\end{figure*}

\begin{figure*}
\centering
\includegraphics[height=0.15\linewidth,trim=400 200 400 200,clip]{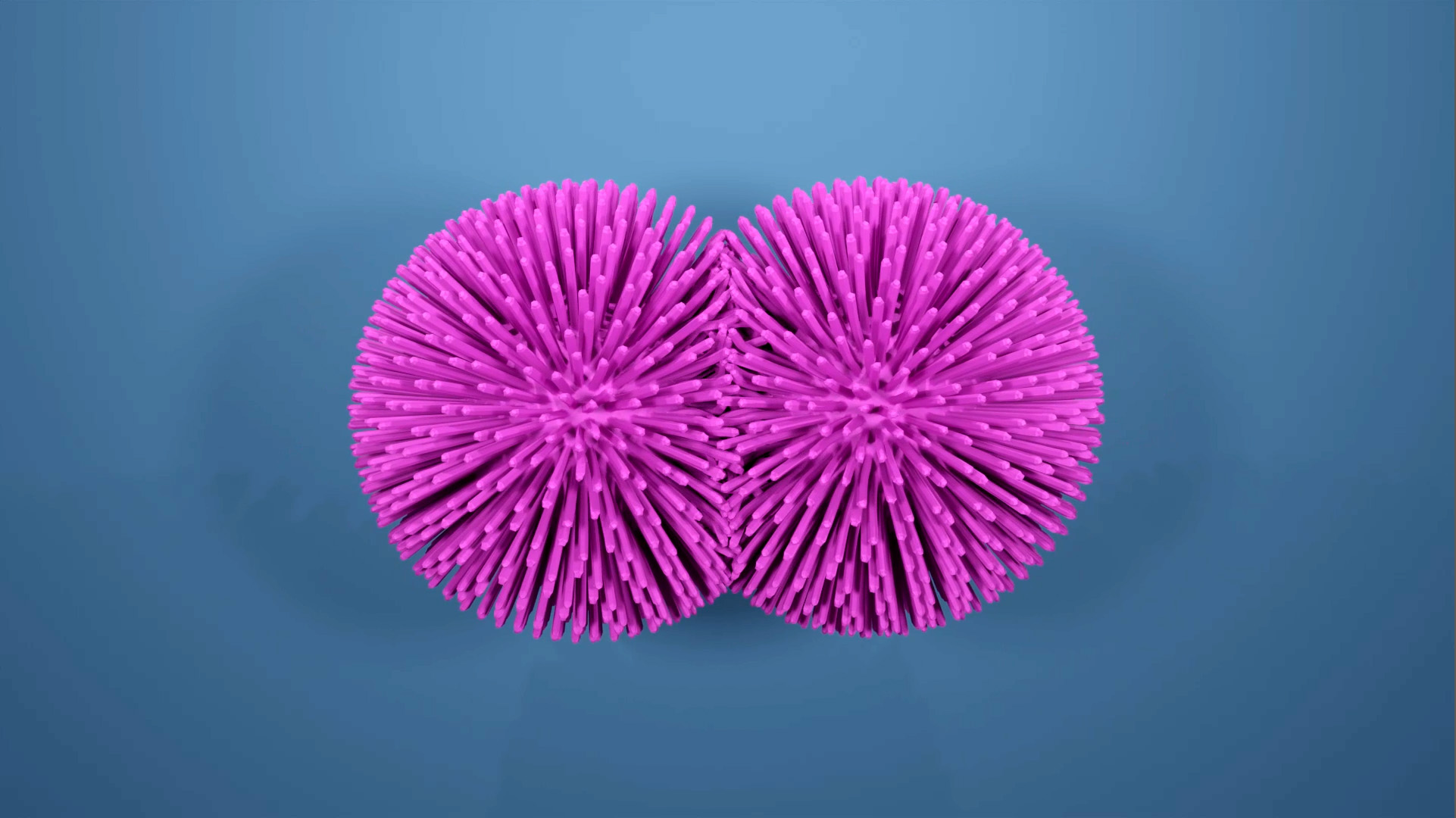}\hfill%
\includegraphics[height=0.15\linewidth,trim=500 200 500 200,clip]{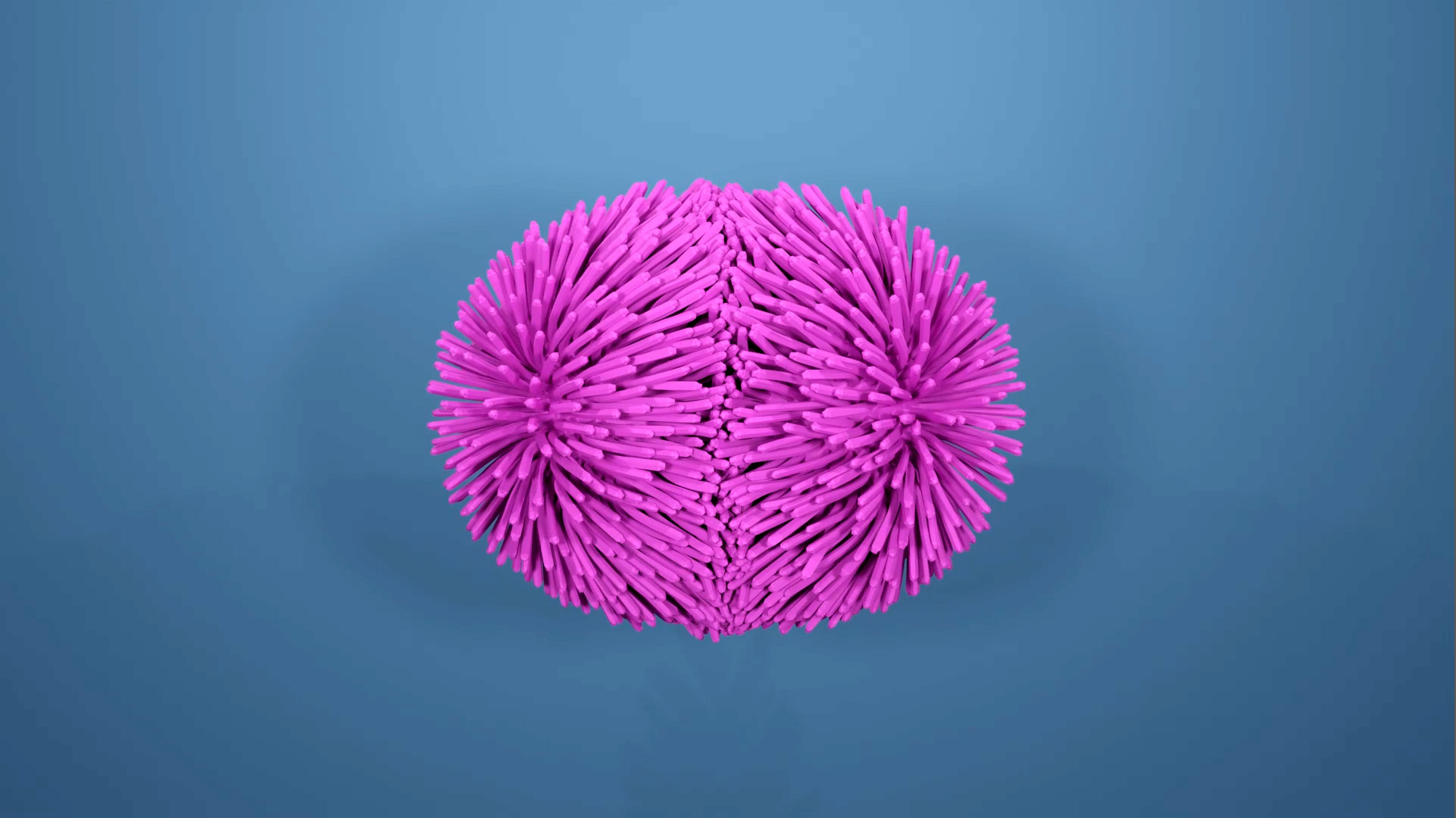}\hfill%
\includegraphics[height=0.15\linewidth,trim=400 200 400 200,clip]{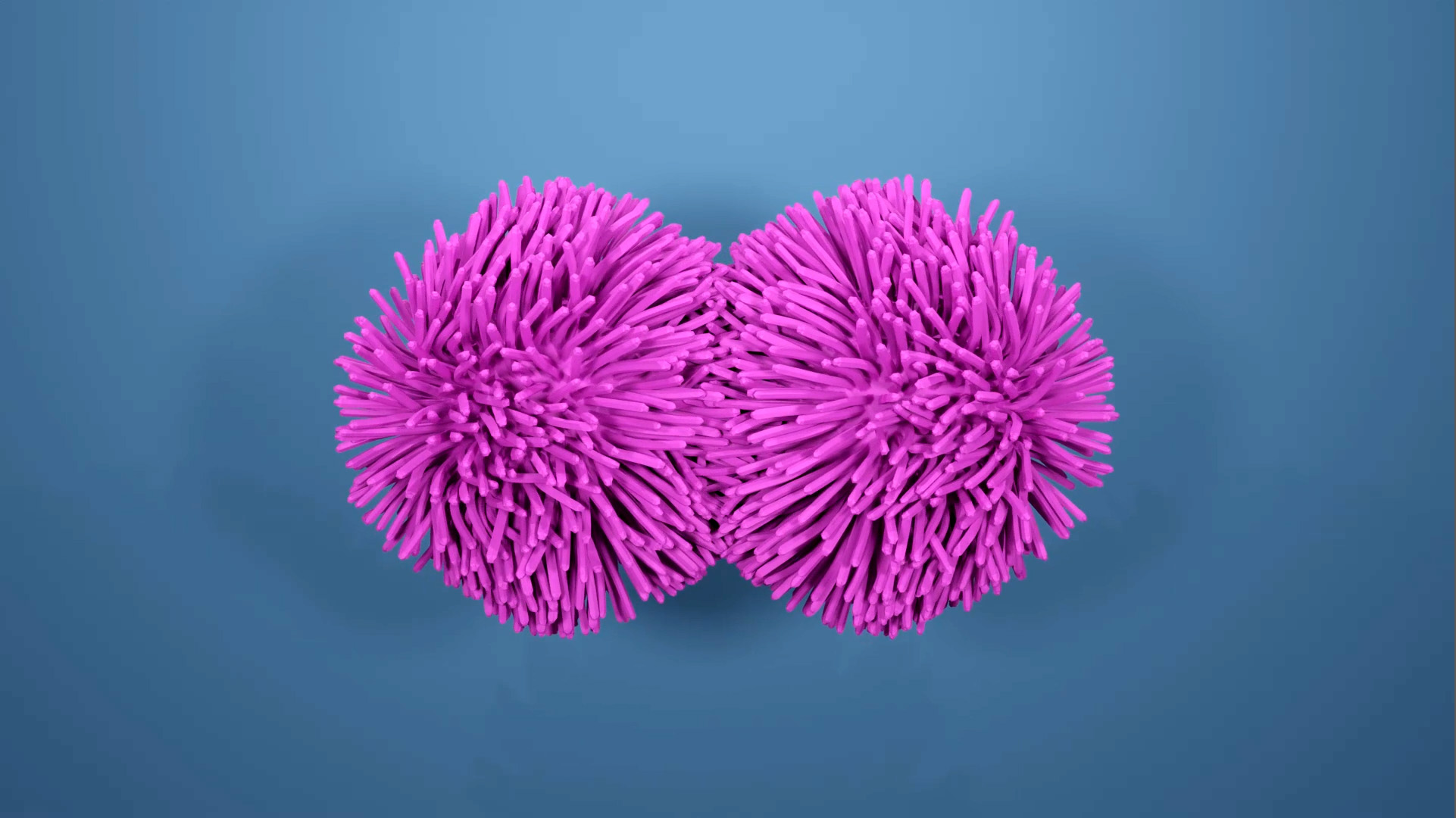}\hfill%
\includegraphics[height=0.15\linewidth,trim=300 200 300 200,clip]{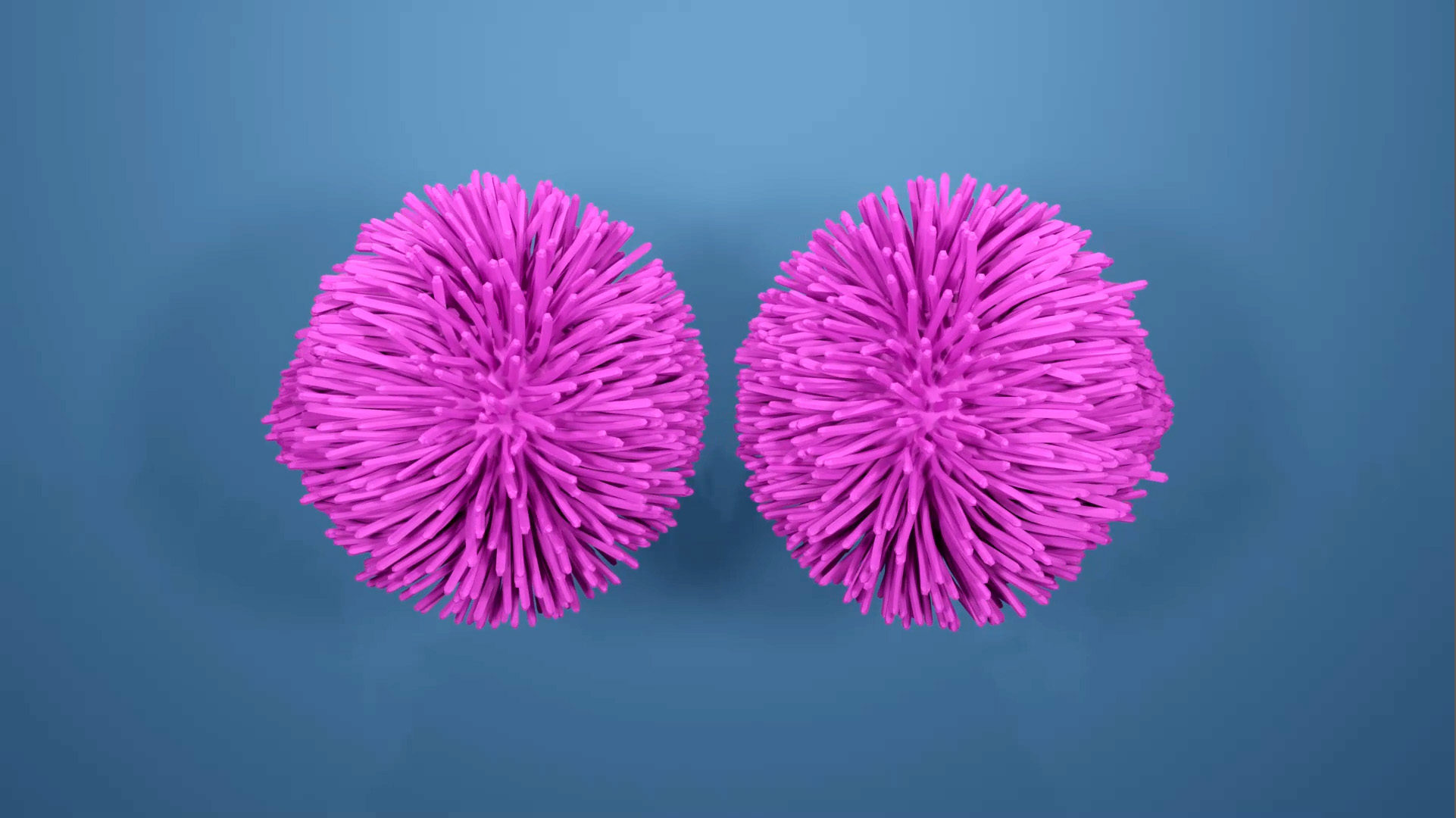}
\caption{Simulation of two squishy balls in head-on collision that come to contact at a relative speed of $50m/s$. Both self-collisions and collisions between the two squishy balls are handled using our method.} 
\label{fig:10_2SquishyballColliding}
\end{figure*}

\add{%
Unlike CCD alone, DCD with our method significantly improves the robustness of collision handling when using a simulation system like XPBD that does not guarantee resolving all collision constraints. This is demonstrated in \autoref{fig:19_CCD_DCD_Comparison}, comparing different collision detection approaches with XPBD. Using only CCD leads to missed collisions when XPBD fails to resolve the collisions detected in previous steps, because CCD can no longer detect them. This quickly results in objects completely penetrating through each other (\autoref{fig:19_CCD_DCD_Comparison}a).
Our method with only DCD effectively resolves the majority of collisions (\autoref{fig:19_CCD_DCD_Comparison}b), but it inherits the limitations of DCD. More specifically, using only DCD with sufficiently large time steps and fast enough motion, some collisions can be missed and deep penetrations can resolve the collisions by moving the objects in incorrect directions, again resulting in object parts passing through each other. Furthermore, our method only provides the closest path to the boundary and properly resolving the collisions is left to the simulation system. Unfortunately, XPBD cannot provide any guarantees in collision resolution, so detected penetrations may remain unresolved. 

We recommend a hybrid solution that uses both CCD and DCD with our method. This hybrid solution performs DCD in the beginning of the time step to identify the preexisting penetrations or collisions that were not properly resolved in the previous time step.
The rest of the collisions are detected by CCD without requiring our method to find the closest surface point. 
The same simulation with this hybrid approach is shown in \autoref{fig:19_CCD_DCD_Comparison}c.
Since all penetrations are first detected by CCD and proper collision constraints are applied immediately, deep penetrations become much less likely even with large time steps and fast motion.
Yet, this provides no theoretical guarantees. The addition of CCD allows the simulation system to apply collision constraints immediately, before the penetrations become deep, and DCD with our method allows it to continue applying collision constraints when it fails to resolve the initial collision constraints. Note that, while this significantly reduces the likelihood of failed collisions, they can still occur if the simulation system keeps failing to resolve the detected collisions.
}


\add{%
The collision handling application of our method is not exclusive to PBD.
Our method can also be used with force-based simulation techniques for defining a penalty force with penetration potential energy
\begin{equation}
    E_c = \frac{1}{2} \, k \, \big( (\Vec{p} - \Vec{s}) \cdot \Vec{n}\big)^2
\end{equation}
where $k$ is the collision stiffness. 
An example of this is shown in \autoref{fig:ImplicitEuler}.
}

%% file: 06_Results.tex
\section{Results}
\label{sec:results}

We use XPBD \cite{macklin2016xpbd} to evaluate our method, because it is one of the fastest simulation methods for deformable objects, providing a good baseline for demonstrating the minor computation overhead introduced by our method.
We use mass-spring or NeoHookean \cite{macklin2021constraint} material constraints implemented on the GPU. We handle the collision detection and handling part on the CPU, including the position updates of the collision constraints.
\add{We use the hybrid collision detection approach that combines CCD and DCD, as explained above.}

We \del{use}\add{implement both collision detection and closest point query on CPU using} Intel's Embree framework \cite{wald2014embree} to create BVH\del{ for both collision detection and closest point query}.
%
%
We generate our timing results on a computer with an AMD Ryzen 5950X CPU, an NVIDIA RTX 3080 Ti GPU, and 64GB of RAM. We acknowledge that our timings are affected by the fact that we copy memory from GPU to CPU every iteration in order to do collision detection and handling, and the whole framework can be further accelerated by implementing the collision detection and shortest path querying on the GPU.
As to the parameters of the algorithm, we set $\epsilon_r$ to $-0.01$.  $\epsilon_i$ is related to the scale and unit of the object, when  the object is at a scale of a few meters, we set $\epsilon_i$ to $1^{-10}$.

\subsection{Stress Tests}


\autoref{fig:SqueezingBall} shows a squishy ball with thin tentacles compressed on two sides and flattened to a thickness that is only $1/20$ of its original radius.
Notice that all collisions, 
including self-intersections of tentacles,
are properly resolved even under such extreme compression.
Also. the model was able to revert to its original state after the 
the two planes compressing it were
removed.


\autoref{fig:10_2SquishyballColliding} shows a high-speed head-on collision of two squishy balls. Though the tentacles initially get tangled with frictional-contact right after the collision, all collisions are properly resolved and the two squishy balls bounce back, as expected.

\begin{figure}
\centering
\includegraphics[width=\linewidth,trim=160 280 200 450, clip]{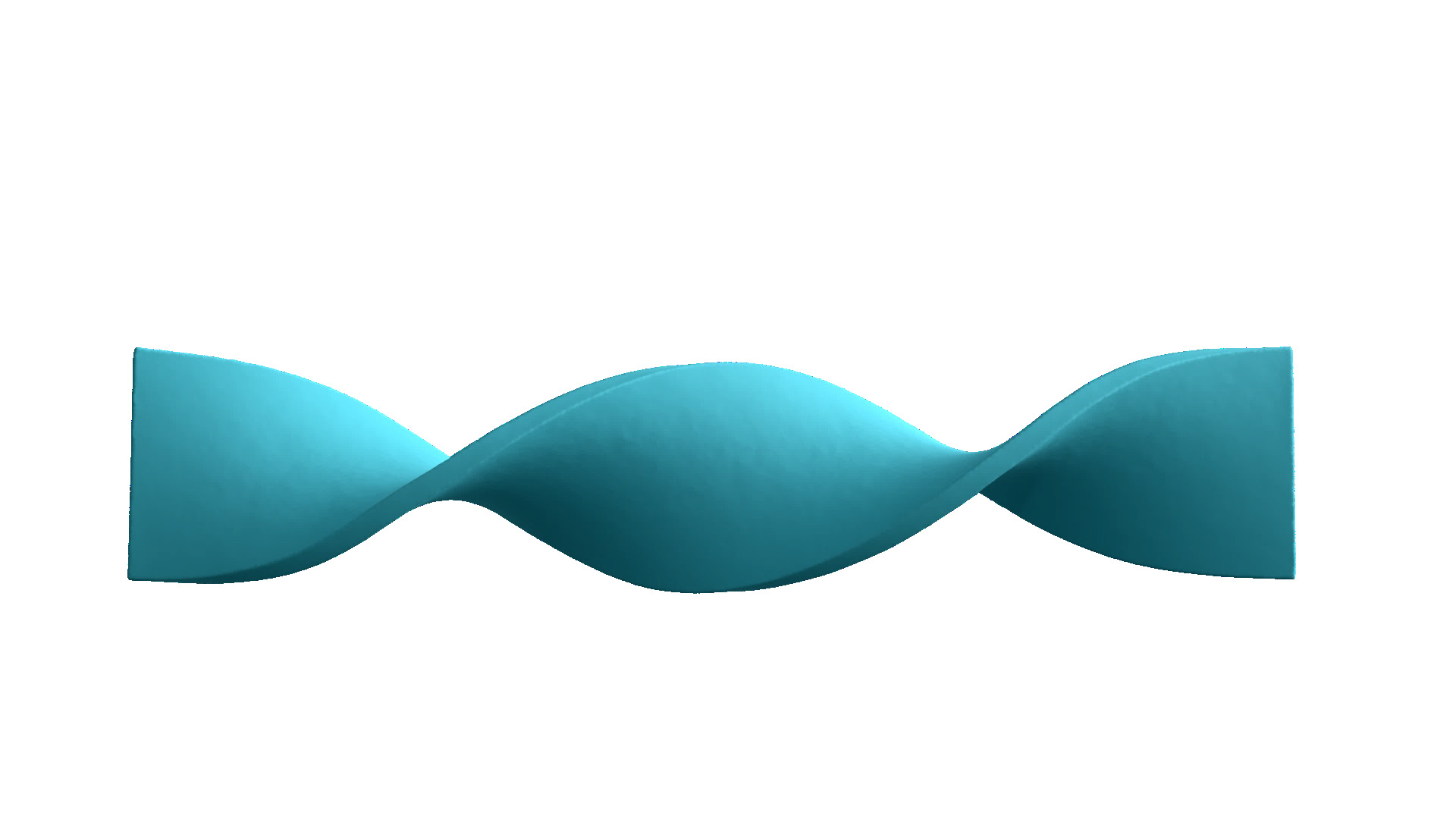}\\
\includegraphics[width=\linewidth,trim=100 300 150 250, clip]{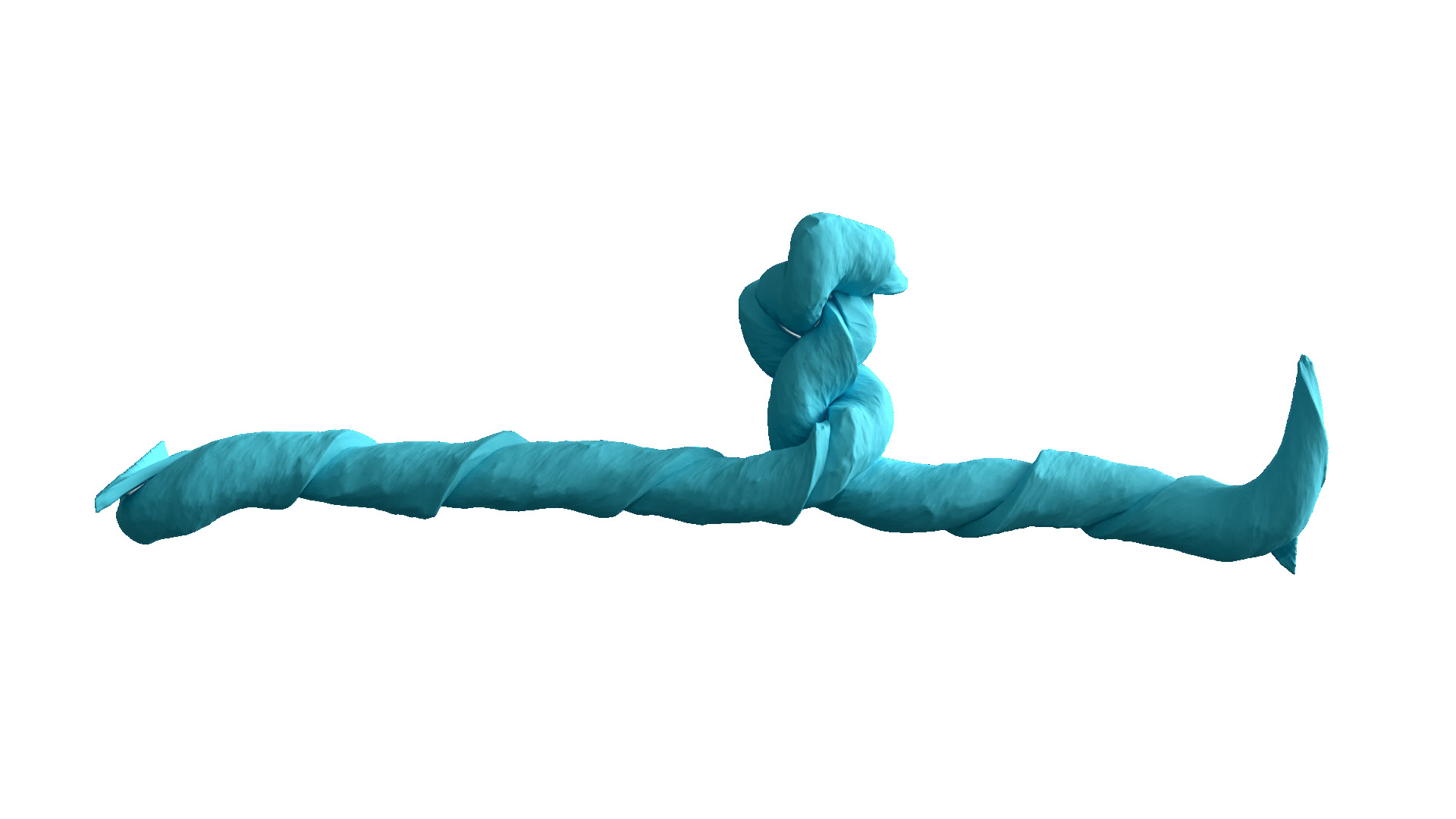}\\
\includegraphics[width=\linewidth,trim=160 450 140 450, clip]{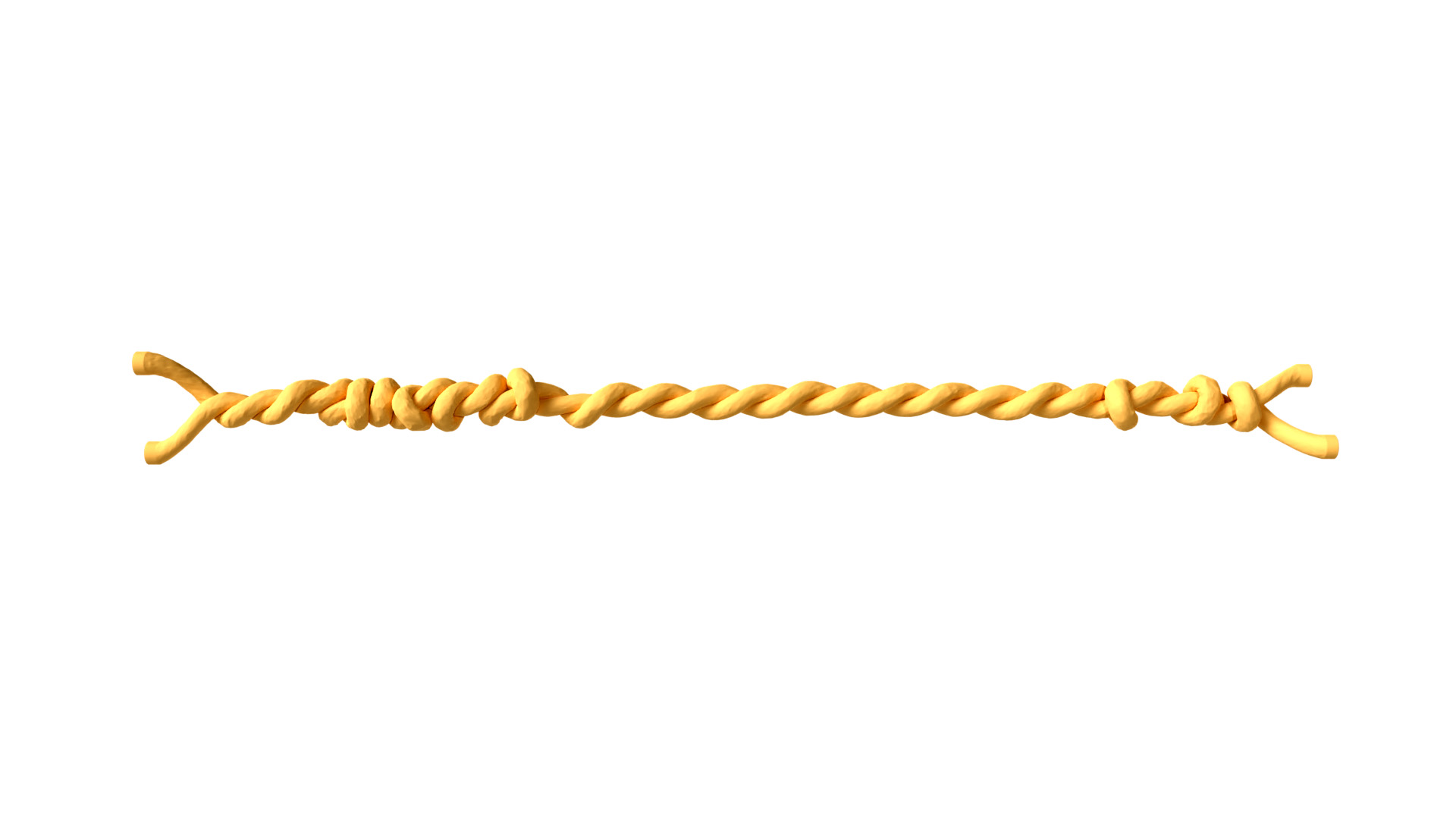}
\caption{Twisting (top-middle)~a thin beam with 400K vertices \& 1.9M tetrahedra, and (bottom)~two elastic rods with 281K vertices \& 1.3M tetrahedra, demonstrating unpredictable self-collisions and buckling.}
\label{fig:08_ExtremeTwist}
\end{figure}

\autoref{fig:08_ExtremeTwist} shows two challenging examples of self-collisions caused by twisting a thin beam and two elastic rods.
Both instances have shown notable buckling after the twisting. A different frame for the same thin beam is also included in \autoref{fig:teaser}.
Such self-collisions are particularly challenging for prior self-collision handling methods that pre-split the model into pieces, since it is unclear where the self-collisions might occur before the simulation.

\begin{figure}
\centering
\includegraphics[height=0.25\linewidth]{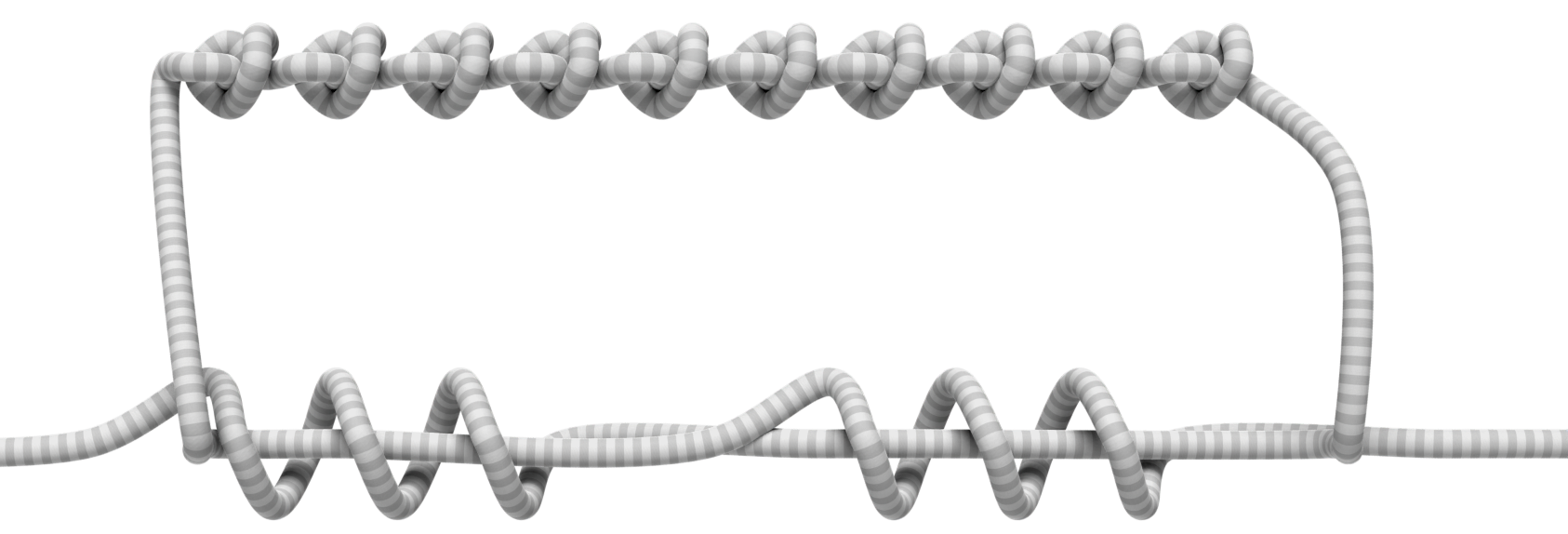}\hfill%
\includegraphics[height=0.25\linewidth]{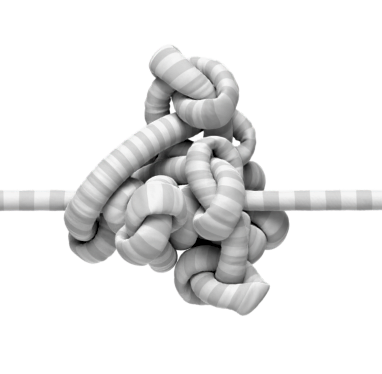}
\caption{Simulation of a nested knot starting from (left)~the initial state to (right)~the final knot.
}
\label{fig:09_Knots}
\end{figure}

Another challenging self-collision case is shown in \autoref{fig:09_Knots}, where nested knots were formed by pulling an elastic rod from both sides.
In this case, there is a significant amount of sliding frictional self-contact, changing pairs of elements that collide with each other.
Though a substantial amount of force is applied near the end, the simulation is able to form a stable knot.

\autoref{fig:10_2SquishyballColliding_traverse:without} shows the same experiment using naive closest boundary point computation for the collisions between the two squishy balls (by picking the closest boundary point on the other object purely based on Euclidean distances), only handling self-collisions with our method. Notice that it includes (temporarily) entangled tentacles between the two squishy balls and visibly more deformations of tentacles elsewhere, as compared to using our method (\autoref{fig:10_2SquishyballColliding_traverse:with}). This is because, in the presence of self-collisions, naively handling closest boundary point queries between different objects is prone to picking incorrect boundary points that do not resolve the collision, resulting in prolonged contact and inter-locking.

\subsection{Solving Existing Intersections}



\begin{figure*}
\centering
\newcommand{\addfig}[2]{%
\begin{subfigure}{0.25\linewidth}
\includegraphics[width=\linewidth,trim=300 0 330 100, clip]{Figures/12_SovlingExistingCollision_Spaghetti/#1}
\caption{#2}\label{fig:12_SovlingExistingCollision_Spaghetti:#1}
\end{subfigure}}
\addfig{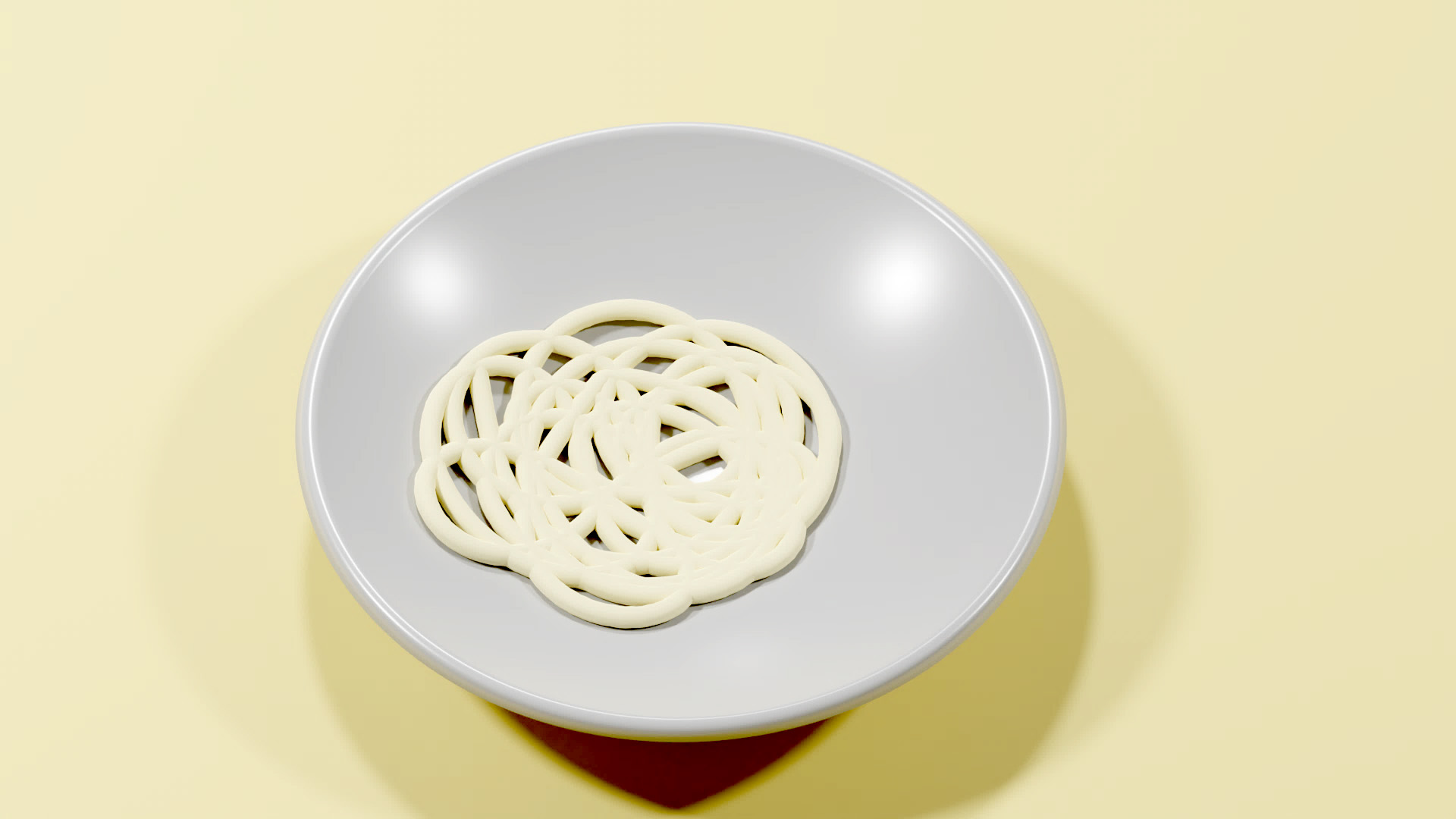}{Initial state}\hfill%
\addfig{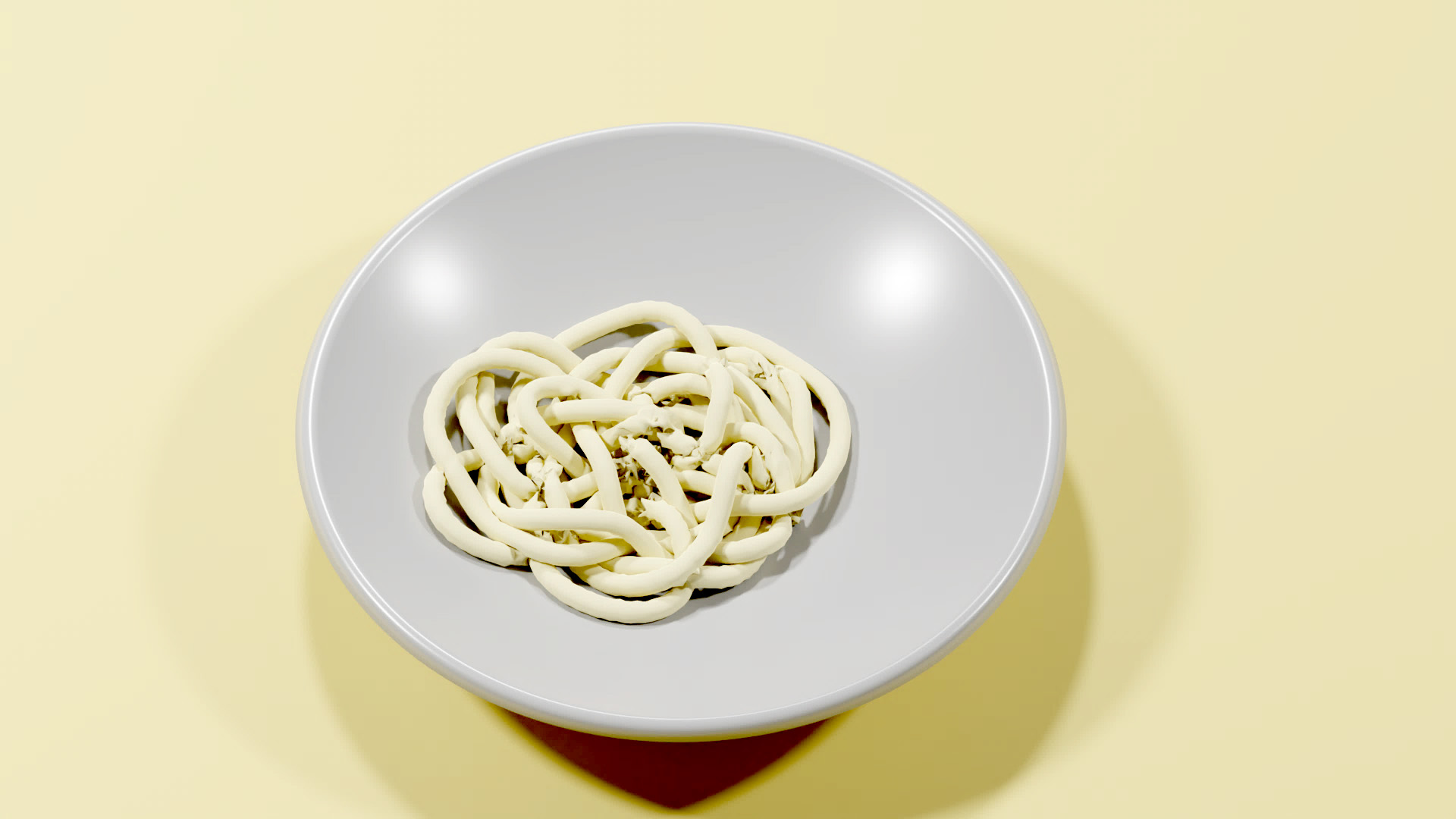}{Collisions enabled}\hfill%
\addfig{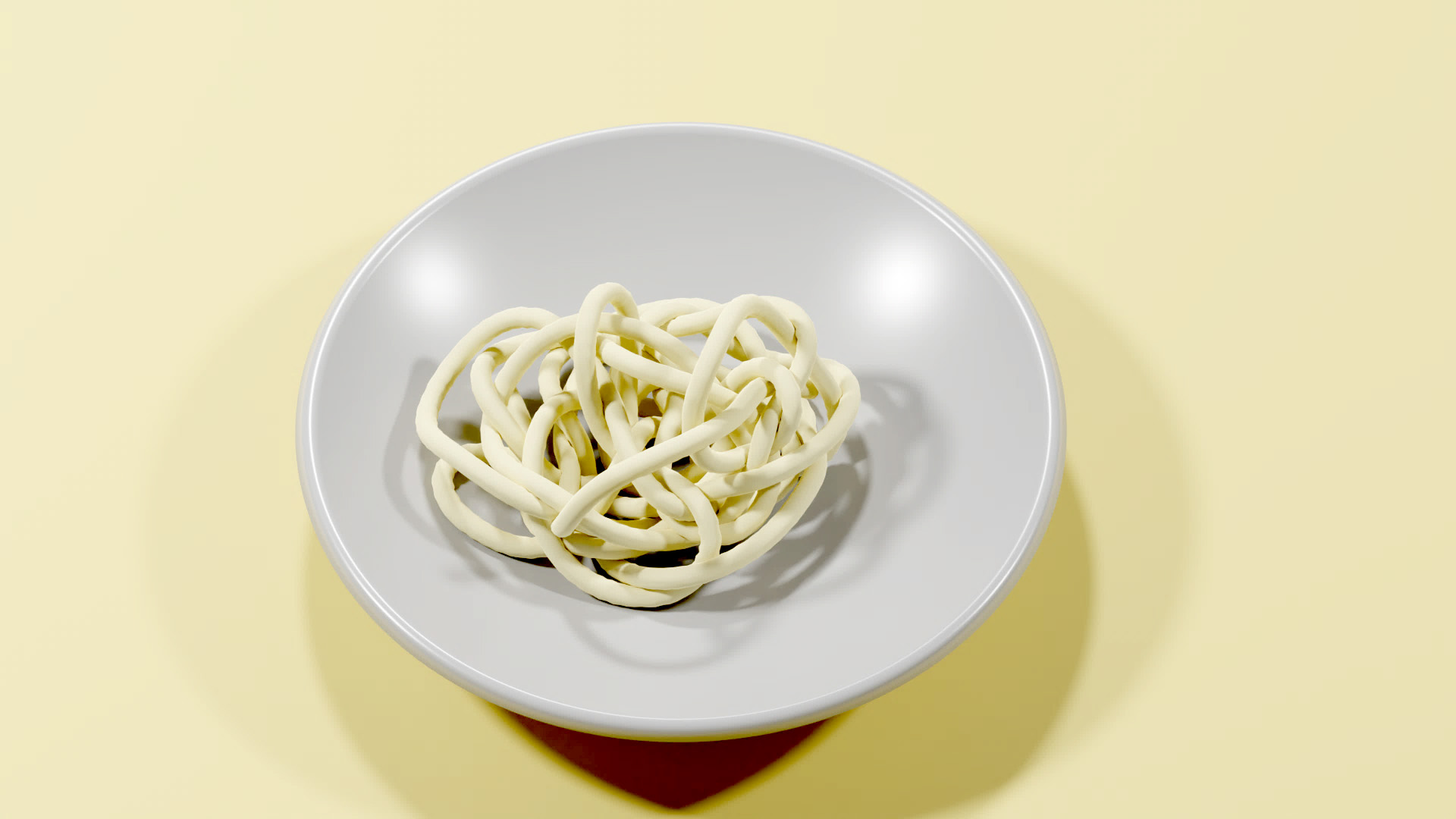}{Collisions resolved}\hfill%
\addfig{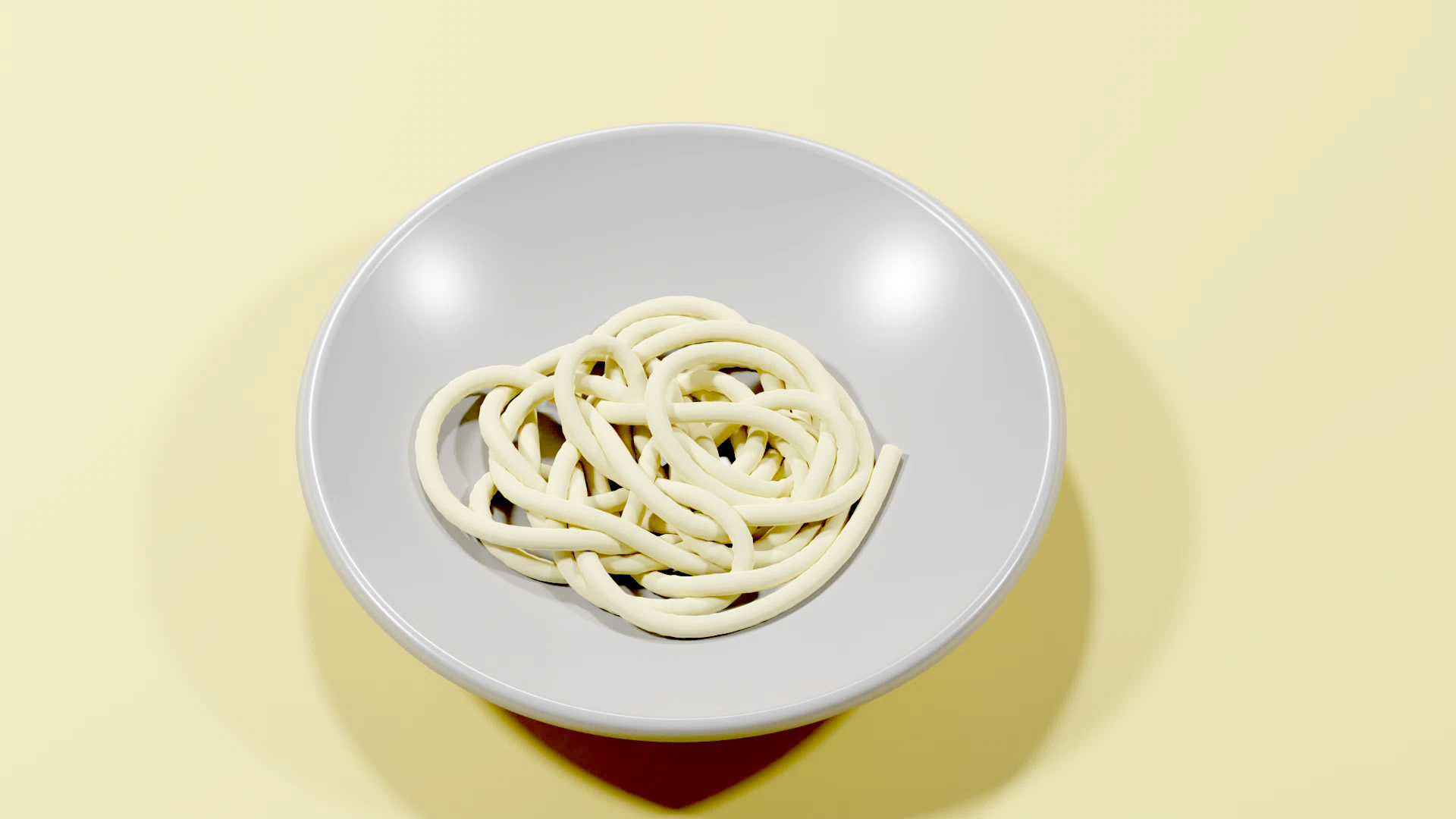}{Final state}\hfill%
\caption{Solving existing collisions. Starting from (a)~an initial state with many self-collisions, (b)~after collision handling is enabled, (c)~our method can quickly resolve them, and (d)~achieve a self-collision-free final state.}
\label{fig:12_SovlingExistingCollision_Spaghetti}
\end{figure*}

Our method can successfully resolve existing self collisions. A demonstration of this is provided in \autoref{fig:12_SovlingExistingCollision_Spaghetti}. In this example, the initial state (\autoref{fig:12_SovlingExistingCollision_Spaghetti:a}) is generated by dropping a noodle model without handling self-collisions. When we turn on self-collisions, all existing self-intersections are quickly resolved within 10 substeps (\autoref{fig:12_SovlingExistingCollision_Spaghetti:b}), resulting in numerous inverted elements due to strong collision constraints. Then, the simulation resolves them (\autoref{fig:12_SovlingExistingCollision_Spaghetti:c}) and finally the model comes to a rest with self-contact (\autoref{fig:12_SovlingExistingCollision_Spaghetti:d}).
In this experiment, we perform collision projections for all vertices (not only for surface vertices) and all tetrahedra's centroids to resolve the intersection in the completely overlapping parts. 
Note that we do not provide a theoretical guarantee to resolve all the existing intersections. In practice, however, in all our tests
all collisions are resolved after just a few iterations/substeps.

\begin{figure}
\centering
\newcommand{\fig}[3]{\begin{subfigure}{0.495\linewidth}\centering
\includegraphics[width=\linewidth,trim=700 400 700 400,clip]{Figures/10_2Squishyball Colliding/#1}
\caption{#2}\label{fig:10_2SquishyballColliding_traverse:#3}
\end{subfigure}}
\fig{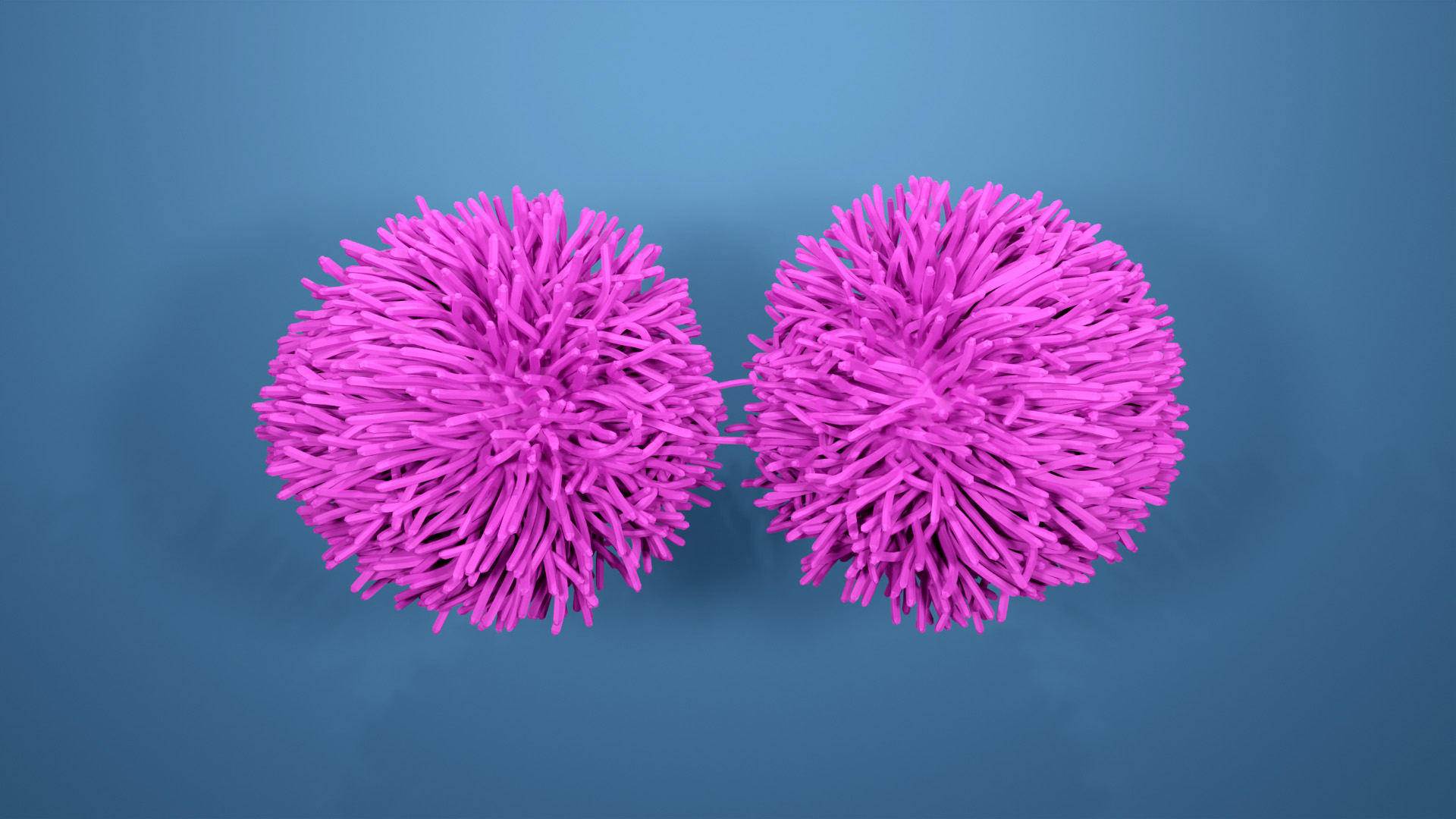}{Naive collisions between objects}{without}\hfill%
\fig{2balls4}{Our inter-object collisions}{with}
\caption{Two squishy balls in head-on collision comparing collision handling between two different objects 
(a)~by naively accepting the Euclidean closest point as the closest boundary point and (b)~with our method.
All self-collisions are handled using our method in both cases.} 
\label{fig:10_2SquishyballColliding_traverse}
\vspace{1em}
\end{figure}

\begin{figure}
\centering
\newcommand{\fig}[2]{\begin{subfigure}{0.495\linewidth}\centering
\fbox{\includegraphics[width=\linewidth,trim=340 100 820 100, clip]{Figures/13_SovlingExistingCollision_Squishyball/#1f}}
\caption{#2}\label{fig:13_SovlingExistingCollision_Squishyball:#1}
\end{subfigure}}
\fig{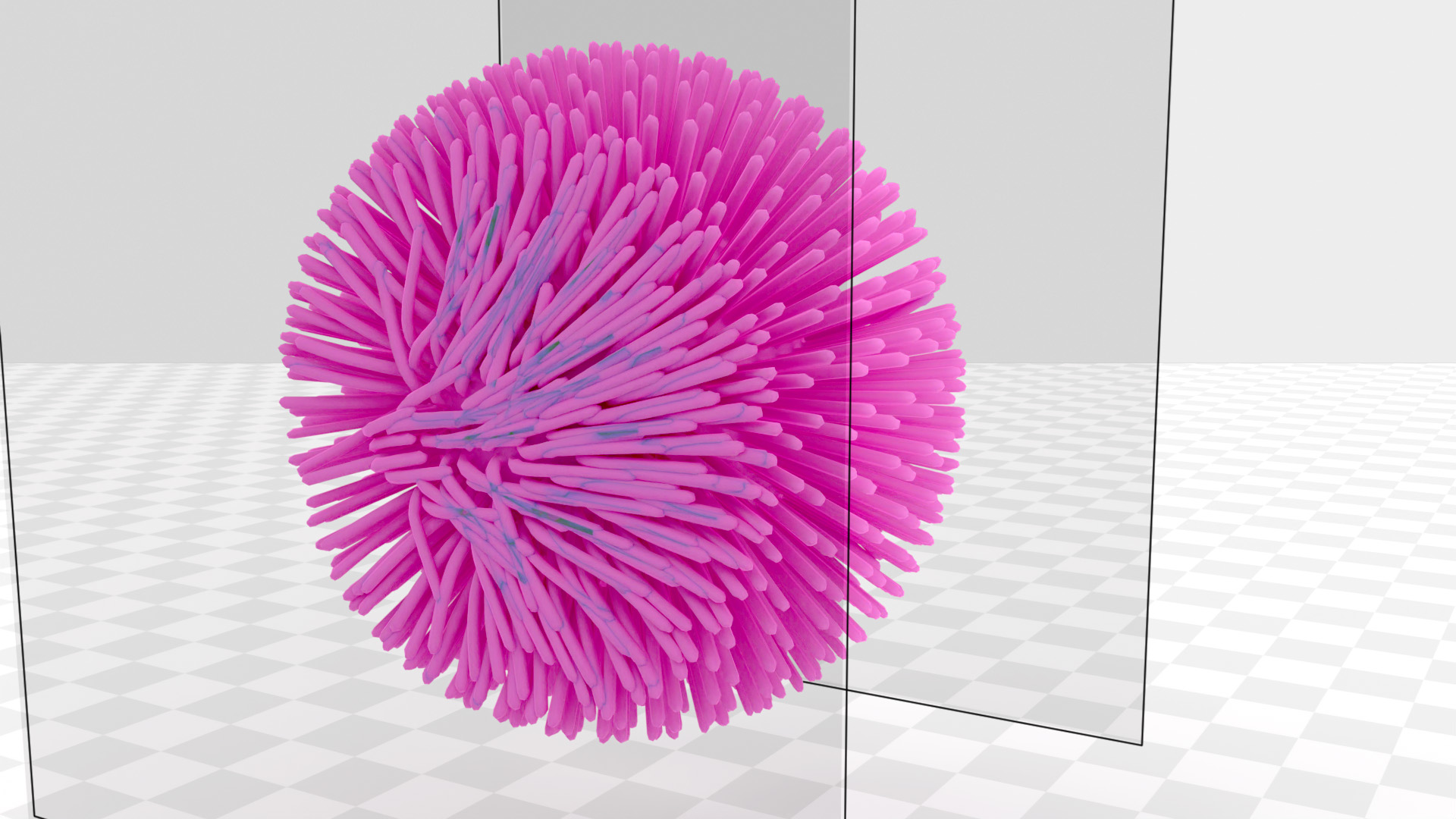}{Self-collisions off}\hfill%
\fig{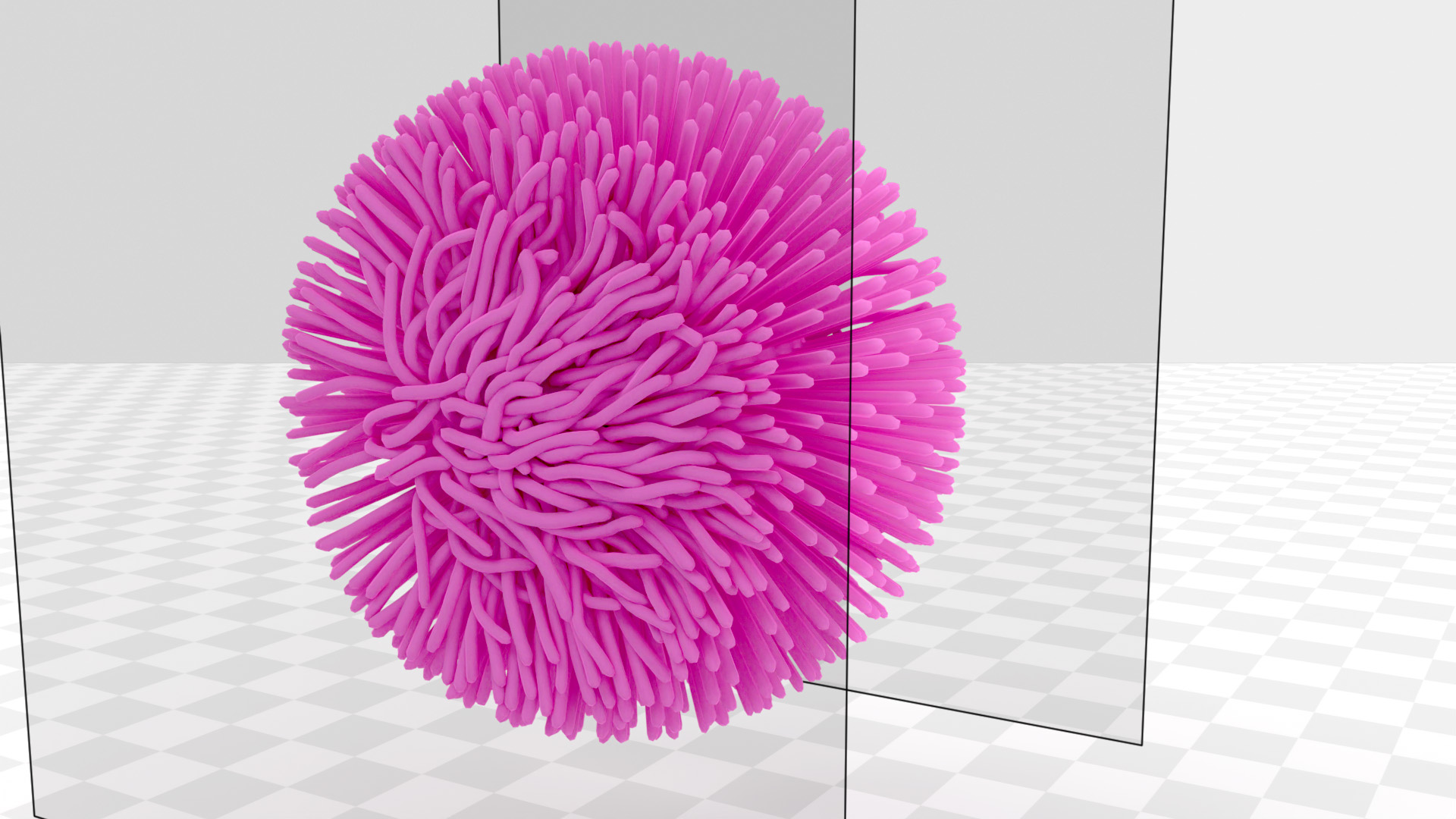}{Self-collisions turned on}
\caption{Simulation of a squishy ball compressed from either side, as in \autoref{fig:SqueezingBall}, (a)~with self-collisions turned off to produce a state with many complex self-collisions, where the blue tint highlights deeper penetrations caused by subsurface scattering artifacts due to intersecting geometry, and (b)~a few frames after self-collisions are turned on, showing that our method with XPBD quickly recovers them.}
\label{fig:13_SovlingExistingCollision_Squishyball}
\end{figure}

\begin{figure*}
\centering
\newcommand{\fig}[1]{\includegraphics[width=0.248\linewidth,trim=400 0 350 0,clip]{Figures/14_Octopi/#1}}
\fig{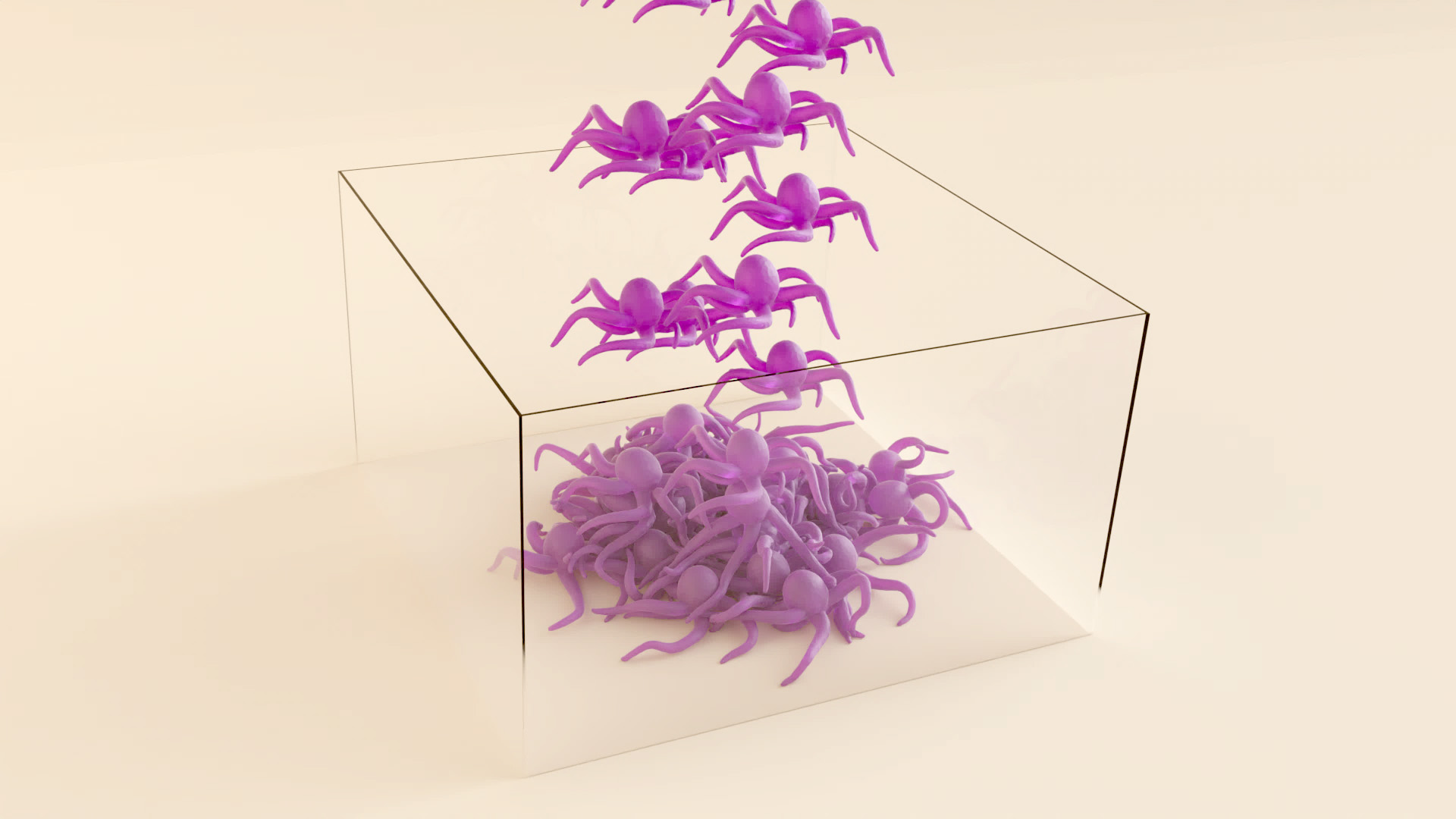}\hfill%
\fig{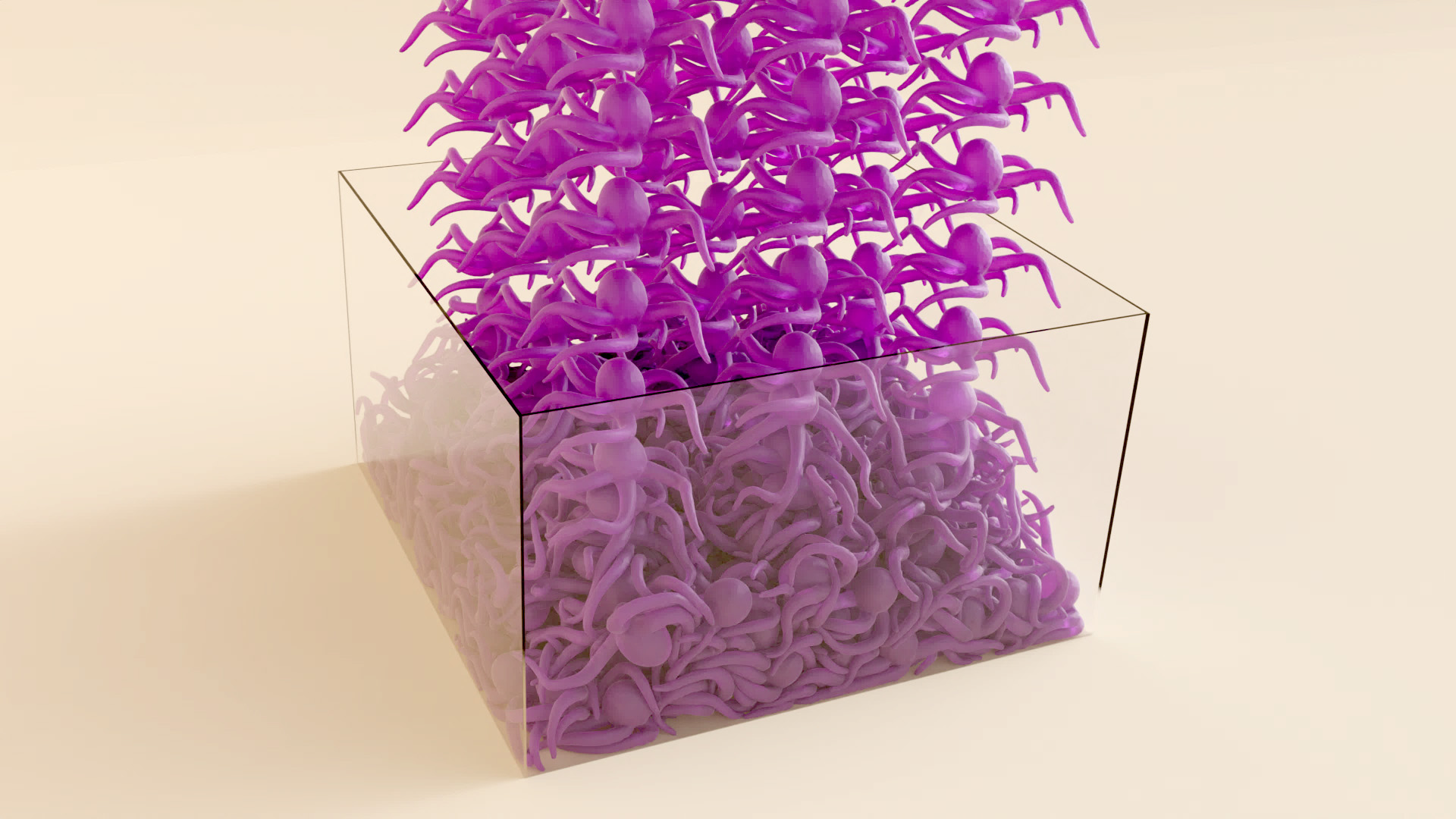}\hfill%
\fig{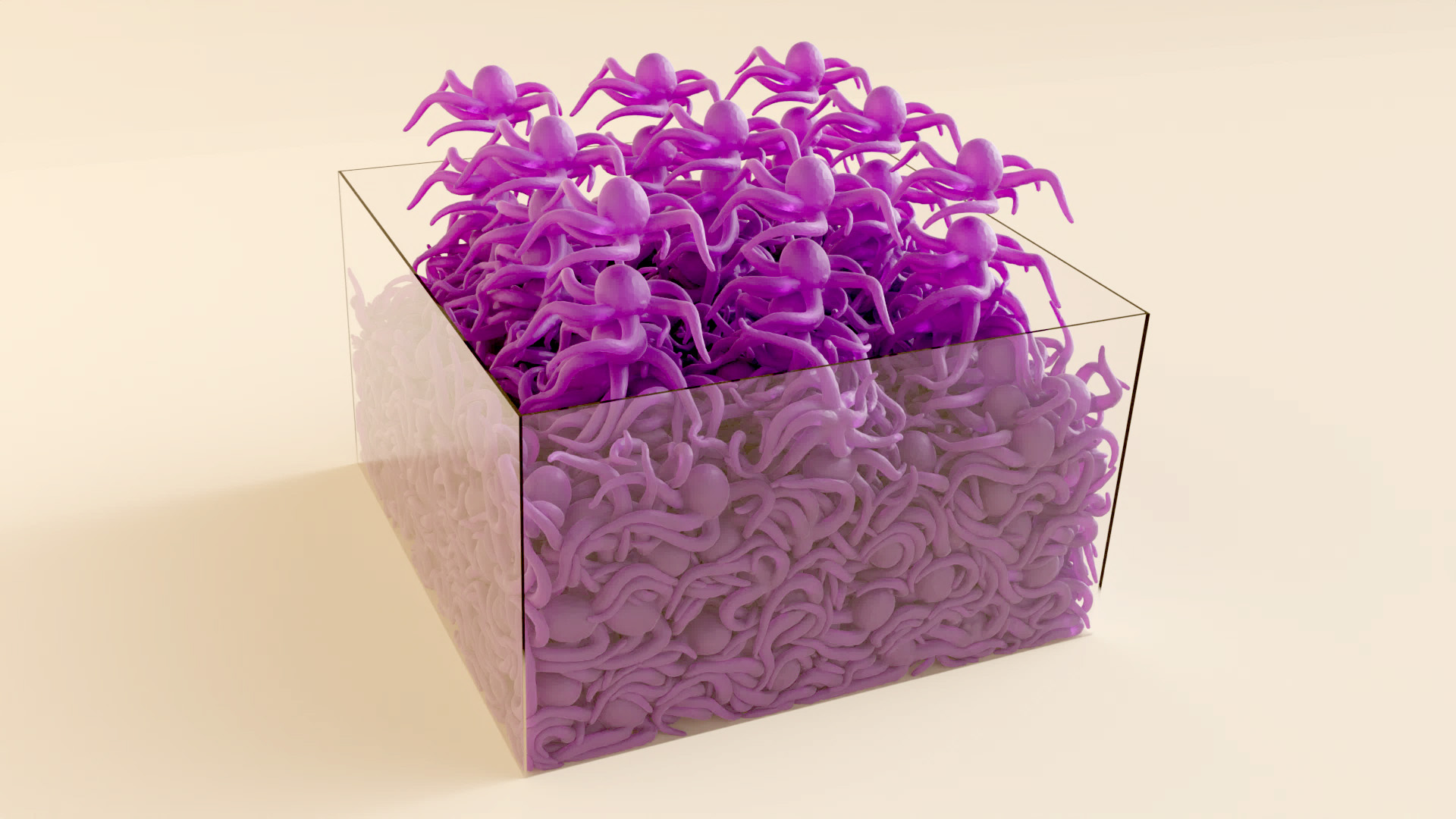}\hfill%
\fig{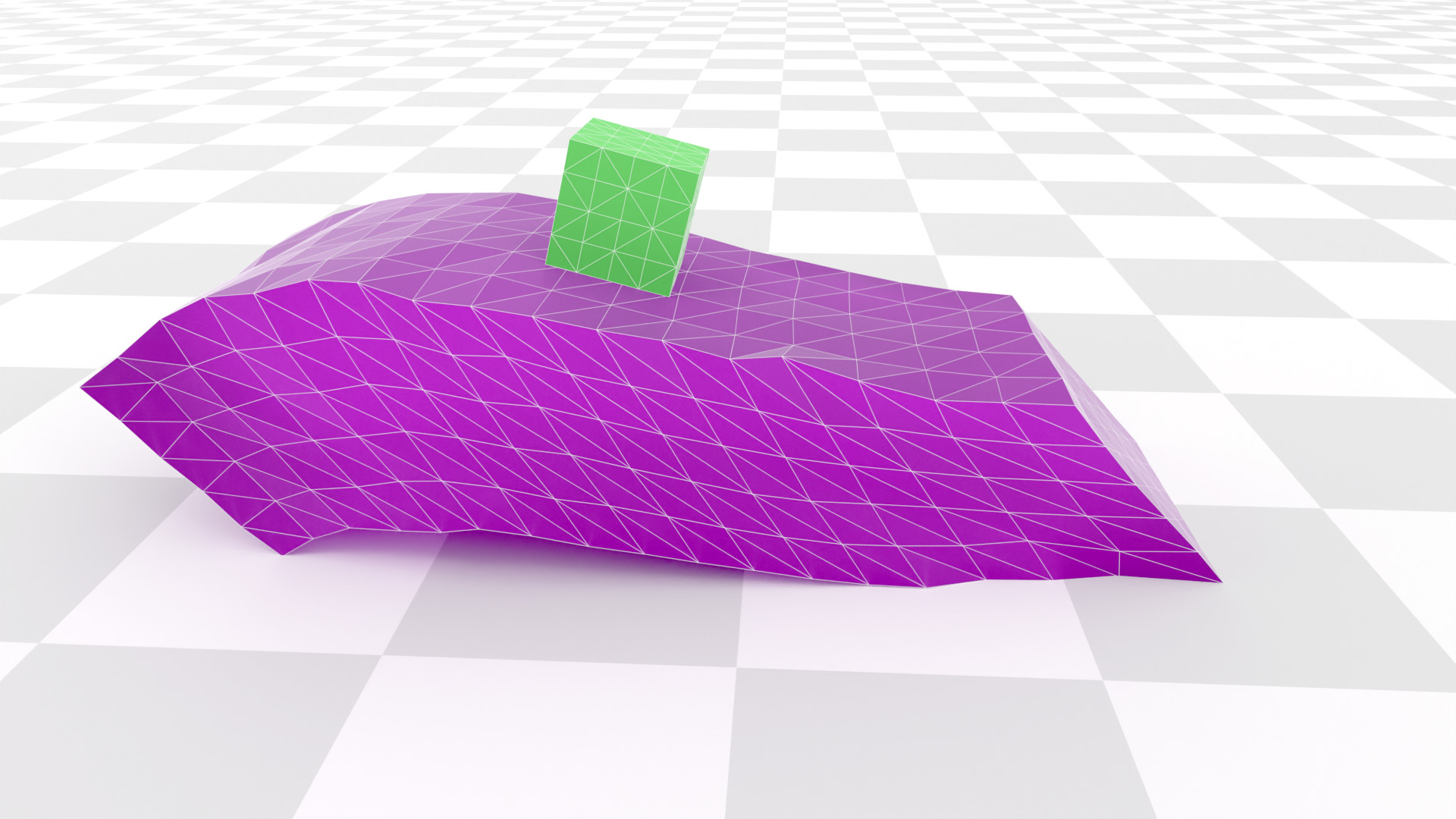}
\caption{\del{Simulating }600 deformable octopus models (3.1M vertices and 8.88M tetrahedra in total) dropped into a container, forming a pile with collisions.}
\label{fig:14_OctopiTest}
\end{figure*}

\begin{figure*}
\centering
\newcommand{\fig}[1]{\includegraphics[width=0.248\linewidth,trim=360 0 400 0, clip]{Figures/15_16SquishyballsTest/#1}}
\fig{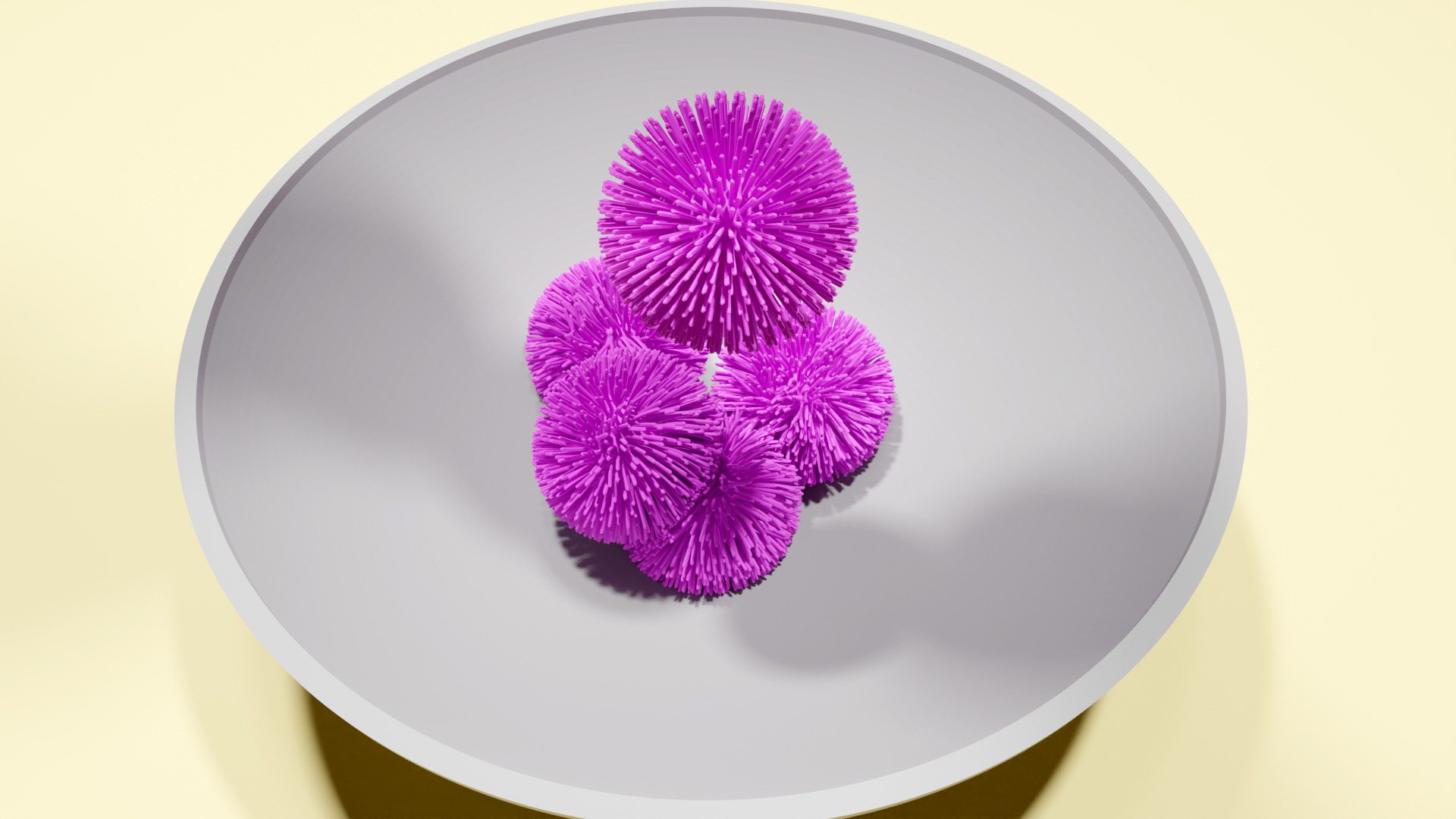}\hfill%
\fig{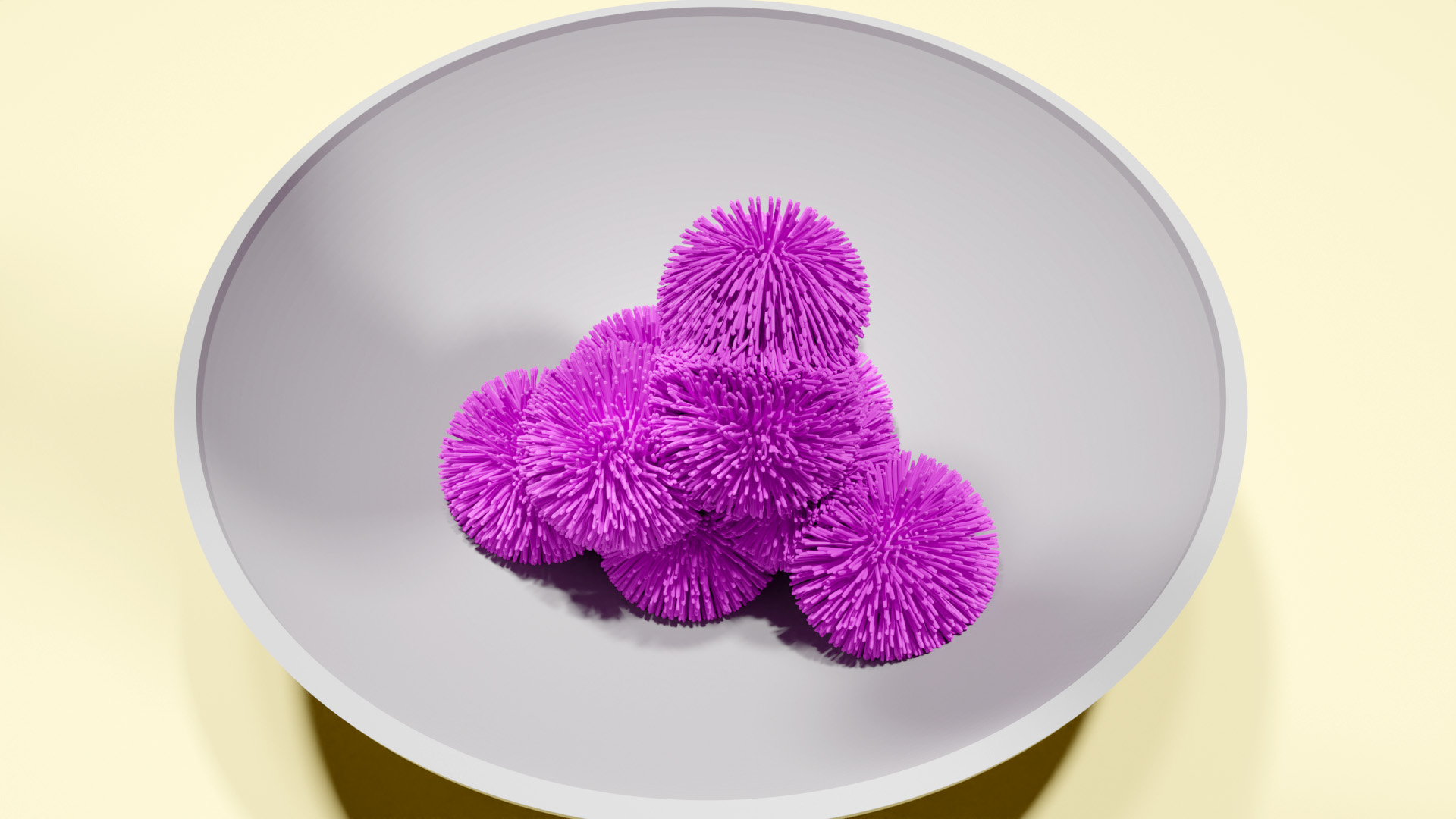}\hfill%
\fig{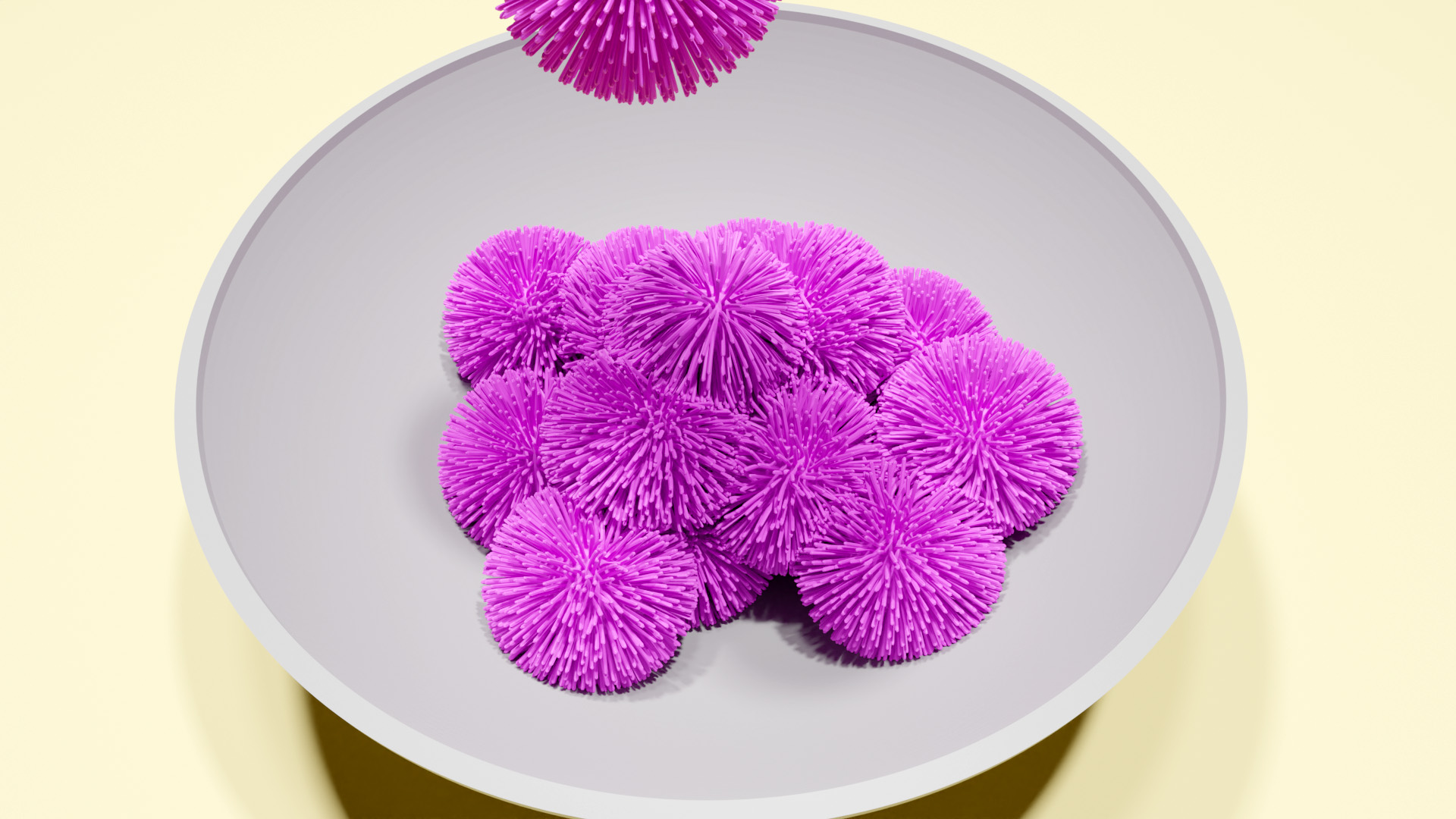}\hfill%
\fig{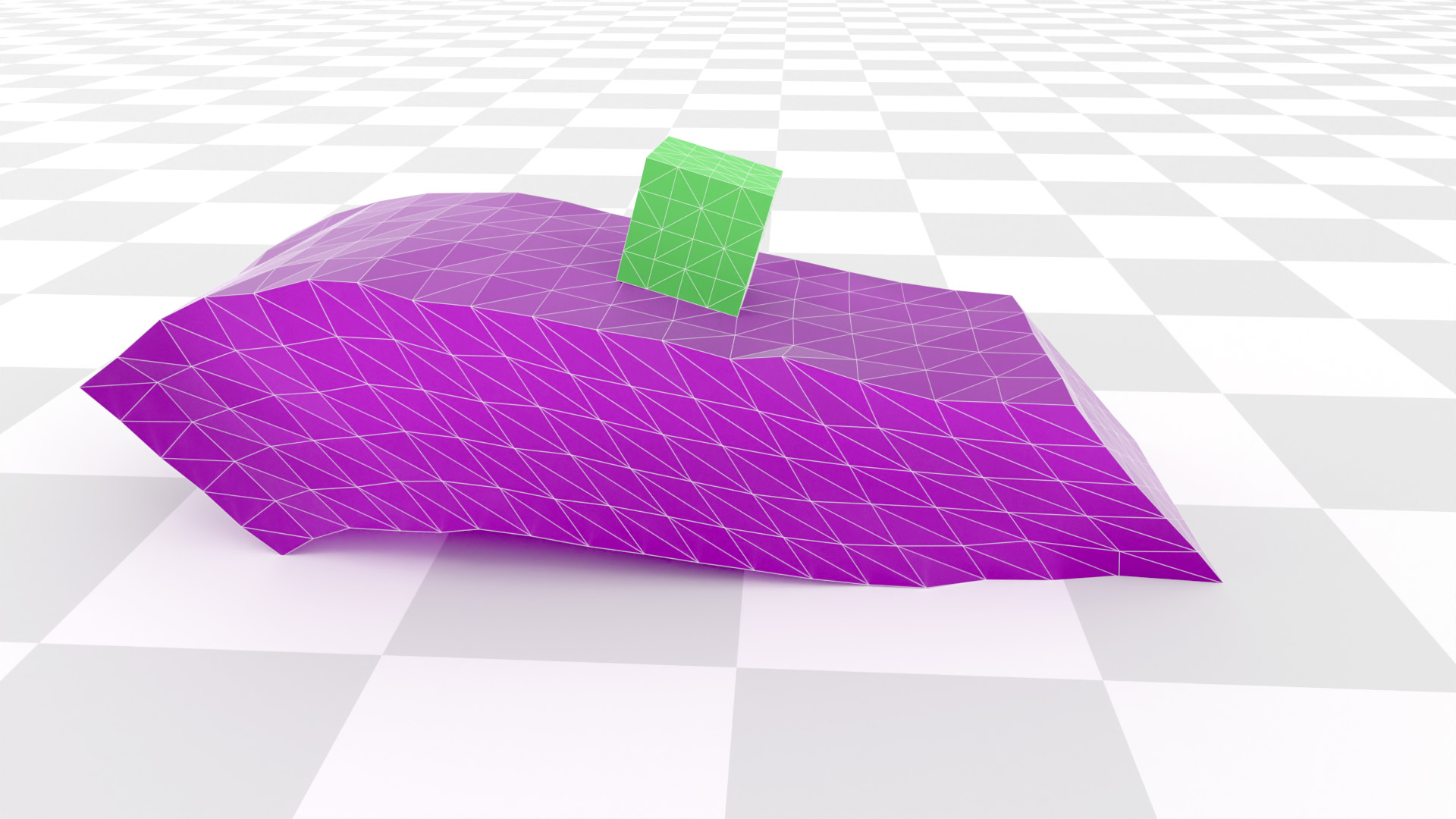}
\caption{Simulation of 16 squishy balls (a total of 11.2 million tetrahedra) dropped into a bowl, forming a stable pile with active collisions.}
\label{fig:15_16SquishyballsTest}
\end{figure*}

\begin{figure*}
\centering
\includegraphics[height=0.205\linewidth]{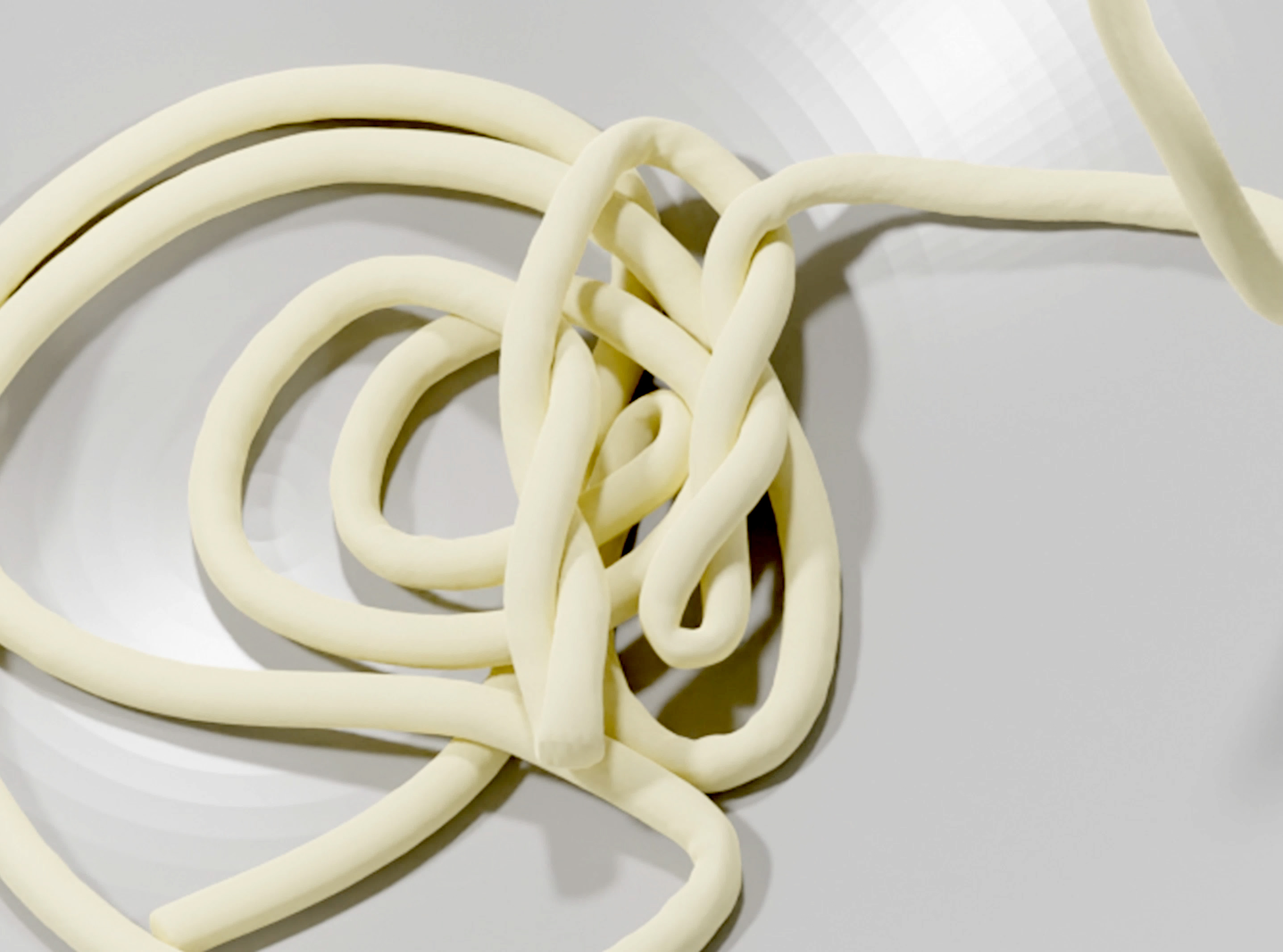}\hfill%
\includegraphics[height=0.205\linewidth]{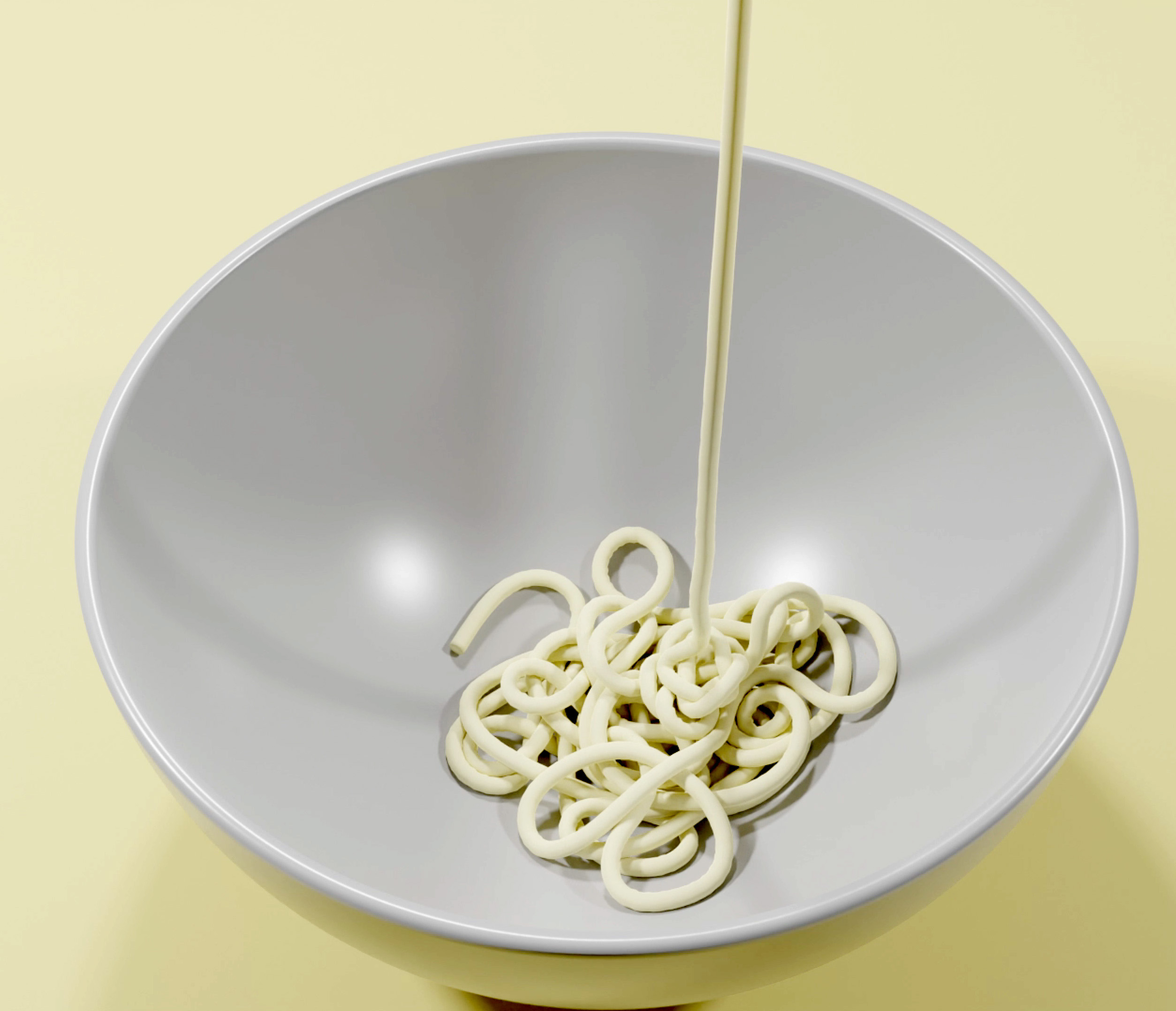}%
\includegraphics[height=0.205\linewidth]{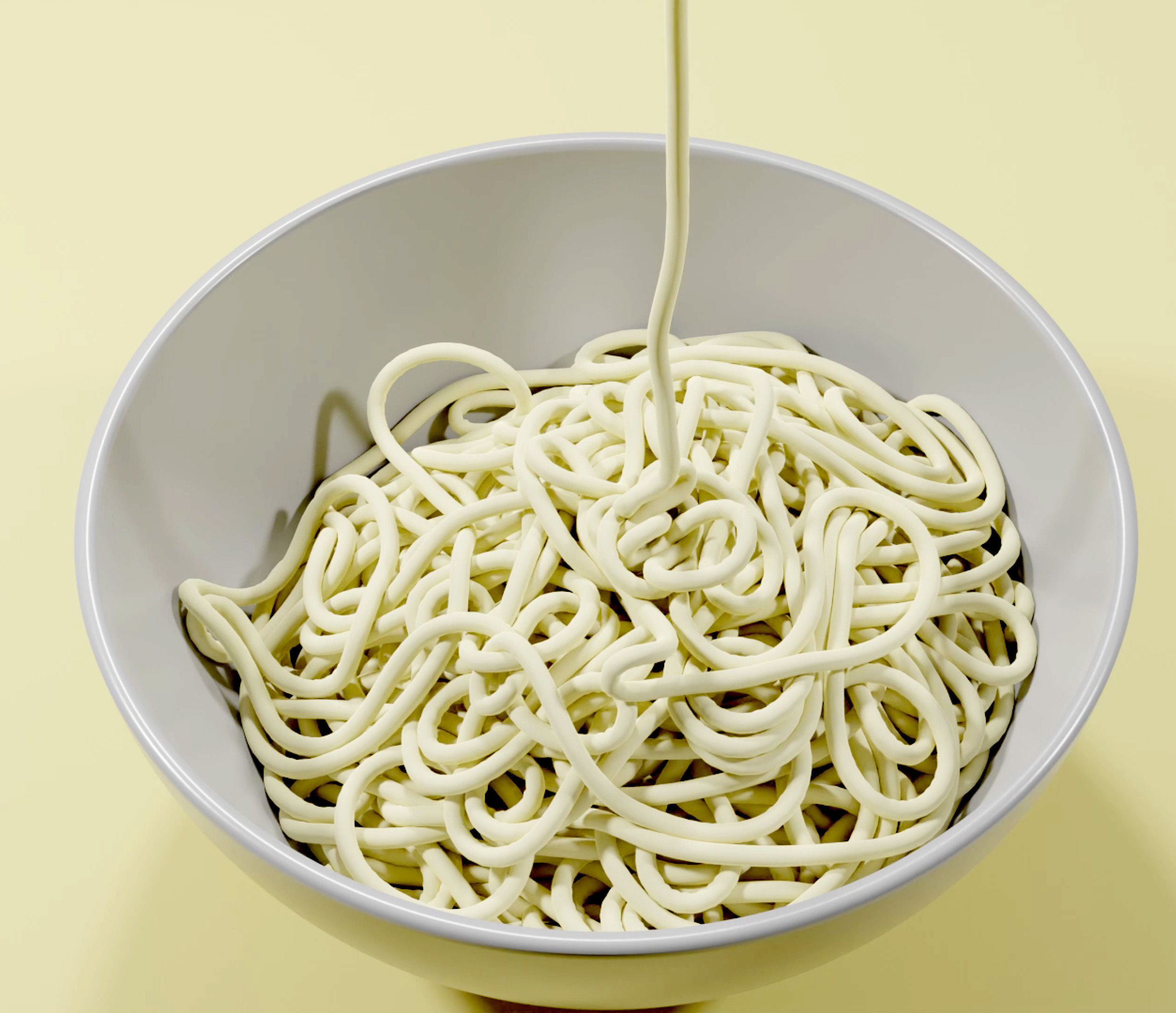}%
\includegraphics[height=0.205\linewidth]{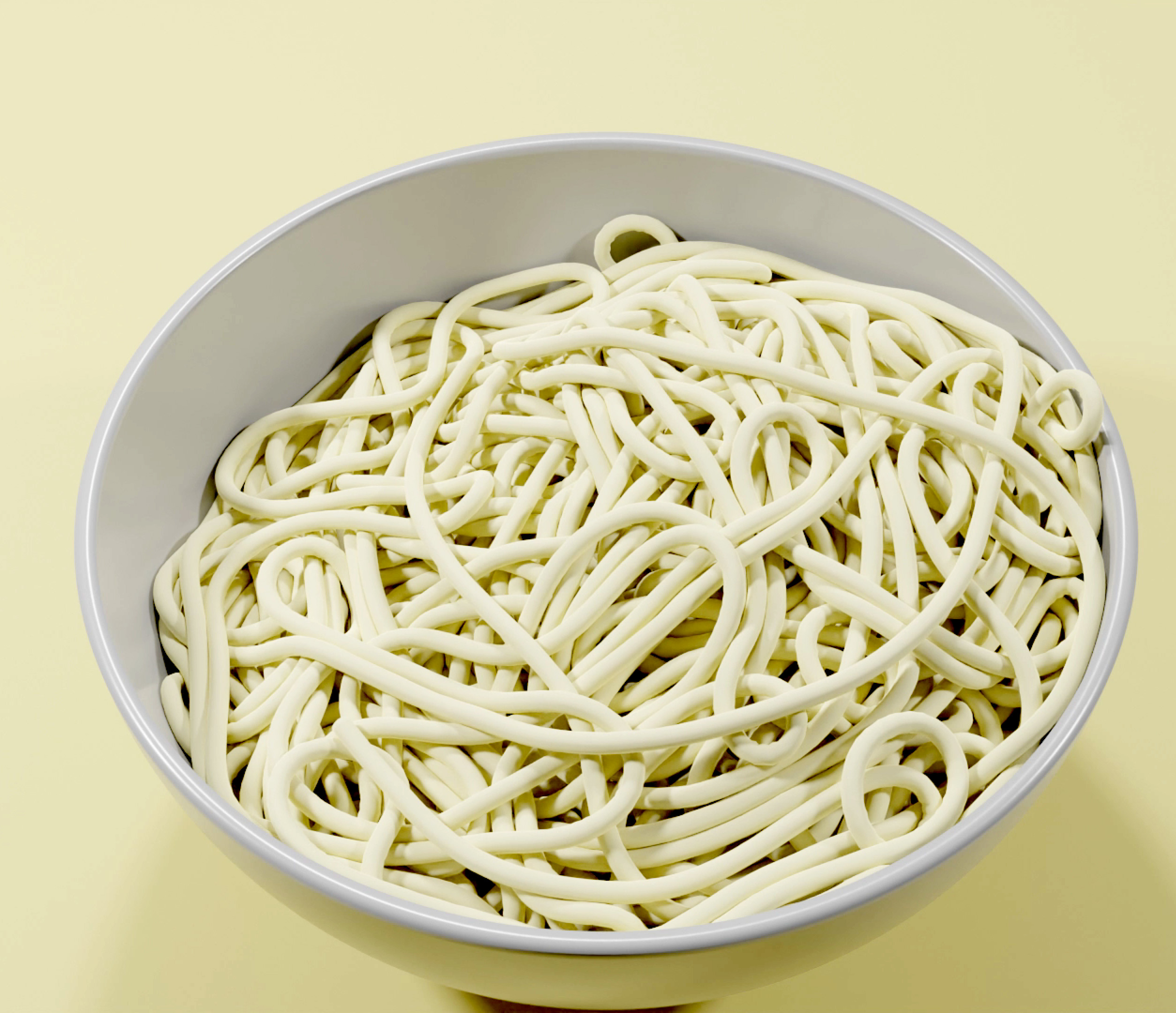}\vspace{-1.5em}
\begin{minipage}{0.27\linewidth}\begin{flushright}\small\textbf{(a)}\end{flushright}\end{minipage}\hfill%
\begin{minipage}{0.71\linewidth}\small\textbf{(b)}\end{minipage}
\caption{Simulation of a long noodle (a)~presenting unpredictable complex self-collisions and (b)~forming a large pile with self-collisions.}
\label{fig:16_200MSpaghetti}
\end{figure*}

Another example is shown in \autoref{fig:13_SovlingExistingCollision_Squishyball}, generated by compressing a squishy ball with two planes on either side, similar to \autoref{fig:SqueezingBall} but without handling self-collisions. This results in a significant number of complex unresolved self-collisions (\autoref{fig:13_SovlingExistingCollision_Squishyball:before}), which are quickly resolved within a few substeps when self-collision handling is turned on (\autoref{fig:13_SovlingExistingCollision_Squishyball:after}).

\subsection{Large-Scale Experiments}

An important advantage of our method is that, by providing a robust collision handling solution, we can use fast simulation techniques for scenarios involving a large number of objects and complex collisions.
%
%
An example of this is demonstrated in \autoref{fig:14_OctopiTest}, showing 600 deformable octopus models forming a pile. 
Due to its complex geometry, the octopus model can cause numerous self-collisions and inter-object collisions. Both collision types are handled using our method. 
At the end of the simulation, a stable pile is formed with 185K active collisions per time step.

\autoref{fig:15_16SquishyballsTest} shows another large-scale experiment involving 16 squishy balls.
Another frame from this simulation is also included in \autoref{fig:teaser}.
At the end of the simulation, the squishy balls form a stable pile and remain in rest-in-contact with active self-collisions (12K) and inter-object collisions (125K) between neighboring squishy balls.

We also include an experiment with a single long noodle piece in \autoref{fig:16_200MSpaghetti} that is dropped into a bowl. This simulation forms numerous complex and unpredictable self-collisions (\autoref{fig:16_200MSpaghetti}a).
At the end of the simulation, we achieve a stable pile with 104K active self-collisions per time step in this example. \autoref{fig:teaser} includes a rendering of this final pile without the bowl and a cross-section view, showing that the interior self-collisions are properly resolved.

\begin{figure}[]
    \centering
    \includegraphics[width=\columnwidth]{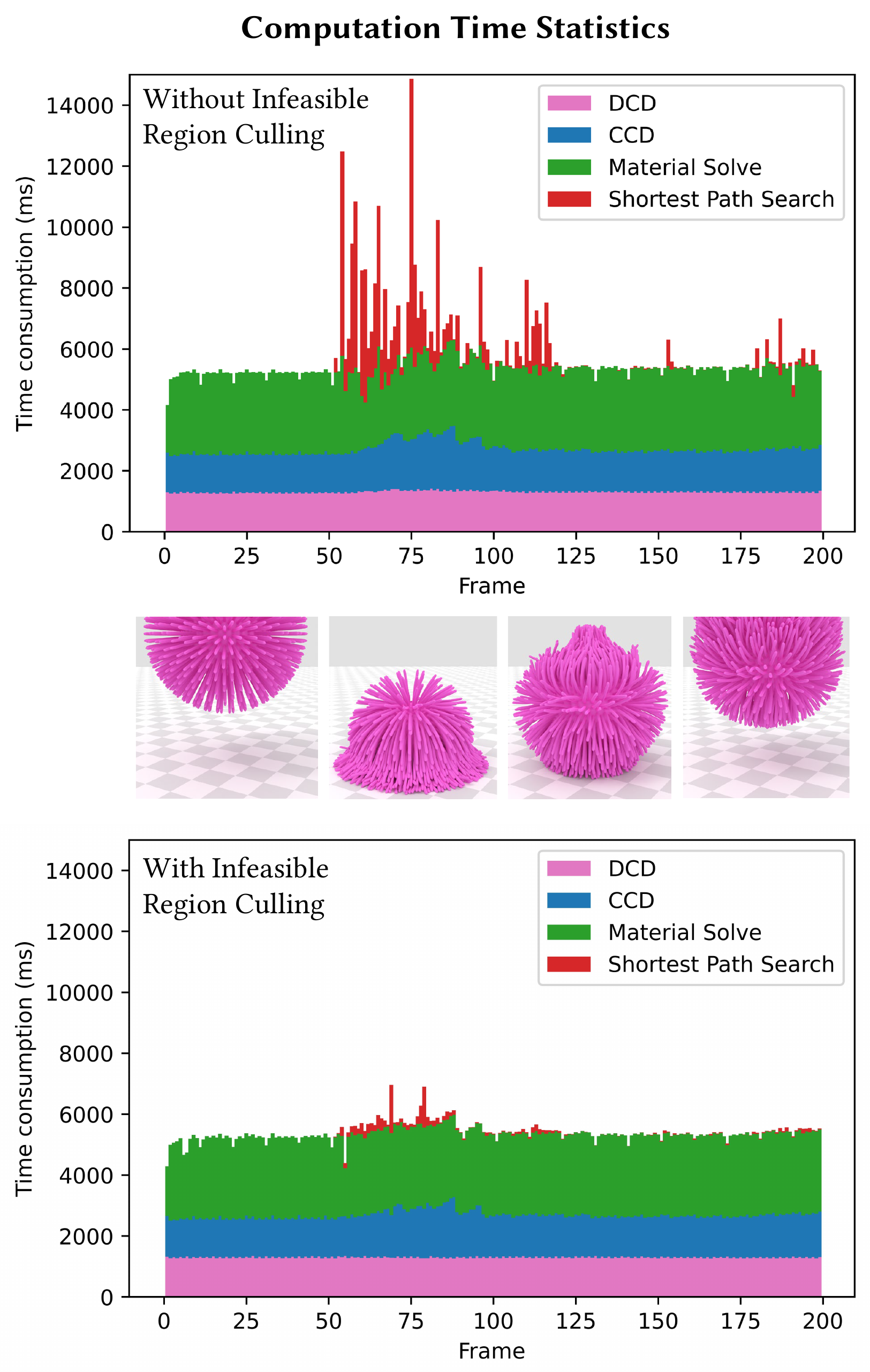}
    \caption{The computation time statistics of each simulation component on stacked bar charts: (top)~without infeasible region culling and (bottom)~with infeasible region culling. The middle row shows the simulation at frames 0, 75, 125, 175 respectively. }
    \label{fig:17_AddmisbleRegionAcceleration}
\end{figure}

\subsection{Performance}

We provide the performance numbers for the experiments above in \autoref{tbl:perf}. Notice that, even though we are using a highly efficient material solver that is parallelized on the GPU, our method provides a relatively small overhead. This includes some highly-challenging experiments, involving a large number of complex collisions. The highest overhead of our method is in experiments in which deliberately disabled self-collisions to form a large number of complex self-collisions. Note that all collision detection and handling computations are performed on the CPU, and a GPU implementation would likely result in a smaller overhead.


We demonstrate the effect of our infeasible region culling 
by simulating a squishy ball dropped to the ground with and without this acceleration.
The computation time breakdown of all frames are visualized 
in \autoref{fig:17_AddmisbleRegionAcceleration}. In this example, using our infeasible region culling, the shortest path query gains a speed-up of 10-30$\times$ for some frames, providing identical results. Additionally, the accelerated shortest path query 
results in a more uniform computation time,
avoiding the peaks 
visible in the graph.

\begin{figure}
\centering
\newcommand{\img}[1]{\fbox{\includegraphics[width=\linewidth,trim=50 250 150 100, clip]{Figures/18.5_RestPoseQueryComparison_cuboid/#1}}}
\newcommand{\imgs}[1]{\img{#12}\vspace{-0.2em}\\\img{#13}}
\newcommand{\sfig}[2]{\begin{subfigure}{0.485\linewidth}#1\vspace{-0.3em}\caption{#2}\end{subfigure}}
\sfig{\img{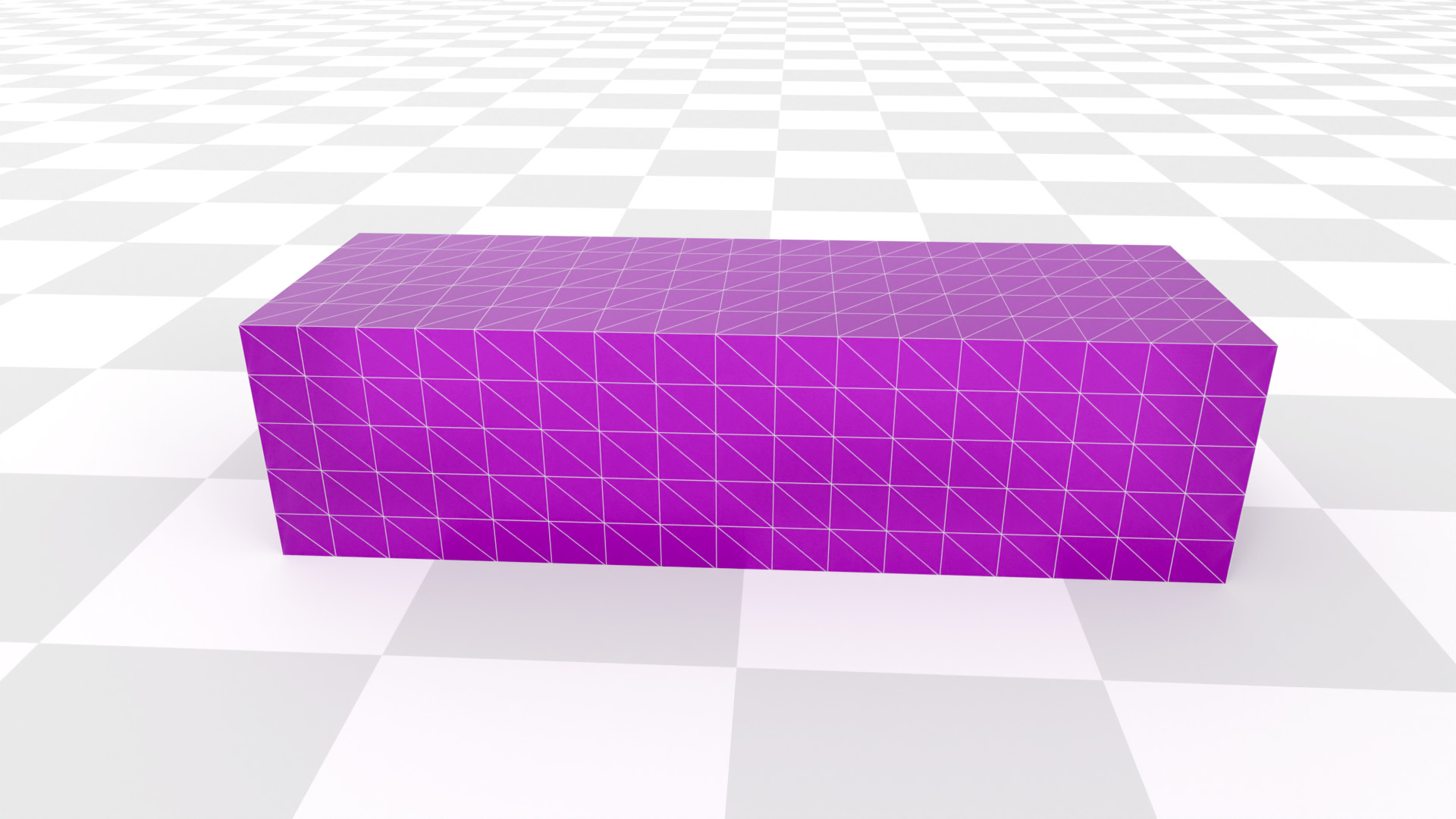}}{Rest \del{pose}\add{shape}}\hfill%
\sfig{\img{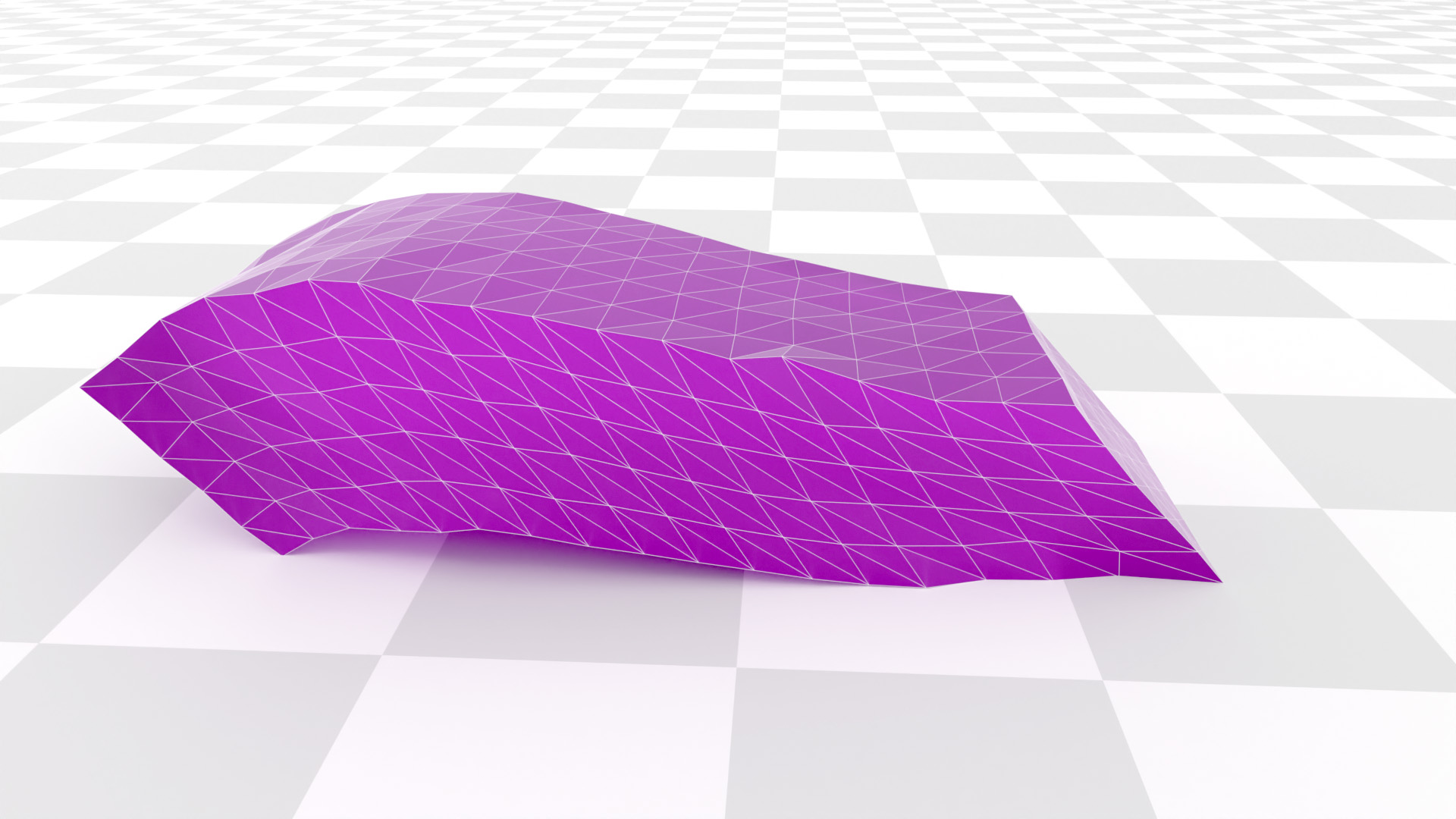}}{Deformed \del{pose}\add{shape}}\vspace{0.5em}
\sfig{\imgs{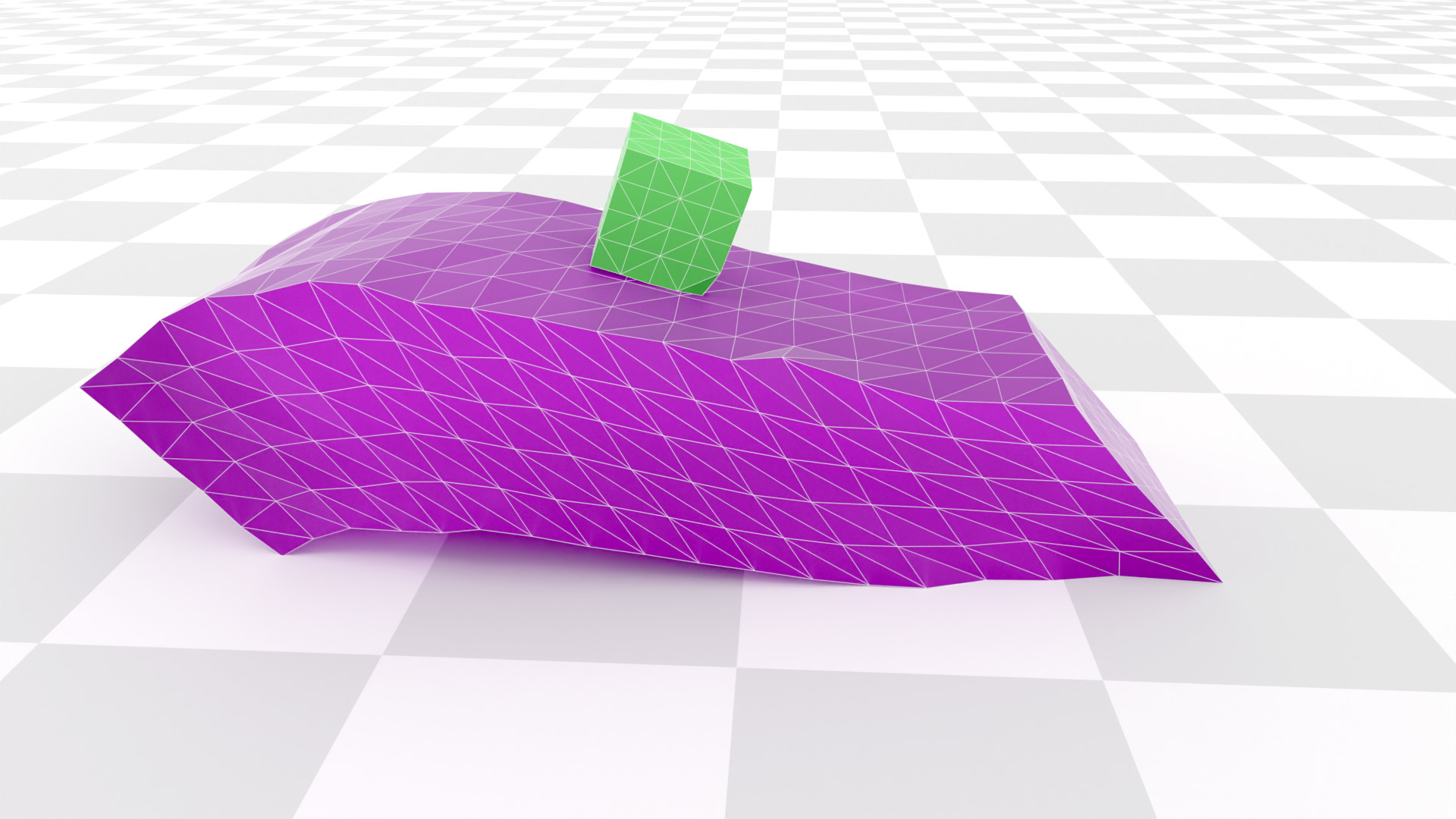}}{Collisions using rest \del{pose}\add{shape}}\hfill%
\sfig{\imgs{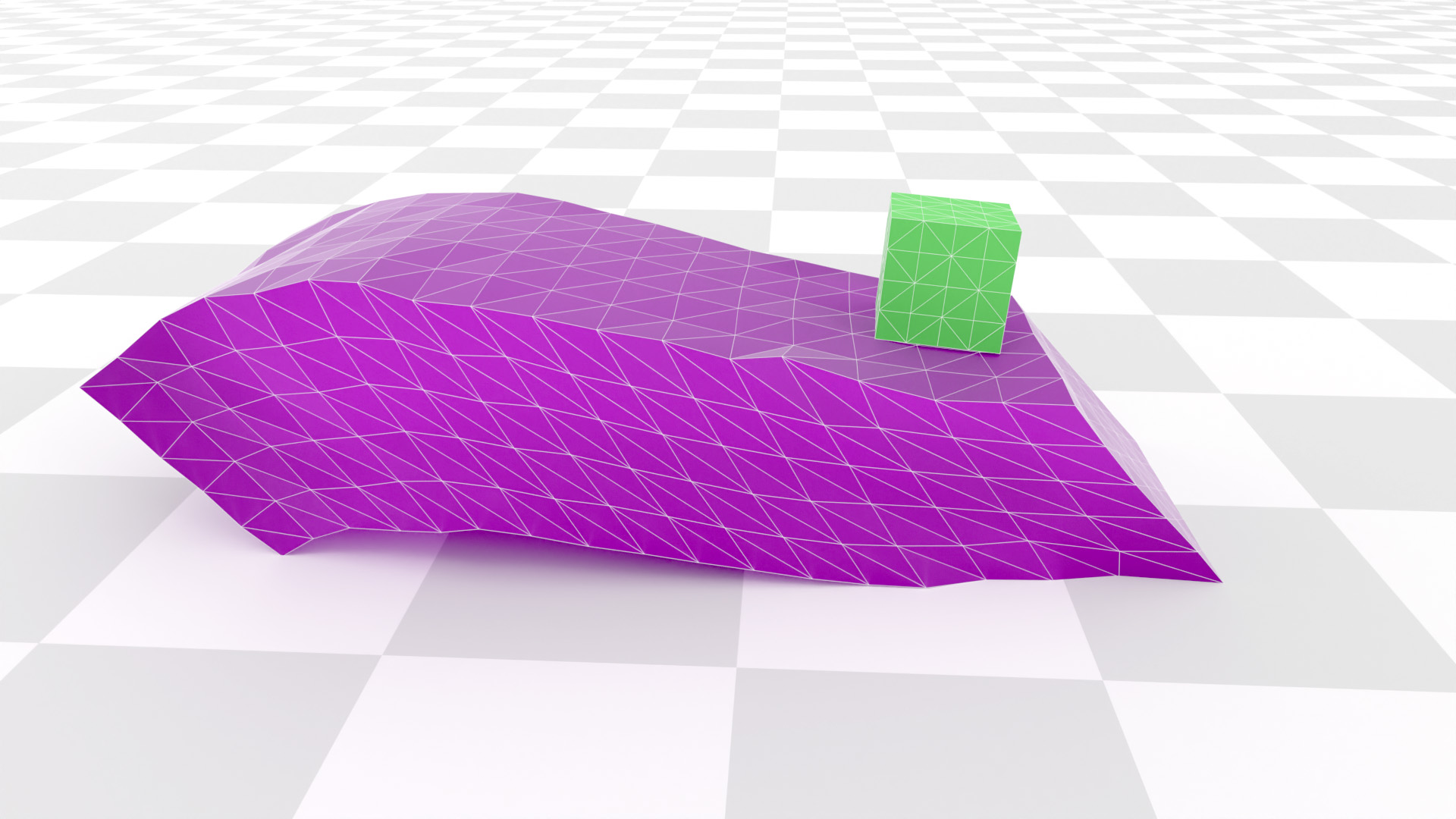}}{Collisions using deformed \add{shape}}
\begin{subfigure}{\linewidth}\centering\includegraphics[width=\linewidth,trim=250 500 550 350, clip]{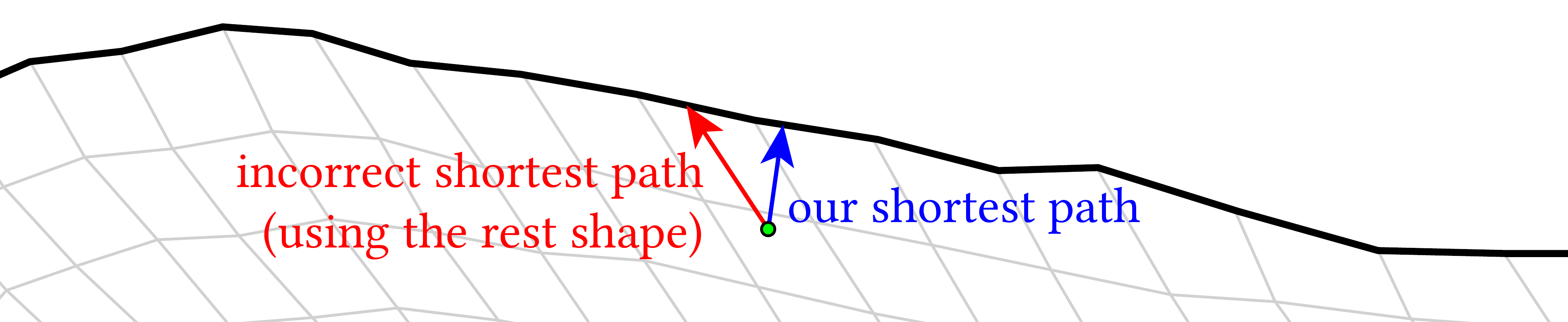}
\caption{The incorrect shortest path using the rest shape vs. ours}
\end{subfigure}
\caption{Comparison between the rest pose closest boundary point and our closest boundary point. (a) The rest pose the cuboid model. (b) We deform the cuboid to a certain shape, then drop a cube on top of it. (c) In the simulation using the rest pose closest boundary point, the cube got incorrectly pulled up. (d) Using our exact closest point, the cube successfully slides down. 
\add{(e)~The shortest path to the surface for an example point, showing that using the closest surface point queried from the rest shape results in an incorrect and longer path.}
}
\label{fig:185_RestPoseQueryComparison_cuboid}
\end{figure}

\begin{figure}
\centering
\includegraphics[width=\columnwidth]{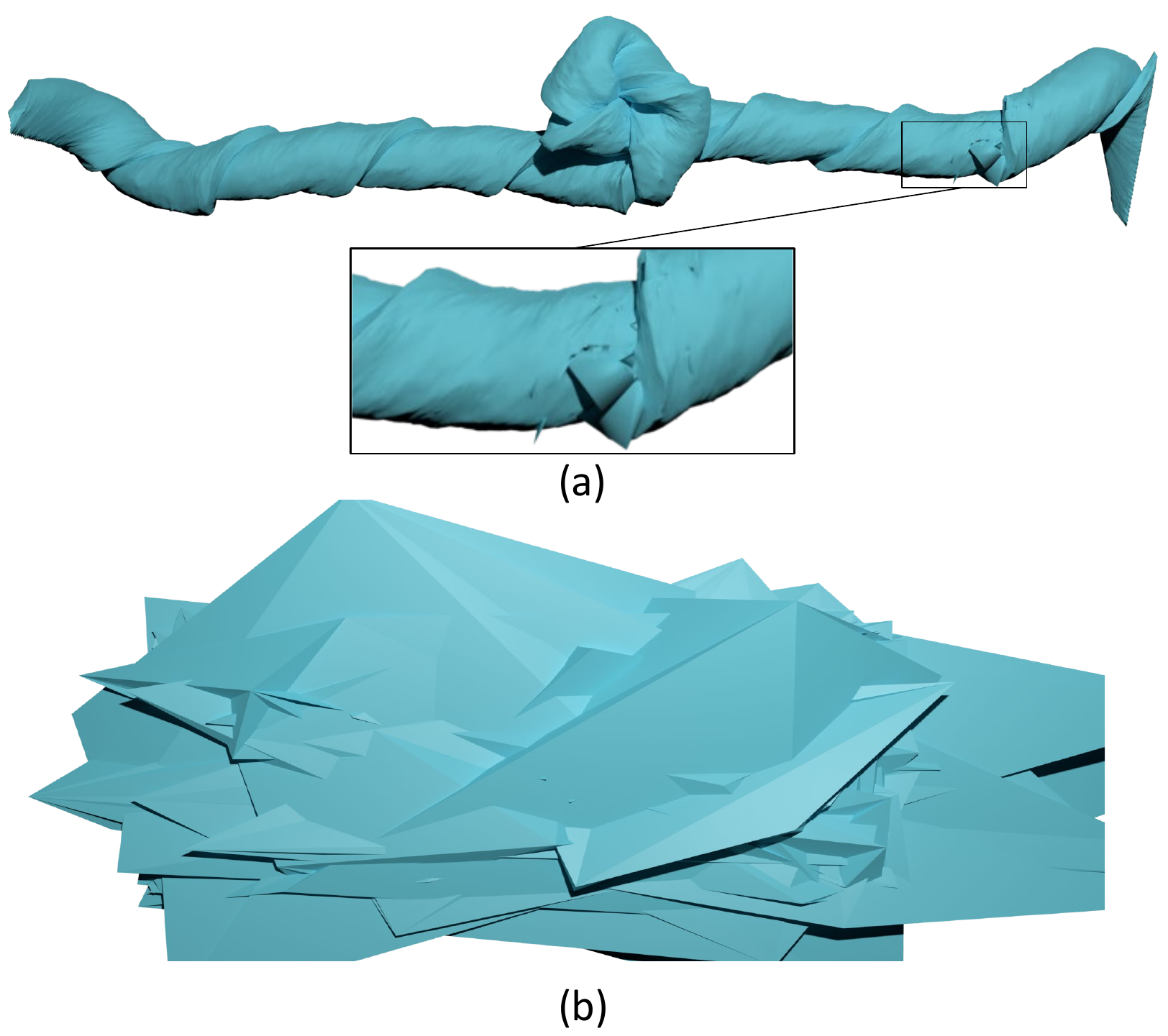}
\caption{Simulation of twisting a thin beam, shown in \autoref{fig:08_ExtremeTwist}, soon after replacing our method with using the rest pose for finding the closest boundary point: (a)~instabilities caused by incorrect closest boundary points found using this approach, and (b)~exploded simulation after a few frames. }
\label{fig:18_RestPoseQueryComparison}
\end{figure}

\input{results_table}

\subsection{Comparisons \add{to Rest Shape Shortest Paths}}

A popular approach in prior work for handling self-collisions is using the rest \del{pose}\add{shape} of the model that does not contain self-collisions for performing the shortest path queries. This makes the computation much simpler, but obviously results in incorrect shortest boundary paths. 
\del{We demonstrate this using two examples: (1)~a simple example in \autoref{fig:185_RestPoseQueryComparison_cuboid} that only involves inter-object collisions and (2)~a complex self-collision example in \autoref{fig:18_RestPoseQueryComparison} that is initially simulated using our method (\autoref{fig:08_ExtremeTwist}) until complex self-collisions are formed. In both cases, incorrect shortest boundary paths found using the rest pose leads to clear instabilities.}\add{With sufficient deformations, these incorrect boundary paths can lead to large enough errors and instabilities.}

\del{In \autoref{fig:185_RestPoseQueryComparison_cuboid}, the}\add{\autoref{fig:185_RestPoseQueryComparison_cuboid} shows a simple example, where a}
small cube \add{is} dropped onto \del{the}\add{a} deformed object\add{. Notice that the rest shape of the object (\autoref{fig:185_RestPoseQueryComparison_cuboid}a) is sufficiently different from the deformed shape (\autoref{fig:185_RestPoseQueryComparison_cuboid}b). With collision handling using this rest shape, the cube} moves against gravity and eventually bounces back \add{(\autoref{fig:185_RestPoseQueryComparison_cuboid}c)}, instead of sliding down the surface, as simulated using our method \add{(\autoref{fig:185_RestPoseQueryComparison_cuboid}d)}. 
\add{
\autoref{fig:185_RestPoseQueryComparison_cuboid}e shows a 2D illustration of example shortest paths generated by both methods. Notice that using the rest shape results in a longer path to the surface that corresponds to higher collision energy. 
In contrast, our method minimizes the collision energy 
by using the actual
shortest path to the boundary.}

\del{In \autoref{fig:18_RestPoseQueryComparison},}\add{\autoref{fig:18_RestPoseQueryComparison} shows a more complex example with self-collisions that is initially simulated using our method (\autoref{fig:08_ExtremeTwist}) until complex self-collisions are formed. When we switch to using the rest shape to find the boundary paths,} the simulation explodes following a number of incorrectly-handled self-collisions. 

In general, using the rest \del{pose}\add{shape} not only generates incorrect shortest boundary paths, but also injects energy into the simulation. This is because an incorrect shortest boundary path is, by definition, longer than the actual shortest boundary path, thereby corresponds to higher potential energy.

%% file: results_table.tex
\begin{table*}\centering
\renewcommand\figureautorefname{Fig.}
\newcommand{\m}[1]{\add{#1}}
\newcommand{\namea}{Flattened Squishy Ball     }\newcommand{\figa}{(\autoref{fig:SqueezingBall})                          }
\newcommand{\nameb}{Twisted Thin Beam          }\newcommand{\figb}{(\autoref{fig:08_ExtremeTwist})                        }
\newcommand{\namec}{Twisted Rods               }\newcommand{\figc}{(\autoref{fig:08_ExtremeTwist})                        }
\newcommand{\named}{Nested Knots               }\newcommand{\figd}{(\autoref{fig:09_Knots})                               }
\newcommand{\namee}{2 Squishy Balls            }\newcommand{\fige}{(\autoref{fig:10_2SquishyballColliding})               }
\newcommand{\namef}{Pre-Intersect. Noodle      }\newcommand{\figf}{(\autoref{fig:12_SovlingExistingCollision_Spaghetti})  }
\newcommand{\nameg}{Pre-Intersect. Squishy Ball}\newcommand{\figg}{(\autoref{fig:13_SovlingExistingCollision_Squishyball})}
\newcommand{\nameh}{600 Octopi                 }\newcommand{\figh}{(\autoref{fig:14_OctopiTest})                          }
\newcommand{\namei}{16 Squishy Balls           }\newcommand{\figi}{(\autoref{fig:15_16SquishyballsTest})                  }
\newcommand{\namej}{Long Noodle                }\newcommand{\figj}{(\autoref{fig:16_200MSpaghetti})                       }
\newcommand{\namek}{8 Octopi CCD Only          }\newcommand{\figk}{(\autoref{fig:19_CCD_DCD_Comparison}a)                 }
\newcommand{\namel}{8 Octopi DCD Only          }\newcommand{\figl}{(\autoref{fig:19_CCD_DCD_Comparison}b)                 }
\newcommand{\namem}{8 Octopi hybrid            }\newcommand{\figm}{(\autoref{fig:19_CCD_DCD_Comparison}c)                 }
\caption{Performance results. \add{Time step size and frame times are given in seconds, where frame times are measured at 60 FPS. Operations Q., Tr., and Tet. represent the number of BVH queries, traversals, and total tetrahedra visited on average per time step, respectively.}}
\label{tbl:perf}
\resizebox{\linewidth}{!}{%
\begin{tabular}{|ll||r|r||r|r||r|r|r||r||r|r||r|r|r|r|}
\hline
&&\multicolumn{2}{c||}{Number of}&\multicolumn{2}{c||}{\m{Avrg. Collisions}}&\multicolumn{3}{c||}{\m{Avrg. Operations}}&\multicolumn{1}{c||}{Time Step}&\multicolumn{2}{c||}{Frame Time}&\multicolumn{4}{c|}{Average Time \%}\\
&&Vert.&Tet.&\m{CCD}&\m{DCD}&\m{Q.}&\m{Tr.}&\m{Tet.}&\multicolumn{1}{c||}{Size $\times$ Iter.}&Avrg.&Max.& XPBD & CCD & DCD & \textbf{Ours} \\
\hline
\namea&\figa&  774 K & 2.81 M & \m{ 16.8 K} & \m{ 7.1 K} & \m{ 56} & \m{ 6.6} & \m{ 5.2} & 3.3e-4 $\times$ 3 & 10.89 & 18.04 & 30.9 \% & 29.0 \% & 31.7 \% & \textbf{ 8.3 \%} \\
\nameb&\figb&  400 K &  1.9 M & \m{  8.3 K} & \m{ 3.1 K} & \m{ 45} & \m{ 5.7} & \m{ 7.2} & 3.3e-4 $\times$ 3 &  8.16 & 15.92 & 29.7 \% & 31.3 \% & 32.1 \% & \textbf{ 6.9 \%} \\
\namec&\figc&  281 K &  1.3 M & \m{  4.8 K} & \m{ 2.6 K} & \m{ 33} & \m{ 5.3} & \m{ 4.7} & 3.3e-4 $\times$ 3 &  5.25 & 11.17 & 42.1 \% & 26.9 \% & 27.0 \% & \textbf{ 4.0 \%} \\
\named&\figd& 38.1 K &  103 K & \m{  3.1 K} & \m{ 0.6 K} & \m{ 31} & \m{ 4.2} & \m{ 4.4} & 5.5e-4 $\times$ 3 &  0.25 &  0.32 & 61.8 \% & 23.3 \% &  9.5 \% & \textbf{ 5.2 \%} \\
\namee&\fige&  418 K &  1.4 M & \m{ 22.4 K} & \m{ 1.3 K} & \m{ 36} & \m{ 9.6} & \m{11.3} & 3.3e-4 $\times$ 3 &  1.96 &  2.87 & 52.2 \% & 28.4 \% & 18.5 \% & \textbf{10.9 \%} \\
\namef&\figf&   40 K &  110 K & \m{    N/A} & \m{15.2 K} & \m{ 65} & \m{12.0} & \m{13.6} & 8.3e-4 $\times$ 3 &  0.21 &  0.45 & 51.6 \% & 17.6 \% & 18.4 \% & \textbf{12.4 \%} \\
\nameg&\figg&  219 K &  704 K & \m{    N/A} & \m{45.8 K} & \m{ 89} & \m{12.0} & \m{14.0} & 3.3e-4 $\times$ 3 &  1.54 &  2.63 & 44.3 \% & 18.2 \% & 19.1 \% & \textbf{18.4 \%} \\
\nameh&\figh&  3.1 M & 8.88 M & \m{104.0 K} & \m{6.4 K} & \m{ 12} & \m{ 3.6} & \m{ 4.1} & 8.3e-4 $\times$ 3 & 16.40 & 17.90 & 68.3 \% & 15.4 \% & 13.4 \% & \textbf{ 2.9 \%} \\
\namei&\figi&  3.5 M & 11.2 M & \m{118.5 K} & \m{8.5 K} & \m{ 29} & \m{ 4.5} & \m{ 6.6} & 3.3e-4 $\times$ 3 & 18.50 & 20.20 & 49.3 \% & 25.0 \% & 21.8 \% & \textbf{ 3.9 \%} \\
\namej&\figj&  860 K & 2.29 M & \m{102.6 K} & \m{6.1 K} & \m{ 11} & \m{ 3.6} & \m{ 3.2} & 8.3e-4 $\times$ 3 &  4.10 &  4.50 & 67.6 \% & 14.8 \% & 14.9 \% & \textbf{ 2.7 \%} \\
\namek&\figk&  40 K  & 118 K  & \m{  2.1 K} & \m{   N/A} & \m{N/A} & \m{ N/A} & \m{ N/A} & 3.3e-3 $\times$ 5 & 0.028 & 0.036 & 86.6 \% & 13.4 \% & N/A     &  N/A \\
\namel&\figl&  40 K  & 118 K  & \m{    N/A} & \m{ 2.4 K} & \m{ 13} & \m{ 3.7} & \m{ 4.1} & 3.3e-3 $\times$ 5 & 0.038 & 0.045 & 79.2 \% &  N/A    & 10.7 \% & \textbf{10.1 \%} \\
\namem&\figm&  40 K  & 118 K  & \m{  2.3 K} & \m{ 0.2 K} & \m{ 11} & \m{ 3.3} & \m{ 3.9} & 3.3e-3 $\times$ 5 & 0.035 & 0.038 & 79.3 \% & 10.1 \% &  9.6 \% & \textbf{ 1.0 \%} \\
\hline
\end{tabular}%
}
\end{table*}

%% file: 07_Limitation.tex
\section{Discussion}

An important advantage of our method is that it can work with simulation systems that do not provide any guarantees about resolving collisions. Therefore, we can use fast simulation techniques like XPBD to handle complex scenarios involving numerous self-collisions, as demonstrated above.

Yet, our method cannot handle all types of self-collisions and it
requires a volumetric mesh. We cannot handle collisions of codimensional objects, such as cloth or strands. 
Our method would also have difficulties handling meshes with thin volumes or no interior elements.

Our method is essentially a shortest boundary path computation method.
It is based on the fact that an interior point's shortest path to the boundary is always a line segment. This assumption always holds for objects like tetrahedral meshes in 3D or triangular mesh in 2D  Euclidean space. 
Therefore, our method cannot handle shortest boundary paths in non-Euclidean spaces, such as geodesic paths on surfaces in 3D.

Using our method for collision handling with DCD inherits the limitations of DCD. For example,
when with large time steps and sufficiently fast motion, penetration can get too deep, and the shortest boundary path may be on the other side of the penetrated model, causing undesirable collision handling. In practice, 
this problem can be efficiently solved by coupling CCD and DCD, 
as we demonstrate with our results above.



%% file: Appendix.tex
\appendix

\section{Implementation}

\subsection{Algorithms}
We show the pseudocode of our algorithm  to determine whether a line segment is a valid path in \autoref{alg:TetrahedralTraverse}, the ray-triangle intersection algorithm in \autoref{alg:ExitFaceSelection}, the shortest path query algorithm in \autoref{alg:ShortestPathToSurface} and the infeasible region culling algorithm in \autoref{alg:FeasibleRegionCheck}. We provide the proof of theorems in \autoref{sec:proof}.

\begin{algorithm}[t]
\LinesNumbered
\SetAlgoNoLine
\KwIn{
 $\mathbf{s}$: A surface point of $M$, 
$t_0$: the tetrahedron that $\mathbf{s}$ belongs to,
$\mathbf{p}$: the target internal point,
$f_0$: surface triangle containing $\mathbf{s}$, the deformed model $M$. 
}
\KwOut{Boolean value representing whether there is a valid tetrahedral traverse between $\mathbf{s}$ and $\mathbf{p}$.}
$\mathcal{T}=\emptyset$ \;
$\mathbf{r}=\mathbf{b}-\mathbf{a}$ \;
$\mathbf{u}, \mathbf{v}=$generateRayCoordinateSystem($\mathbf{r}$) \cite{duff2017building}\;
$F=$exitFaceSelection($t_0$, $f_0$, $\mathbf{u}$, $\mathbf{v}$, $\mathbf{s}$);\; 
$T=\emptyset$ \;
\For{$f$ in $F$}{
    $t=$the tetrahedron on the other side of $f$\;
    $T$.push\_back($t$)\;
}

\While{$F\neq \emptyset$}{
    $f= F$.back()\;
    $F$.pop\_back()\;
    $t=T$.back()\;
    $T$.pop\_back()\;

    \uIf{$t \in \mathcal{T}$}{
        continue\;
    }\uElse{
        $\mathcal{T}=\mathcal{T}\cup \{t\}$\;
    
    }
    \uIf{$\mathbf{p} \in t$}{
        return True\;
    }
    
    \uIf {$f$ is a surface triangle}{
        return False;
    }

    $\mathbf{c} = $the intersecting point on $f$\;
    \uIf{intersection free}{
        \uIf{$| \mathbf{c} - \mathbf{s}|$  > $| \mathbf{p} - \mathbf{s}|$} {
            return False;
        }
    }

    $F_{\text{new}}$ = ExitFaceSelection($t$, $f$, $\mathbf{u}$, $\mathbf{v}$, $\mathbf{s}$)\;
    \For{$f'$ in $F_\text{new}$}{
        $t=$the tetrahedron on the other side of $f'$\;
        $T$.push\_back($t$)\;
        $F$.push\_back($f'$)\;
        
    }    
    return False;
}
\caption{TetrahedralTraverse}
\label{alg:TetrahedralTraverse}

\end{algorithm}
\begin{algorithm}[t]
\LinesNumbered
\SetAlgoNoLine
\KwIn{$t$: the current tetrahedron; $f$: the incoming face of the ray; $\mathbf{u}$, $\mathbf{v}$: coordinate basis; $\mathbf{s}$: ray origin. }
\KwOut{possible exiting faces of the ray.}
$\mathbf{p}_0, \mathbf{p}_1, \mathbf{p}_2$ = 3 vertices on $f$, ordered according to $f$'s orientation\;
$\mathbf{p}_3$ = the vertex of $t$ that is not on $f$\;
$\mathbf{p'}_0, \mathbf{p'}_1, \mathbf{p'}_2, \mathbf{p'}_3$ = projectTo2D($\mathbf{p}_0\sim \mathbf{p}_3$, $\mathbf{u}$, $\mathbf{v}$, $\mathbf{s}$)\;

sign = signOf(($\mathbf{p'}_1 - \mathbf{p'}_0)\times(\mathbf{p'}_2 - \mathbf{p'}_0)$)\;
$F_\text{exit} = \emptyset$ \;
\uIf{sign$\cdot$det($\mathbf{p'}_3, \mathbf{p'}_1)\geq -\epsilon_i$ and sign$\cdot$det($\mathbf{p'}_3, \mathbf{p'}_2)\leq \epsilon_i$} {
    $F_\text{exit} = F_\text{exit} \cup \{ f_0 \}$
}
\uIf{sign$\cdot$det($\mathbf{p'}_3, \mathbf{p'}_2)\geq -\epsilon_i$ and sign$\cdot$det($\mathbf{p'}_3, \mathbf{p'}_0)\leq \epsilon_i$} {
    $F_\text{exit} = F_\text{exit} \cup \{ f_1 \}$
}
\uIf{sign$\cdot$det($\mathbf{p'}_3, \mathbf{p'}_0)\geq -\epsilon_i$ and sign$\cdot$det($\mathbf{p'}_3, \mathbf{p'}_1)\leq \epsilon_i$} {
    $F_\text{exit} = F_\text{exit} \cup \{ f_2 \}$
}

return $F_\text{exit}$
\caption{ExitFaceSelection}
\label{alg:ExitFaceSelection}
\end{algorithm}

\subsubsection{Explanation of the Tetrahedral Traverse Algorithm}
\begin{figure}
    \centering
    \includegraphics[width=\columnwidth]{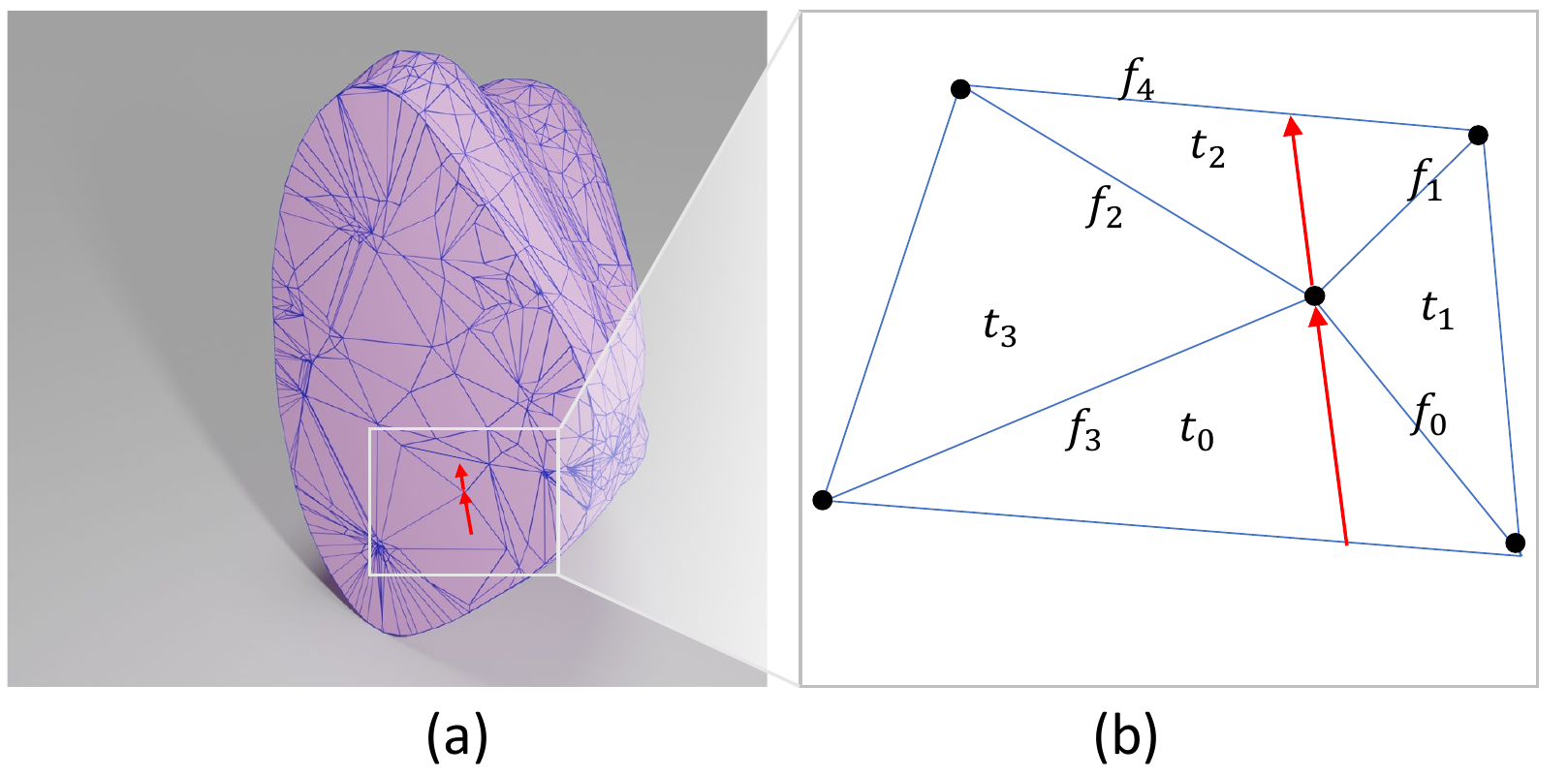}
    \caption{(a) A clip view of a tetrahedral mesh, where a ray (marked by red arrows) is passing exactly through a vertex of a tetrahedron. (b) Zoom in to the section where the ray-vertex intersection happens.}
    \label{fig:TetrahedralTraverseInfLoop}
\end{figure}
Here we discuss some details of \autoref{alg:TetrahedralTraverse}. We import some techniques from the field of tetrahedral traverse based volumetric rendering to accelerate the tetrahedral traverse procedure \cite{aman2022compact}, in which they construct a 2D coordinate system for each ray \cite{duff2017building} and determine the ray-triangle intersection based on it. 
This drastically reduced the number of arithmetic operations. 
Nevertheless, there are some robustness issues associated with tetrahedral traverse that are still unsolved, such as dead ends and infinite loops.
In \cite{aman2022compact}, they just discard the ray if it forms a loop because one ray does not matter much among billions of rays running in parallel. In our case though, we can not do that because that very ray may lead to the actual global geodesic path we are looking for. Instead, we try to recover from an earlier state and get out of the loop the other way.
In addition, since we need to handle the case of the inverted tetrahedron and the case of the ray going backward, we modified their 2D ray-triangle intersecting algorithm to take the orientation of the incoming into face consideration, see \autoref{alg:ExitFaceSelection}.

We found that the tetrahedral traverse only forms a loop when the ray is hitting near a vertex or edge of the tetrahedron.
As illustrated by \autoref{fig:TetrahedralTraverseInfLoop}, the ray is trying to get out of tetrahedron $t_0$. 
However, the ray is intersecting with a vertex of $t_0$, and the ray-triangle intersection algorithm is determining the ray intersecting with $f_0$. Then the algorithm will go to tetrahedron $t_1$, and similarly it goes to $t_2$ and $t_3$ through $f_1$ and $f_2$. However, when at $t_3$, the ray-triangle intersection algorithm can determine the ray intersects with $f_3$ and put the algorithm back to $t_0$ and thus forms an infinite loop: $t_0 \rightarrow t_1  \rightarrow t_2  \rightarrow t_3  \rightarrow t_0 \rightarrow \dots$, and the algorithm will be stuck at this vertex.
To determine whether the traverse has formed a loop, we maintain a set $\mathcal{T}$ that records all the traversed tetrahedra, as shown in line 1 of \autoref{alg:TetrahedralTraverse}, and do loop check every iteration.

After the loop is detected, we need to be able to  recover from an earlier state where the loop has not been formed. 
This is achieved by managing a candidate intersecting face stack $F$, see \autoref{alg:TetrahedralTraverse}.
As shown in \autoref{alg:ExitFaceSelection}, we use a constant positive parameter $\epsilon_i$ to relax the ray-triangle intersection. We regard a ray as intersecting with a face if its close enough to its boundary. Thus a ray can intersect with multiple faces of a tet.
For each tetrahedron we traverse through, we find all its intersecting faces except the incoming face and put them to the stack $F$, as shown in line 28\textasciitilde32 of \autoref{alg:TetrahedralTraverse}. We also manage another stack $T$ to store the tetrahedron on the other side of the face, which always has the same size as $F$. When the ray intersects with one of the vertices of the tetrahedron, we can end up putting more than one face to $F$.
Then we have a while loop that pops a face at each iteration from the end of $F$ and goes to the tetrahedron on the other side of $f$ by popping the tetrahedron at the end of $T$, and we repeat this procedure to add more faces and tetrahedra to the stack.
Since each time the algorithm will only pop the newest element from the stack, the traverse will perform a depth-first-search-like behavior when no loop is detected. 

In the case of a loop is detected, the algorithm will pick an intersecting face from $F$ and a tetrahedron from $T$ which are added from a previously visited tetrahedron and continue going, see lines 11\textasciitilde14 of \autoref{alg:TetrahedralTraverse}. 
Since the ray is not allowed to go back to a tetrahedron that it has visited, this guarantees the algorithm will not fall into an infinite loop.

\subsubsection{Finding the Shortest Path to the Surface.} 
We propose an efficient algorithm that searches for the shortest path to the surface for an interior point $\mathbf{p}$ inside a tetrahedral mesh, see \autoref{alg:ShortestPathToSurface}. With a spatial partition algorithm, we can partition all the surface elements (i.e., all the surface triangles) of $M$ using a spatial partition structure (e.g., bounding box hierarchy). Then we can start a point query centered at $\mathbf{p}$ using an infinite query radius, see line 1 and line 4 of \autoref{alg:ShortestPathToSurface}, which will return all the surface elements sequentially in an approximately close-to-far order. For each surface triangle $f$, we can compute $\mathbf{p}$'s Euclidean closest point on $f$, which we denote as $\mathbf{s}$. Then we can try to build a tetrahedral traverse from  $\mathbf{s}$ to  $\mathbf{p}$ using \autoref{alg:ShortestPathToSurface}. Once we successfully find a tetrahedral traverse, there is no need to look for a surface point further that $| \mathbf{s} - \mathbf{p} |$. Thus we can reduce the query radius to be $| \mathbf{s} - \mathbf{p} |$ every time we find a valid tetrahedral traverse for a surface point $\mathbf{s}$ that has a smaller distance to $\mathbf{p}$ than the current query radius, and continue querying, see the line 9 of \autoref{alg:ShortestPathToSurface}. The query will stop after all the surface elements within the query radius are examined. 
When the query stops, the last surface point that triggers the query radius update is the closest surface point to $\mathbf{p}$ that exists a tetrahedral traverse to allow $l(\mathbf{s}, \mathbf{p})$ to embedded to. According to Theorem 1 in the paper, $l(\mathbf{s}, \mathbf{p})$ is the $\mathbf{p}$'s shortest path to boundary.

When $\Psi$ is causing no inversions and non-degenerate, \autoref{alg:ShortestPathToSurface} is guaranteed to find the shortest path to the surface in a tetrahedral mesh with self-intersection.
\begin{algorithm}[t]
\LinesNumbered
\SetAlgoNoLine
\KwIn{An interior point $\mathbf{p}$, a penetrated surface point $\mathbf{s}$ that is overlapping with  $\mathbf{p}$. }
\KwOut{$\mathbf{p}$'s shortest path to the surface.}
$r = inf$\;
$\mathcal{S} = \{\text{all the surface triangles} f\}$\;

\While{$S\neq \emptyset$}{
    $f =$ do primitives query for $\mathcal{S}$ centered at $\mathbf{p}$\ with radius $r$;

    $\mathbf{s'} =$ $\mathbf{p}$'s Euclidean closest point on $f$\;
    \uIf{$\mathbf{s}=\mathbf{s}'$}{
        continue\;
    }
    \uIf{TetrahedralTraverse($\mathbf{s'}$, $\mathbf{p}$, $f$)}{
        $r=\|\mathbf{s'}-\mathbf{p}\|$\;
        $\mathcal{S} = \mathcal{S} \setminus \{f\} $ \;
        $\mathbf{s}_\text{closest} = \mathbf{s}$ \;
    }

}

return $l(\mathbf{s}_\text{closest}, \mathbf{p})$
\caption{ShortestPathToSurface}
\label{alg:ShortestPathToSurface}
\end{algorithm}

\begin{algorithm}[t]
\LinesNumbered
\SetAlgoNoLine
\KwIn{A surface point $\mathbf{s} \in M$, the closest type of $\mathbf{s}$, an interior point $\mathbf{p}$. }
\KwOut{Boolen value representing whether there $\mathbf{p}$ is at $\mathbf{s}$'s feasible region.}
\Switch {ClosestType}
{
    \Case {At the interior}{
        return True;
    }
    \Case {On an edge}{
        find two endpoints of the edge: $\mathbf{a, b}$;\\
        find two neighbor faces of the edge: $f_1, f_2$;\\

        \uIf {$(\mathbf{p} - \mathbf{a})(\mathbf{b}-\mathbf{a})< \epsilon_r$ }{
                return False;
        }\uElseIf {$(\mathbf{p} - \mathbf{b})(\mathbf{a}-\mathbf{b})<\epsilon_r$ } {
                return False;
        }
        $\mathbf{n}_1 = \text{normal}(f_1)$;\\
        $\mathbf{n}_1^\perp =  \mathbf{(b-a)} \times \mathbf{n}_1$;  \\
        \uIf {$(\mathbf{p} - \mathbf{a})\mathbf{n}_1^\perp < \epsilon_r$ }{
                return False;
        }
        $\mathbf{n}_2 = \text{normal}(f_2)$;\\
        $\mathbf{n}_2^\perp =  \mathbf{(a-b)} \times \mathbf{n}_2$;  \\
        \uIf {$(\mathbf{p} - \mathbf{a})\mathbf{n}_1^\perp < \epsilon_r$ }{
                return False;
    
        }

    }
    \Case{Vertex}{
        \For { $\mathbf{a} \in N_s(\mathbf{s})$}{
            \uIf {$\mathbf{(p-s)(s-a)} < \epsilon_r$}{
                return False;
            }
        }
    }
    return True;
}
\caption{FeasibleRegionCheck 
}
\label{alg:FeasibleRegionCheck}
\end{algorithm}
\subsection{Implementation and Acceleration}
We have two implementations of the tetrahedral traverse algorithm: the static version and the dynamic version.
In the static version of the implementation, the candidate intersecting face stack $F$, the candidate next tetrahedron stack $T$, and the traversed tetrahedra list $\mathcal{T}$ are all aligned static array  on stack memory, which best utilizes the cache and SMID instructions of the processor. 
Of course, the static version of the algorithm will fail when the members in those arrays have exceeded their capacity. In those cases, the dynamic version will act as a fail-safe mechanism. The dynamic version of the traverse algorithm supports dynamic memory allocation thus those arrays can change size on the runtime. By carefully choosing the size of those static arrays, we can make sure most of the tetrahedral traverse is handled by the static implementation, optimizing efficiency and memory usage.

We also add some acceleration tricks to the implementations. First, we find that all the infinite loops encountered by the tetrahedral traverse only contain a few tetrahedra. Actually, the size of the loop is unbounded by the max number of tetrahedra adjacent to a vertex/edge. In practice, the number is even much smaller than that. Thus it is unnecessary to keep all the traversed tetrahedra in an array and check them at every step for the loop. Instead, we only keep the newest 16 traversed tetrahedra. In all our experiments, we have never encountered a loop with a size larger than 16.
$\mathcal{T}$ is implemented as a circular array and the oldest member will be automatically overwritten by the newest member. The array of size 16 can be efficiently examined by the SIMD instructions. This modification reduced about 10\% of the running time of our method.
In addition, in the presence of inversions, we will stop the ray after it has passed 2 times of the distance between $\vec{p}$ and $\vec{s}$, instead of waiting for it to reach the boundary. This procedure reduced 20\% of our running time.

We would like to point out that the shortest path query for each penetrated point is embarrassingly parallel because the process can be done completely independently. All the querying threads will share the BVH structure of the surface and the topological structure of the tetrahedral mesh. During the execution no modification to those data is needed, thus no communication is needed between those threads. Also, our shortest path querying algorithm, especially the version with static arrays, can be efficiently executed on GPU.

\subsection{Collision Handling Framework} 
\add{
We provide the pseudocode of our XPBD framework (see \autoref{alg:XPBDFramework}) and our implicit Euler framework \autoref{alg:XPBDFramework}. 
At the beginning of both frameworks, we separate all the surface points (edges can also be included) into two categories: initially penetrated points $\mathcal{P}_\text{in}$ and initially penetration-free points $\mathcal{P}_\text{out}$.
For the XPBD framework, we apply collision constraints at the end of each time step, after the object has been moved by the material and the external force solving. Thus another DCD must be applied to every point in $\mathcal{P}_\text{in}$ to re-detect the tetrahedra including it for the purpose of shortest path query. 
In the implicit Euler framework, we only need to do DCD once because the collisions are built into the system as a penalty force and solved along with other forces. We use friction power that points to the opposite direction of the velocity and is proportional to the contact force. 

Note that when solving models with existing significant self-intersections, we turn off CCD and set $\mathcal{P}_\text{out}=\emptyset$. $\mathcal{P}_\text{in}$ will not only contains surface points, but also all the interior vertices and all tetrahedra's centroids to resolve the intersection in the completely overlapping parts. 
\begin{algorithm}[t]
\LinesNumbered
\SetAlgoNoLine
\caption{One XPBD Time Step with Collision Handling}
\LinesNumbered
\SetAlgoNoLine
\KwIn{Position of the previous step: $\vec{x}_0$, velocity of the the previous step: $\vec{v}_0$, external force $\vec{f}$.
}
\KwOut{Position $\vec{x}$ and $\vec{v}$ velocity of the new time step.}

update $\mathcal{P}_\text{in}$ and $\mathcal{P}_\text{out}$ by DiscreteCollisionDetection\;

$\vec{v} = \vec{v}_0 + \Delta t \vec{f}$ \;
$\vec{x} = \vec{x}_0 + \Delta t \vec{v}$ \;

XPBD material solve \;

$\mathcal{C}=\emptyset$ \;

\SetKwBlock{DoParallel}{do in parallel}{end}
    \DoParallel{
        \For{$\vec{s} \in \mathcal{P}_\text{in}$}{
            \uIf { DiscreteCollisionDetection($\vec{s}$) }{
                $\vec{p}$ = $\vec{s}$'s overlapping interior point\;
                $\vec{s}_\text{closest}$ =  ShortestPathToSurface($\vec{p}$, $\vec{s}$)\;
                $\mathcal{C}$.add(CollisionConstraint($\vec{s}$, $\vec{s}_\text{closest}$)) \;
            }
        }
        \For{$\vec{s} \in \mathcal{P}_\text{out}$}{
            \uIf { ContinuousCollisionDetection($\vec{s}$) }{
                $\vec{s}_\text{collide}$ =  the surface point $\vec{s}$ collide with\;
                $\mathcal{C}$.add(CollisionConstraint($\vec{s}$, $\vec{s}_\text{collide}$)) \;
            }
        }
    }
    
\For{$c \in \mathcal{C}$}{
    project constraint c\;
}

$\vec{v} = (\Vec{x} - \vec{x}_0)/\Delta t$

\label{alg:XPBDFramework}
\end{algorithm}
}

\add{

\begin{algorithm}[t]
\LinesNumbered
\SetAlgoNoLine
\caption{One Implicit Euler Time Step with Collision Handling}
\LinesNumbered
\SetAlgoNoLine
\KwIn{Position of the previous step: $\vec{x}_0$, velocity of the the previous step: $\vec{v}_0$, external force $\vec{f}$.

}
\KwOut{Position $\vec{x}$ and $\vec{v}$ velocity of the new time step.}
update $\mathcal{P}_\text{in}$ and $\mathcal{P}_\text{out}$ by DiscreteCollisionDetection\;

$\mathcal{C}=\emptyset$ \;

\SetKwBlock{DoParallel}{do in parallel}{end}
    \DoParallel{
        \For{$\vec{s} \in \mathcal{P}_\text{in}$}{
            $\vec{p}$ = $\vec{s}$'s overlapping interior point\;
            $\vec{s}_\text{closest}$ =  ShortestPathToSurface($\vec{p}$, $\vec{s}$)\;
            $\mathcal{C}$.add(CollisionConstraint($\vec{s}$, $\vec{s}_\text{closest}$)) \;
        }
        \For{$\vec{s} \in \mathcal{P}_\text{out}$}{
            \uIf { ContinuousCollisionDetection($\vec{s}$) }{
                $\vec{s}_\text{collide}$ =  the surface point $\vec{s}$ collide with\;
                $\mathcal{C}$.add(CollisionConstraint($\vec{s}$, $\vec{s}_\text{collide}$)) \;
            }
        }
    }
    
Evaluate penetration force and its Jacobian (i.e. the Hessian of the penetration energy) \;
Evaluate internal force  and its Jacobian (i.e. the Hessian of the elasticity energy)\;

Evaluate external forces \;
Build and solve implicit Euler system to obtain $\vec{x}$ and $\vec{v}$ \cite{baraff1998large} \;

\label{alg:ShortestPathToSurface}
\end{algorithm}
}

\section{Proof of Theorems}
\label{sec:proof}
\begin{figure}
    \centering
    \includegraphics[width=\columnwidth]{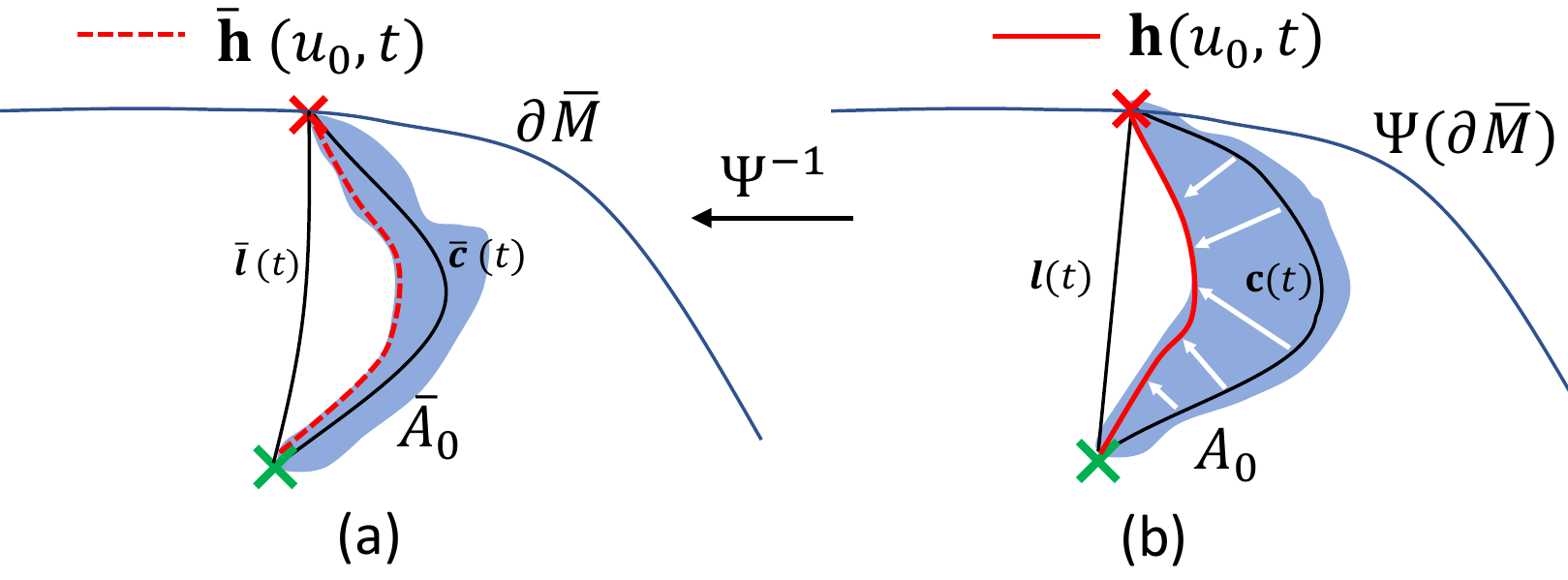}
    \caption{Constructing the undeformed pose curve. (a) The undeformed pose. (b) The deformed pose. }
    \label{fig:Sup_01_CurveCluster}
\end{figure}
\subsection{Theorem 1}
We first prove this lemma:
\begin{lemma}
\label{thm:lineSeg}
For any point $\vec{p}\in M$, its shortest path to the boundary as $\rest{\vec{p}}$ is a line segment.
\end{lemma}
\begin{proof}
Let's assume that $\vec{p}$'s shortest path to boundary as $\rest{\vec{p}}$ is not a line segment, and  $\vec{p}$'s closest boundary point as as $\rest{\vec{p}}$ is $\vec{s}$ (as $\rest{\vec{s}}$).
Then there must be a curve on the undeformed pose: $\rest{\vec{c}}(t): I \mapsto \rest{M}, \rest{\vec{c}}(0)=\rest{\vec{p}},  \rest{\vec{c}}(1)=\rest{\vec{s}}$, s.t., $\vec{c}(t) = \Psi(\rest{\vec{c}}(t))$ is $\vec{p}$'s shortest path to boundary.

Since $\nabla\Psi>0$, we know that $\Psi$ is locally bijective. Thus, according to Heine–Borel theorem \cite{borel1895quelques}, we can have a limited number of open sets $\mathcal{A}$ covering $\rest{M}$, such that $\Psi$ is bijective on each open set in $ \mathcal{A}$.
We can select a subset $\mathcal{A}_0\subseteq \mathcal{A}$, such that $\rest{\vec{c}}(I)$ is totally contained by  $\rest{A}_0$= the union of $\mathcal{A}_0$, see \autoref{fig:Sup_01_CurveCluster}a.

We then construct a cluster of curves: $\vec{h}(u, t)=u\vec{c}(t) + (1-u)\vec{l}(t): I\times I \mapsto \rest{M}$, such that, $\vec{h}(0, t) = \vec{c}(t), \vec{h}(1, t) = \vec{l}(t), \forall t\in I$, where $\vec{l}(t)$ is the line segment from $\vec{p}$ to $\vec{s}$, see \autoref{fig:Sup_01_CurveCluster}b.
 $\vec{h}(u, t)$ will smoothly deformed from $\vec{c}(t)$ to $\vec{l}(t)$ as $u$ changes from 0 to 1. Note that any moment $u$, the length of the curve: $\vec{c}_u(t)=\vec{h}(u, \cdot )$ must be shorter than $\vec{c}(t)$.

Since $\vec{c}(I)\in A_0 = \Psi(\rest{A}_0)$, there must exist a $u_0>0$, such that, $\vec{h}([0, u_0], I)\in A_0$. Additionaly, because $\Psi$ is bijective on $A_0$, we can define a undeformed pose curve cluster: $\rest{\vec{h}}(u, t) = \Psi^{-1}(\vec{h}(u, t)): [0, u_0]\times I \mapsto \rest{A}_0$, as shown in  \autoref{fig:Sup_01_CurveCluster}a. 

We can then select another group open set $\mathcal{A}_1$ containing $\rest{\vec{h}}(u_0, I)$, and repeat the above procedure. This will give us another cluster of curves: $\rest{\vec{h}}(u, t) = \Psi^{-1}(\vec{h}(u, t)): [u_0, u_1]\times I \mapsto \rest{A}_0$.  During such process, the curve will not touch the boundary of the model, otherwise, there will be a shorter curve connecting $\rest{\vec{p}}$ and the boundary, which violates our assumption. Since $ \mathcal{A}$ is a limited set, we can eventually obtain a $\rest{\vec{h}}(u, t): [u_k,1]\times I\mapsto \rest{M}$ within a limited $k+1$ steps, such that $\Psi(\rest{\vec{h}}(1, t))=\vec{l}(t), \forall t\in I$, $\rest{\vec{h}}(1, 0)=\rest{\vec{p}}$ and $\rest{\vec{h}}(1, 1)=\rest{\vec{s}}$.

Here we have proved that $\vec{l}(t)$ is a valid path, which must be shorter than $\vec{c}(t)$ due to Euclidean metrics. Hence creating a contradiction. 
\end{proof}

With Lemma \autoref{thm:lineSeg}, the proof of Theorem 1 becomes trivial. Since the shortest path to the boundary must be a valid path, of course it should be the \textit{shortest} valid line segment to boundary.

\subsection{Theorem 2}
The proof of Theorem 2 is similar to Theorem 1. 
Say the closest boundary point is $\vec{s}^\prime \in f$ (as  $\rest{\vec{s}}^\prime$) and the Euclidean closest boundary point on $f$ is $\vec{s}$  (as  $\rest{\vec{s}}$), where $f$ is a boundary face. 
We also construct a cluster of curves:  $\vec{h}(u, t)=u\vec{l}(t) + (1-u)\vec{l}^\prime(t) I\times I \mapsto \rest{M}$, where $\vec{l}^\prime(t)$ and $\vec{l}(t)$ are the line segment from $\vec{p}$ to $\vec{s}^\prime $ and  $\vec{s}$, respectively.
This $\vec{h}(u, t)$ also holds the property that at any given moment $u$, the length of the curve: $\vec{c}_u(t)=\vec{h}(u, \cdot )$ must be shorter than $\vec{l}^\prime(t)$.

Note that instead of fixing two ends, we only fix one end of $\vec{h}(u, t)$, as it deforms from  $\vec{l}^\prime(t)$ to $\vec{l}(t)$. Because $\Psi$ is bijective on each boundary face, we can explicitly construct the line segment on the undeformed pose that goes from  $\rest{\vec{s}}^\prime $ and  $\rest{\vec{s}}$, this allows us to move the position of the end point.

Similar to Lemma 1, we can induce a cluster of curves on the undeformed pose:  $\rest{\vec{h}}(u, t)$, which will give us the pre-image of  $\vec{l}(t)$ as $\rest{\vec{h}}(1, t)$, connecting $\rest{\vec{p}}$ and  $\rest{\vec{s}}$.
Hence we have proven that $\vec{l}(t)$ is also a valid path from  $\rest{\vec{p}}$ to  $\rest{\vec{s}}$. This contradicts the assumption that $\vec{s}^\prime$ is the closest boundary point.

\subsection{Element Traverse and Valid Path}
\label{appendix:TetrahedralTraverse}
We can give an equivalent definition of a curve being a valid path in the discrete case.
\begin{theorem}
\label{thm:TetrahedralTraverse}
 A line segment connecting $\vec{a}\in e_a$ and $\vec{b}\in e_b$ in a mesh, is a valid path if and only it is included by element traverse from  $e_a$ to $e_b$. 
\end{theorem}
\begin{proof}

\textit{Sufficiency}.
If there exists such a element traverse $\mathcal{T}(\vec{a}, \vec{b})=(e_1, e_2, e_3, \dots, e_{k-1}, e_k) $, s.t., $e_{0} = e_a$ and $e_k = e_b$, we can explicitly construct a continuous piece-wise linear curve $\rest{\vec{c}}(t)$ defined on them, whose image is $\vec{c}(t)$. We do this by making a division of $I$: $I=[t_0,t_1]\cup[t_1,t_2]\cup[t_2,t_3]\cup \dots \cup [t_{k-1},t_k] $, where $t_0=0$, $t_k=1, t_0 \leq t_1 \leq t_2 \leq \dots \leq t_k$.
The division can be obtained by making $\frac{t_i}{t_{i+1}}=\frac{|\rest{\vec{r}}_i-\rest{\vec{r}}_{\rest{\vec{r}}+1}|}{|\rest{\vec{r}}_{i+1}-\rest{\vec{r}}_{i+2}|}, \forall i=1,2,\dots, k+1$, where $\rest{\vec{r}}_i=\Psi|_{e_i}^{-1}(\vec{r}_i)$ is the preimage of the line segment's exit point from $e_i$.
The curve can be constructed as:
\begin{equation}
    \rest{\vec{c}}(t)= \frac{t-t_i}{t_{i+1} - t_i} \rest{\vec{r}}_i + (1- \frac{t-t_i}{t_{i+1} - t_i}) \rest{\vec{r}}_{i+1}, \text{if } t \in [t_i, t_{i+1}]
\end{equation}

\textit{Necessity}.
Suppose we have a curve on the undeformed pose $\rest{\vec{c}}(t)$ connecting $\restbf{a}, \restbf{b}$, whose image under $\Psi$ is a line segment.
If  $\rest{\vec{c}}(t)$ passes no vertex of $\rest{M}$, we directly obtain an element traversal by enumerating the elements that $\rest{\vec{c}}(t)$ passes by as $t$ continuously changes from 0 to 1.

When $\rest{\vec{c}}(t)$ passes a vertex $\rest{\vec{v}}$ of $\rest{M}$, assume it goes from $e_i$ to $e_{i+1}$ at that point. According to the definition of a manifold, we can search around that  $\rest{\vec{v}}$ and guarantee to have an element traversal from $e_i$ to $e_{i+1}$ formed by elements adjacent to  $\rest{\vec{v}}$. 

The case of $\rest{\vec{c}}(t)$ passing an edge can be proved similarly.

\end{proof}

\subsection{Further Discussion on Inverted Elements}
As we can see from Figure 6b of the paper, the line segment $\mathbf{s} \mathbf{p}$ is only a subset of such a path constructed by our algorithm.
In fact, in this case, the length of  path $\vec{c}(t)$  is evaluated by this formula:
\begin{equation}
    \int_{0}^{1} sign((\mathbf{s} -\mathbf{p})r'(t) |r'(t)| dt
    \label{eq:InversionMetric}
\end{equation}
which means, in the presence of inverted tetrahedra, that the length of $\vec{c}(t)$ grows as it goes in the direction of $\mathbf{s} -\mathbf{p}$, and decreases if it goes in the opposite direction, which happens when it passes through the inverted tetrahedron. This is understandable because when solving the self-intersection, the penetrated point does not need to go back and forth, it only needs to pass through the overlapping part once. Thus the length of the overlapping part should only count once.
With inverted tetrahedra, \autoref{alg:ShortestPathToSurface} is actually constructing the shortest line segment  connecting $\mathbf{p}$ and a surface point under the metrics introduced by Eq.\ref{eq:InversionMetric}.
